\documentclass[final,3p,times,sort&compress]{elsarticle}

\usepackage{graphicx}   % For including images
\usepackage{amsmath}   % For mathematical symbols and equations
\usepackage{float}
\usepackage[dvipsnames]{xcolor}  % For color definitions
\usepackage{booktabs} % For better table formatting
\usepackage{url}
\usepackage{algorithm}
\usepackage{algorithmic}
\usepackage{tikz} % For drawing diagrams
\usetikzlibrary{arrows.meta}  % Load the arrows.meta library
\usetikzlibrary{shapes.geometric}
\usetikzlibrary{positioning}  % for 'right=of' and layout spacing
\usepackage{pgfplots}
\pgfplotsset{compat=1.18}
\usepackage[labelfont=bf]{caption}
\usepackage{subcaption}

\usepackage{hyperref}
\hypersetup{colorlinks,
  linkcolor=blue,
  citecolor=blue,
  urlcolor=blue}
\newcommand{\cref}[3]{\hyperref[#2]{#1~\ref*{#2}{#3}}}
\newcommand{\colref}[2]{\hyperref[#2]{#1~\ref*{#2}}}

\newcommand{\eqnref}[1]{\colref{Eq.}{#1}}
\newcommand{\figref}[1]{\colref{Figure}{#1}}

\newcommand{\secref}[1]{\colref{Section}{#1}}
\newcommand{\tableref}[1]{\colref{Table}{#1}}
\newcommand{\coloredref}[2]{\hyperref[#2]{#1~\ref*{#2}}}
\newcommand{\coloredsubref}[3]{\hyperref[#2]{#1~\ref*{#2}{#3}}}

\graphicspath{{./}}
\definecolor{myblockfill}{RGB}{231,219,219}  % Light reddish-brown from beaver block

\newif\ifhighlightrevisions
\highlightrevisionsfalse % highlighted copy: true | clean copy: false

\definecolor{revIIcolor}{RGB}{0,102,204}    % Reviewer 2 -- blue
\definecolor{revIIIcolor}{RGB}{204,0,102}   % Reviewer 3 -- magenta

\ifhighlightrevisions
  \newcommand{\RII}[1]{{\color{revIIcolor}#1}}    % Reviewer 2 revisions
  \newcommand{\RIII}[1]{{\color{revIIIcolor}#1}}  % Reviewer 3 revisions
\else
  \newcommand{\RII}[1]{#1}
  \newcommand{\RIII}[1]{#1}
\fi

\makeatletter
\def\bstctlcite{\@ifnextchar[{\@bstctlcite}{\@bstctlcite[@auxout]}}
\def\@bstctlcite[#1]#2{\@bsphack
  \@for\@citeb:=#2\do{%
    \edef\@citeb{\expandafter\@firstofone\@citeb}%
    \if@filesw\immediate\write\csname #1\endcsname{\string\citation{\@citeb}}\fi}%
  \@esphack}
\makeatother

\begin{document}
\bstctlcite{BSTcontrol}

\begin{frontmatter}
\title{
Neural-Network-based Viscosity Closure for Non-Newtonian Multiphase Flows
}

\author[trac]{Suresh Murugaiyan}
\author[ucsb]{Claire L. Nelson}
\author[isu]{Dhruv Gamdha}
\author[ucsb]{Austin Cunniff}
\author[isu]{Cheng-Hau Yang}
\author[ucb]{Abraham Wiletsky}
\author[ucsb]{Kaitlyn W. Dilley}
\author[ucsbme]{Patrick Babb}
\author[ucsb]{Andrew Rhode}
\author[ucsb]{Christopher M. Bates}
\author[ucsb]{Angela A. Pitenis}
\author[ucsb]{\\Michael L. Chabinyc}
\author[trac,isu]{Adarsh Krishnamurthy\texorpdfstring{\corref{cor1}}}
\author[trac,isu]{Baskar Ganapathysubramanian\texorpdfstring{\corref{cor1}}}

\address[trac]{Translational AI Center, Iowa State University, Ames, IA}
\address[isu]{Department of Mechanical Engineering, Iowa State University, Ames, IA}
\address[ucsb]{Materials Department, University of California, Santa Barbara, CA}
\address[ucsbme]{Department of Mechanical Engineering, University of California, Santa Barbara, CA}
\address[ucb]{Department of Chemical and Biological Engineering, University of Colorado Boulder, Boulder, CO}

\cortext[cor1]{Corresponding Authors}

\begin{abstract}
Materials used in polymer-based additive manufacturing processes, such as Digital Light Processing (DLP) and direct ink writing (DIW), typically exhibit non-Newtonian rheology. Carreau--Yasuda and power-law models describe basic shear-thinning and shear-thickening behavior well, but applying them to a new material requires choosing a functional form, deriving it, and re-implementing it inside the flow solver. We present a deployment workflow in which a neural network trained on experimental rheometry data serves as the viscosity closure inside a Cahn--Hilliard--Navier--Stokes (CHNS) finite element solver. \RIII{The learned closure is generalized Newtonian: viscosity depends only on the local shear rate at the current instant.} Lipschitz regularization during training produces smooth viscosity predictions, and the trained network is exported in the Open Neural Network Exchange (ONNX) format and queried by the solver at runtime via the ONNX runtime, without solver modification or network reimplementation. The framework is built on a parallel octree-based adaptive mesh refinement infrastructure that concentrates resolution at the fluid interface. We validate the CHNS solver against benchmark shear-thinning bubble-rise cases from the literature, reproducing reported bubble shapes across varying power-law indices and Weber numbers. We characterized two silicone ink formulations, recorded their rise dynamics in perfluorodecalin on high-speed video, and used the resulting data to test the full workflow. Simulated rise velocities fall within the experimentally measured spread, and the simulated steady-state droplet shape agrees with the observed one. This work contributes to a growing body of literature on integrating neural constitutive closures into multiphysics simulations, and demonstrates a practical path for deploying experimentally trained rheological surrogates inside finite element solvers.
\end{abstract}

\begin{keyword}
Data driven constitutive models\sep
Non-Newtonian fluids\sep
Phase field equations\sep
Adaptive mesh refinement 
\end{keyword}

\end{frontmatter}

\section{Introduction}\label{sec:Introduction}

Multiphase flows involving non-Newtonian materials are central to a broad class of polymer-based additive manufacturing processes. In Digital Light Processing (DLP), patterned light from a projector cures a photosensitive resin layer by layer; the resin must flow into place for each new layer while solidifying under exposure. In direct ink writing (DIW) and embedded 3D printing, a non-Newtonian ink is extruded into a support bath, and the interaction between the ink and the surrounding fluid governs filament shape, placement, and print fidelity. In both cases, the coupling between flow, solidification, and the shear-dependent rheology of the printing material determines print resolution and the mechanical properties of the final part. This coupling is difficult to probe directly during printing. As a controlled proxy, we analyze the buoyant rise of a non-Newtonian silicone ink droplet through a viscous fluid, a configuration that imposes shear histories comparable to those encountered during extrusion- and resin-based printing.

Many empirical constitutive laws have been developed to describe non-Newtonian viscosity. The Carreau--Yasuda~\citep{yasuda1981shear} and power-law~\citep{tadros2011rheology} formulations are the standard tools for shear-thinning and shear-thickening behavior in computational studies, capturing the rheology of many polymeric and soft-matter systems with a small number of interpretable parameters. The fixed, low-dimensional form that gives these models their interpretability also limits them: extending them to viscoelastic, thixotropic, or yield-stress behavior generally requires deriving and implementing a new constitutive law inside the flow solver, and refitting parameters for each new system. \citet{freund2015quantitative} used Bayesian inference to show that good fits to rheological data are insufficient grounds for model selection; models with fewer, well-constrained parameters are more credible than empirical forms that achieve closer fits through added flexibility. The same concern applies to data-driven constitutive surrogates, including neural networks, which carry genuine model-selection risk and require appropriate regularization when used as replacements for parameterized constitutive forms.

The literature on bubble rise in non-Newtonian fluids and on data-driven constitutive modeling has grown substantially, but the two have rarely been coupled: few studies deploy constitutive surrogates trained on experimental rheometry inside high-fidelity multiphase solvers. The remainder of this section reviews three relevant areas: numerical modeling of bubble dynamics in two-phase flow solvers, bubble dynamics in non-Newtonian fluids with established constitutive relations, and recent work on data-driven or machine-learning-assisted constitutive modeling. Bubble dynamics in immiscible fluids has been studied extensively with interface-capturing methods. Two-phase simulations have progressed from qualitative shape comparisons to quantitative benchmarks that track bubble circularity, center-of-mass height, and mean rise velocity, providing reproducible targets for code verification in capillary-sensitive flows~\citep{hysing2009quantitative}. Three-dimensional studies~\citep{bunner1999direct,sussman2000coupled,hua2008numerical,pivello2014fully} have captured complex deformation, instability, and breakup behavior under varied flow conditions. \citet{tripathi2015dynamics} mapped the transitions between axisymmetric rise, skirt formation, path instability, and breakup as functions of the Galilei and E\"otv\"os numbers, and the same group subsequently showed that thermocapillary effects in self-rewetting fluids can reverse or arrest bubble motion~\citep{tripathi2015non}. The Cahn--Hilliard--Navier--Stokes (CHNS) formulation~\citep{chella1996mixing,jacqmin1999calculation,badalassi2003computation} has become a standard diffuse-interface approach for interfacial evolution, coalescence, and breakup with thermodynamic consistency~\citep{khanwale2021energy,khanwale2023projection,rabeh2024modeling}. CHNS solvers paired with adaptive mesh refinement~\citep{berger1989local,schmidmayer2019adaptive,yang2025simulating,yang2025octree} resolve steep gradients at phase boundaries efficiently~\citep{khanwale2021energy}.

The studies above mostly concern Newtonian fluids. In non-Newtonian systems~\citep{chhabra2023bubbles,tsamopoulos2008steady,clift1978drops,zhang2010numerical}, viscosity varies with local shear rate and affects bubble deformation, rise velocity, and interfacial stability. \citet{premlata2017dynamics} studied an air bubble rising in shear-thinning and shear-thickening liquids with the Carreau--Yasuda model, finding that stronger shear-thinning raises rise velocity and reduces deformation, and that higher Galilei numbers increase shape distortion. \citet{zhang2010numerical} combined experiment and simulation for a single bubble rising in shear-thinning Carreau fluids and reported that stronger shear-thinning lowers local viscosity near the bubble surface, enhances rise velocity, and produces oblate shapes with high-viscosity wakes. \citet{pang2018numerical} examined bubbles rising in shear-thinning power-law fluids under varied gravity conditions and showed that a lower power-law index produces a detached high-viscosity region behind the bubble, while higher Galilei and E\"otv\"os numbers produce larger wake structures and stronger deformation. \citet{hu2022numerical} added that the liquid-to-bubble viscosity ratio also affects rise velocity, with high ratios inducing unsteady deformation at low power-law indices.

Data-driven constitutive modeling has emerged as an alternative to fixed parametric forms when the rheology is difficult to capture in closed form. Physics-informed neural networks (PINNs) have been used to infer unknown constitutive relations and material parameters from sparse or indirect flow observations. \citet{tartakovsky2018learning} showed that PINNs can recover nonlinear diffusion and transport relations from limited state measurements without prescribing a functional form. \citet{reyes2021learning} extended this approach to non-Newtonian fluids, learning shear-dependent viscosity distributions from velocity data in three-dimensional flows. For viscoelastic fluids, where stress depends on both kinematics and material memory, \citet{thakur2024viscoelasticnet} introduced ViscoelasticNet, which infers stress fields and material parameters from velocity data alone, even with sparse or noisy inputs.

\citet{mahmoudabadbozchelou2021rheology} introduced Rheology-Informed Neural Networks (RhINNs), which embed systems of ordinary differential equations describing thixotropic and viscoelastic behavior into the network architecture. The same model predicts stress evolution under varied flow protocols and infers hidden material parameters from limited experimental data, providing accurate forward and inverse solutions for thixotropic elasto-viscoplastic fluids even under sparse or noisy inputs. A later extension~\citep{mahmoudabadbozchelou2024unbiased} combines sparse regression with automatic differentiation to construct closed-form constitutive relations from experimental data without relying on a predefined model, identifying the dominant physical terms in a material's stress response and producing compact, thermodynamically consistent equations that generalize across flow protocols. \citet{tucny2024learning} used PINNs to recover effective viscosity functions in rarefied gas flows governed by the generalized Stokes model, learning transport coefficients consistent with kinetic theory in regimes where continuum assumptions fail. \citet{lardy2025inferring} inferred viscoplastic constitutive laws directly from velocity field data, identifying parameters and selecting between Herschel--Bulkley and Carreau models even under noisy or sparse sampling.

The present work develops a deployment workflow for using neural-network viscosity closures inside a Cahn--Hilliard--Navier--Stokes multiphase flow solver. We train a Lipschitz-regularized neural network on experimental rheometry data to serve as a surrogate for shear-dependent viscosity; the regularization ensures the learned function remains smooth enough to use as a constitutive closure in a flow solver. The trained network is exported to the Open Neural Network Exchange (ONNX) format and queried at each iteration of the finite element solver, allowing neural constitutive models to be integrated into existing multiphase flow frameworks without reimplementation or solver modification. We characterized two silicone ink formulations, recorded their rise dynamics in perfluorodecalin on high-speed video, and used the resulting data to test the full workflow; simulated rise velocities fall within the experimentally measured spread across four independent cases.

The paper is organized as follows. \secref{sec:GoverningEquations} presents the governing Cahn--Hilliard--Navier--Stokes equations and their nondimensional form. \secref{sec:NumericalMethodology} describes the solution strategy and the neural-network-based viscosity model, including its integration into the finite element solver via ONNX. \secref{sec:ExperimentalSetUp} details the experimental setup and rheological characterization of the two silicone ink formulations. \secref{sec:Results} presents the validation results and the comparison against experimental droplet rise dynamics. \secref{sec:ConclusionAndFutureWork} summarizes the findings and outlines directions for future work. The image processing pipeline used to extract rise velocities from the high-speed video is described in \ref{sec:ImageProcessingFramework}.

\section{Governing Equations}\label{sec:GoverningEquations}

The governing Cahn-Hilliard-Navier-Stokes~\citep{khanwale2021energy} equations are given in equations~\eqref{eq:1a} to~\eqref{eq:3c}. In these equations, 
$u [ms^{-1}]$ is the velocity vector, 
$\rho [kgm^{-3}]$ is the mixture density,
$J [kgm^{-2}s^{-1}]$ is the diffusive flux,
$\sigma [Nm^{-1}]$ is the surface tension,
$\epsilon [m]$ is the interfacial thickness,
$\phi$ is the dimensionless phase field,
$p [Nm^{-2}]$ is the pressure,
$\eta [N\,s\,m^{-2}]$ is the mixture viscosity,
$g [ms^{-2}]$ is the gravitational acceleration,
$m [kg^{-1} m^3 s]$ is the mobility and
$\mu [Nm^{-2}]$ is the chemical potential.
The quantities $\rho^{+}$ and $\rho^{-}$ denote the densities of the two fluid
phases, while $\eta^{+}$ and $\eta^{-}$ represent their corresponding 
viscosities. The phase field $\phi$ distinguishes the two fluids by taking 
the value $+1$ in one phase and $-1$ in the other, and varies continuously 
across the interface between them. The index $i$ and $j$ denote the
number of spatial dimensions. In the present work, $\eta^{+}$ denotes the viscosity of the non-Newtonian fluid, while $\eta^{-}$ corresponds to that of the Newtonian phase. The non-Newtonian viscosity is evaluated using a neural network trained on rheological data (\eqnref{eq:3a}). The network takes the local shear rate ($\dot{\gamma}[s^{-1}]$) as input and predicts the corresponding viscosity (\eqnref{eq:3b}). The shear rate, in turn, is computed from the strain-rate tensor ($D_{ij}[s^{-1}]$) as defined in \eqnref{eq:3c}.

\begin{subequations}
\begin{align}
	\label{eq:1a} 
	\frac{\partial u_i}{\partial x_i} &= 0 \\[10pt]
	\label{eq:1b}
	\frac{\partial (\rho u_i)}{\partial t} +
	\frac{\partial (\rho u_i u_j)}{\partial x_j} +
	\frac{\partial (J_j u_i)}{\partial x_j} +
	\sigma \epsilon \frac{\partial}{\partial x_j}
     \left( \frac{\partial \phi}{\partial x_i}
     \frac{\partial \phi}{\partial x_j}\right) +
	\frac{\partial p}{\partial x_i} -
	\RIII{\frac{\partial}{\partial x_j}
    \left[ \eta \left( \frac{\partial u_i}{\partial x_j} +
    \frac{\partial u_j}{\partial x_i} \right) \right]}-
	\rho g_i &= 0
\end{align}
\end{subequations}

\begin{subequations}
\begin{align}
	\label{eq:2a}
	\qquad \qquad \qquad \qquad \qquad \qquad \qquad \qquad \qquad \qquad \qquad
	\frac{\partial \phi}{\partial t} + 
	\frac{\partial (u_i \phi)}{\partial x_i} -
	\frac{\partial}{\partial x_i} 
	\left( m \frac{\partial \mu}{\partial x_i}\right) &= 0 \\[10pt]
	\label{eq:2b} 
	\frac{\sigma}{\epsilon} \phi^3 - 
	\frac{\sigma}{\epsilon} \phi -
	\sigma \epsilon \frac{\partial}{\partial x_i}
	\left( \frac{\partial \phi}{\partial x_i}\right) 
	-\mu &= 0
\end{align}
\end{subequations}

\begin{equation*}
	\text{Here,}	\qquad 
	J_j = \frac{(\rho^{-} - \rho^{+})}{2} m \frac{\partial \mu}{\partial x_j}, 
	\qquad
    \rho = \frac{\rho^{+} + \rho^{-}}{2} + 
    \frac{\rho^{+} - \rho^{-}}{2} \phi \qquad \text{and} \qquad
    \eta = \frac{\eta^{+} + \eta^{-}}{2} +
    \frac{\eta^{+} - \eta^{-}}{2} \phi
\end{equation*}

\begin{subequations}
\begin{align}
		\label{eq:3a}
		\eta^{+} &= \mathcal{N}_{\theta}\!\left(\dot{\gamma}\right) \\[10pt]
		\label{eq:3b}
		\dot{\gamma} &= \sqrt{2 \, D_{ij} D_{ij}}\\[10pt]
		\label{eq:3c}
		D_{ij} &= \frac{1}{2} \left( \frac{\partial u_i}{\partial x_j} +
		\frac{\partial u_j}{\partial x_i} \right)
\end{align}
\end{subequations}

\RII{In \eqnref{eq:3a}, $\mathcal{N}_{\theta}$ denotes the
trained feed-forward neural network with parameters $\theta$, which maps the local
scalar shear rate $\dot{\gamma}$ to the viscosity of the non-Newtonian phase. Its
architecture, training procedure, and export format are described in
\secref{sec:NumericalMethodology}.}
\RIII{\eqnref{eq:3a} is a generalized-Newtonian closure: the viscosity depends
only on the instantaneous local shear rate.}

\subsection{Dimensionless form of Cahn–Hilliard–Navier–Stokes equations}\label{sec:NondimensionalForm}

The dimensionless form of the Cahn–Hilliard–Navier–Stokes equations is 
obtained by introducing 
length scale $l_r$,
time scale $t_r$,
velocity scale $u_r$,
density scale $\rho_r$,
diffusive flux scale $J_r$,
% surface tension scale $\sigma_r$,
% interfacial thickness scale $\epsilon_r$,
% phase field scale $\phi_r$,
pressure scale $p_r$,
viscosity scale $\eta_r$,
gravitational acceleration scale $g_r$
% mobility scale $m_r$ 
and chemical potential scale $\mu_r$.
Surface tension $\sigma$, 
interfacial thickness $\epsilon$ 
and mobility $m$ are taken as constants. The following are the derived scales, 
non-dimensional numbers and hence the dimensionless form of the governing
equations, where $Re$ is the Reynolds number, $Fr$ is the Froude number,
$We$ is the Weber number, $Pe$ is the Peclet number and $Cn$ is the Cahn 
number.

\begin{equation*}
	t_r = \frac{l_r}{u_r}, \qquad
	p_r = \frac{\sigma}{l_r}, \qquad
	\mu_r = \frac{\sigma}{l_r}, \qquad
	J_r = \frac{\rho_r m \sigma}{l_r^2}
\end{equation*}

\begin{equation*}
Re = \frac{u_r \rho_r  l_r}{\eta_r}, \qquad
Fr = \frac{u_r}{\sqrt{g_r l_r}}, \qquad
We = \frac{\rho_r \, u_r^2 \, l_r}{\sigma}, \qquad
Pe = \frac{u_r \,l_r^2}{m \, \sigma}, \qquad
Cn = \frac{\epsilon}{l_r}
\end{equation*}

 \begin{subequations}
	\begin{align}
		\label{eq:4a}
		\frac{\partial u_i}{\partial x_i} &= 0 \\[5pt]
		\label{eq:4b}
		\frac{\partial (\rho u_i)}{\partial t} + 
		\frac{\partial (\rho u_i u_j)}{\partial x_j} + 
		\frac{1}{Pe}
		\frac{\partial (J_j u_i)}{\partial x_j} + 
		\frac{Cn}{We} \frac{\partial}{\partial x_j}
		 \left( \frac{\partial \phi}{\partial x_i} 
		 \frac{\partial \phi}{\partial x_j}\right) +
		 \frac{1}{We}
		\frac{\partial p}{\partial x_i} -
		\RIII{\frac{1}{Re}
		\frac{\partial}{\partial x_j}
		\left[ \eta \left( \frac{\partial u_i}{\partial x_j} +
		\frac{\partial u_j}{\partial x_i} \right) \right]}-
		\frac{1}{Fr^2}
		\rho g_i &= 0
\end{align}
\end{subequations}

 \begin{subequations}
	\begin{align}
	\qquad \qquad \qquad \qquad \qquad \qquad \qquad \qquad \qquad \qquad 
	\qquad \qquad \qquad \qquad
		\label{eq:5a}
		\frac{\partial \phi}{\partial t} + 
		\frac{\partial (u_i \phi)}{\partial x_i} -
		\frac{1}{Pe}
		\frac{\partial}{\partial x_i} 
		\left(\frac{\partial \mu}{\partial x_i}\right) &= 0 \\[5pt]
		\label{eq:5b}
		\frac{1}{Cn}
		\phi^3 - 
		\frac{1}{Cn}
		 \phi -
		Cn \frac{\partial}{\partial x_i}
		\left( \frac{\partial \phi}{\partial x_i}\right) 
		-\mu &= 0
\end{align}
\end{subequations}

\begin{equation*}
	\text{Here,}	\qquad 
	J_j = \frac{(\rho^{-} - \rho^{+})}{2} \frac{\partial \mu}{\partial x_j}, 
	\qquad
    \rho = \frac{\rho^{+} + \rho^{-}}{2} + 
    \frac{\rho^{+} - \rho^{-}}{2} \phi \qquad \text{and} \qquad
    \eta = \frac{\eta^{+} + \eta^{-}}{2} +
    \frac{\eta^{+} - \eta^{-}}{2} \phi
\end{equation*}

\begin{equation}\label{eq:6} 
\begin{aligned}
		\eta^{+} &= \mathcal{N}_{\theta}\!\left(\dot{\gamma}\right)\\[5pt]
		\dot{\gamma} &= \sqrt{2 \, D_{ij} D_{ij}}\\[5pt]
		D_{ij} &= \frac{1}{2} \left( \frac{\partial u_i}{\partial x_j} +
		\frac{\partial u_j}{\partial x_i} \right)\\
\end{aligned}
\end{equation}

\section{Numerical Methodology}\label{sec:NumericalMethodology}

\subsection{Solution Strategy}

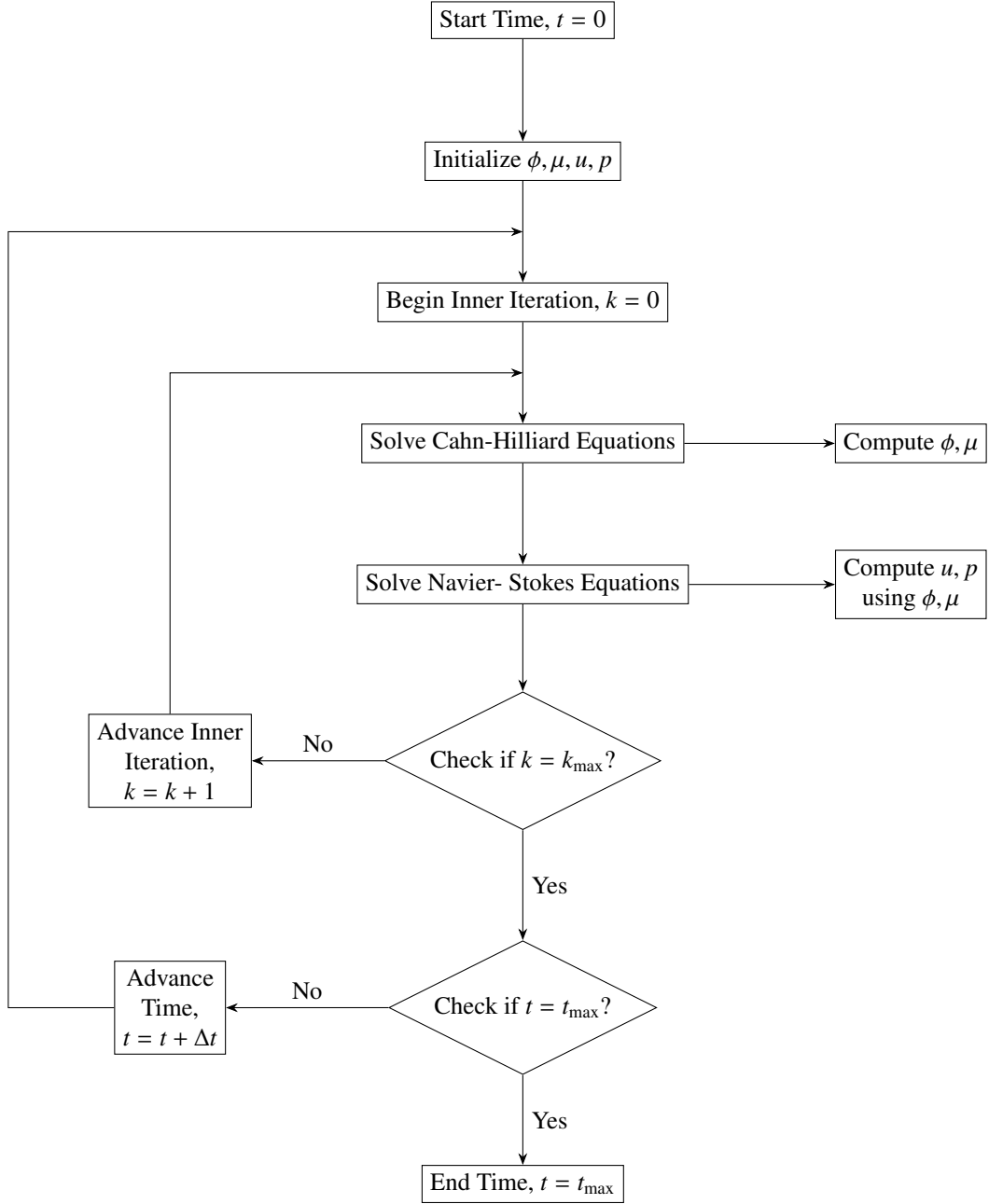
\begin{figure}[t!]
\centering
\begin{tikzpicture}[scale=0.75,>=Stealth,node distance=2cm]

\node[draw, rectangle] (start) {Start Time, $t = 0$};

\node[draw, rectangle] (init) [below of=start] {Initialize $\phi, \mu, u, p$};

\draw[->] (start) -- (init);

\node[draw, rectangle] (startitr) [below of=init] {Begin Inner Iteration, $k = 0$};

\draw[->] (init) -- (startitr) coordinate[midway] (midarrow1);

\node[draw, rectangle] (solvech) [below of=startitr] {Solve Cahn-Hilliard Equations};

\draw[->] (startitr) -- (solvech) coordinate[midway] (midarrow2);

\node[draw, rectangle] (solvens) [below of=solvech] {Solve Navier- Stokes Equations};

\draw[->] (solvech) -- (solvens);

% new node to the right
\node[draw, rectangle, right of=solvens, xshift=3.5cm, align=center] (obtainup) {Compute $u, p$ \\ using $\phi, \mu$ };

\draw[->] (solvens.east) -- (obtainup.west);

\node[draw, rectangle, right of=solvech, xshift=3.5cm, align=center] (obtain) {Compute $\phi, \mu$};
\draw[->] (solvech.east) -- (obtain.west);

% Decision diamond: check k
\node[draw, diamond, aspect=2] (checkk) [below of=solvens, yshift=-0.5cm] {Check if $k = k_{\max}$?};

\draw[->] (solvens) -- (checkk);

% No branch (left) → Advance Inner Iteration, then loop back
\node[draw, rectangle, align=center] (advanceitr) [left of=checkk, xshift=-3cm] {Advance Inner \\ Iteration, \\ $k = k + 1$};
\draw[->] (checkk.west) -- (advanceitr.east) node[midway,above] {No};
\draw[->] (advanceitr.north) |- (midarrow2);

% Yes branch (down) → Check time
\node[draw, diamond, aspect=2] (checkt) [below of=checkk, yshift=-1.5cm] {Check if $t = t_{\max}$?};
\draw[->] (checkk.south) -- (checkt.north) node[midway,right] {Yes};

% No branch (left) → Advance time step → loop back
\node[draw, rectangle, align=center] (advance) [left of=checkt, xshift=-3cm] {Advance \\ Time, \\ $t = t + \Delta t$};
\draw[->] (checkt.west) -- (advance.east) node[midway,above] {No};
\draw[->] (advance.west) -- ++(-2,0) |- (midarrow1);

% Yes branch (down) → End time
\node[draw, rectangle] (end) [below of=checkt, yshift=-0.5cm] {End Time, $t = t_{\max}$};
\draw[->] (checkt.south) -- (end.north) node[midway,right] {Yes};

\end{tikzpicture}
\caption{Flow Chart for the Solution Strategy}\label{fig:solution_strategy}
\end{figure}

The solver builds on the in-house Cahn--Hilliard--Navier--Stokes (CHNS) framework developed by \citet{khanwale2023projection}. The coupled system is advanced with a block-iterative time-stepping strategy: at each time step, the Cahn--Hilliard equations are solved first, followed by the Navier--Stokes equations (\figref{fig:solution_strategy}). The coupling variables from one equation are held constant during the solve of the other.

The Cahn--Hilliard equation is treated as a nonlinear system and solved with the Newton-Raphson method. For the Navier--Stokes system, a projection-based semi-implicit scheme~\citep{khanwale2023projection} is used, with the convection term $\partial_j(u_i u_j)$ linearized~\citep{khara2025semi} using the advecting velocity from the previous time step. The same block-iterative philosophy applies within the Navier--Stokes solve: the velocity prediction, pressure Poisson, and velocity correction equations are solved as separate sub-problems. The projection procedure computes an intermediate velocity using the current pressure, solves a pressure Poisson equation, and corrects the velocity to enforce incompressibility. This decoupling avoids the cost of the fully-coupled saddle-point solve, while preserving second-order accuracy, energy stability, and mass conservation. The residual-based variational multiscale (RBVMS) formulation~\citep{khara2025semi} provides stabilization in advection-dominated regimes. The framework is parallelized with octree-based adaptive mesh refinement (AMR), concentrating resolution near the moving fluid interface. For highly deforming cases such as a rising bubble, the AMR mesh provides a substantial speedup over a uniform fine mesh~\citep{khanwale2023projection}.

\RIII{All fields are discretized with conforming continuous Galerkin finite elements on the octree mesh, using equal-order linear basis functions for velocity, pressure, phase field, and chemical potential. Time integration is second-order semi-implicit: Crank--Nicolson averaging with the advecting velocity extrapolated from the two previous time steps, and a fully implicit treatment of the Cahn--Hilliard equation~\citep{khanwale2023projection}. Two block iterations are performed per time step. A time-step size of $\Delta t = 10^{-3}$ is used for all simulations. The mesh is refined near the interface so that approximately $11$ elements span the diffuse interface in the two-dimensional benchmark cases and $3$ in the three-dimensional droplet-rise cases, and the mesh is re-adapted at every time step.}
%%%%%%%%%%%%%%%%%%%%%%%%%%%%%%%%%%%%%%%%%%%%%%%%%%%%%%%%%%%%%%%%%%%%%%%%%%%%%%%%%%%%%
\subsection{Neural Network for Viscosity Prediction}

A neural network learns the viscosity-shear-rate relation directly from data, without committing to a functional form such as power-law or Carreau--Yasuda. To keep the learned function smooth and bounded in derivative, Lipschitz regularization~\citep{negrini2021system,liu2022learning} is applied during training. Lipschitz continuity bounds how rapidly a function can vary with respect to its input. A function $f(x)$ is Lipschitz continuous if there exists a constant $\lambda_L > 0$ such that
\begin{equation*}
    \|f(x_1) - f(x_2)\| \leq \lambda_L \, \|x_1 - x_2\|.
\end{equation*}

\begin{figure}[b!]
\centering
\begin{subfigure}[t]{0.4\linewidth}
	\centering
	\includegraphics[width=\linewidth]{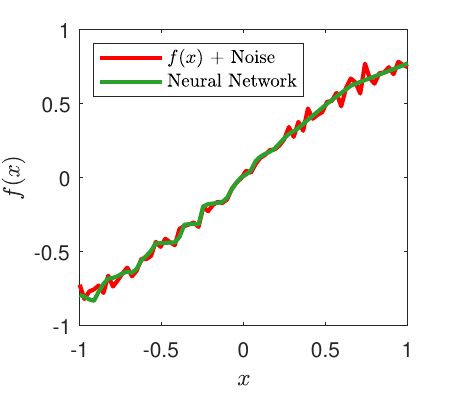}
	\caption{Regular neural network}\label{fig:regular_nn}
\end{subfigure}
\begin{subfigure}[t]{0.4\linewidth}
	\centering
	\includegraphics[width=\linewidth]{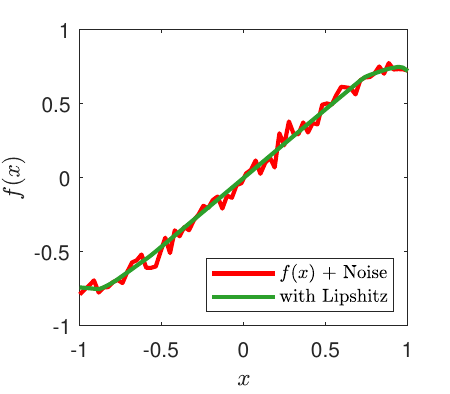}
	\caption{Lipschitz regularized neural network}\label{fig:lipschitz_nn}
\end{subfigure}
\caption{Comparison between regular and Lipschitz regularized neural networks in predicting a noisy function.}
\end{figure}

For differentiable functions, the Mean Value Theorem (\eqnref{eq: Lipschitz_MVT}) relates this to the norm of the derivative, so a bounded derivative (\eqnref{eq: Lipschitz_derivative}) is sufficient for Lipschitz continuity with constant $\lambda_L$.
\begin{equation} \label{eq: Lipschitz_MVT}
    \frac{\|f(x_1) - f(x_2)\|}{\|x_1 - x_2\|} = \|f'(x)\|
\end{equation}
\begin{equation} \label{eq: Lipschitz_derivative}
    \|f'(x)\| \leq \lambda_L,
\end{equation}
Mean squared error (MSE) loss alone is often insufficient when the training data contains noise: the network can overfit by learning highly oscillatory functions with large derivatives (\figref{fig:regular_nn}), which then generalize poorly to unseen inputs and produce non-physical viscosity predictions. Lipschitz regularization addresses this by adding a penalty on the input gradient, \RIII{penalizing rapid changes of} the network output with respect to its input (\figref{fig:lipschitz_nn}). The total loss is given in \eqnref{eq: Lipschitz_loss}, where $\alpha$ is the regularization weight and $\nabla_x f_\theta(x)$ is the gradient of the network output with respect to its input.

\begin{equation} \label{eq: Lipschitz_loss}
    \mathcal{L}_{\text{total}} =
    \underbrace{\text{MSE}}_{\text{data fit}} +
    \underbrace{\alpha \left\| \nabla_x f_\theta(x) \right\|}_{\text{smoothness}},
\end{equation}

\RIII{Training is carried out with $x = \log_{10}\dot{\gamma}$ and
$f_\theta = \ln\eta^{+}$, so the quantity penalized in
\eqnref{eq: Lipschitz_loss} is
$\mathrm{d}\ln\eta^{+}/\mathrm{d}\log_{10}\dot{\gamma}$ rather than
$\mathrm{d}\eta^{+}/\mathrm{d}\dot{\gamma}$.}

We use a fully connected feedforward neural network with two hidden layers of $64$ neurons each and hyperbolic-tangent (Tanh) activations. The network takes the scalar shear rate $\dot{\gamma}$ as input and returns the scalar viscosity $\eta^{+}$. The shear-rate and viscosity data span several decades, so training is carried out in log--log space, with $\log_{10}\dot{\gamma}$ as input and $\ln\eta^{+}$ as output; the transforms are absorbed into a thin wrapper at export time, so the deployed ONNX model accepts \RIII{non}dimensional shear rate and returns \RIII{non}dimensional viscosity directly. \RIII{The data are nondimensionalized before training using the reference scales of \secref{sec:NondimensionalForm}.} Training uses the Adam optimizer with a learning rate of $10^{-3}$, full-batch updates, and $50{,}000$ epochs. The loss function is the Lipschitz-regularized mean-squared error in \eqnref{eq: Lipschitz_loss}, with regularization weight $\alpha = 10^{-4}$ selected by manual tuning. The benchmark networks (\secref{sec:ValidationwithBenchmarkCases}) are trained on shear-rate/viscosity pairs generated from the analytical Carreau--Yasuda formula with the parameters in \tableref{tab:nn_sim_params}; one network per parameter set. The droplet-rise networks (\secref{sec:DropletRiseSimulations}) are trained on $31$ shear-rate/viscosity pairs per material, obtained by averaging three independent rheometer sweeps over $0.1$--$100~\mathrm{s}^{-1}$ as described in \secref{sec:ExperimentalSetUp}. In both settings the data are split $80/20$ into training and validation subsets, and the weights that minimize validation loss are retained for export. Reducing the hidden-layer width from $64$ to $16$ neurons produces an essentially identical $\eta(\dot{\gamma})$ curve, so the result is insensitive to network width over this range. On the silicone-ink rheometer data, where measurement noise is modest, the Lipschitz regularization functions less as a denoising mechanism and more as a smoothness prior.

\begin{figure}[t!]
\centering

% --- Define number of layers and neurons here ---
\newcommand{\numLayers}{4}
\newcommand{\numNeurons}{5}
\newcommand{\hspaceneuron}{4}

% --- Define custom colors for different layers ---
\definecolor{inputcolor}{named}{YellowOrange}  
\definecolor{hiddencolor}{RGB}{0,63,92} % Dark teal 
\definecolor{outputcolor}{named}{SpringGreen} 

% --- Compute vertical spacing and center ---
\pgfmathsetmacro{\neuronSpacing}{2}
\pgfmathsetmacro{\totalHeight}{(\numNeurons-1)*\neuronSpacing}
\pgfmathsetmacro{\yCenter}{\totalHeight/2}

\begin{tikzpicture}[scale=0.4, 
	inputneuron/.style={circle, draw=black, fill=inputcolor,   minimum size=10pt, inner sep=2pt},
	hiddenneuron/.style={circle, draw=black, fill=hiddencolor, minimum size=10pt, inner sep=2pt},
	outputneuron/.style={circle, draw=black, fill=outputcolor, minimum size=10pt, inner sep=2pt},
	>=latex
]
        
% === Input neuron aligned to center ===
\node[inputneuron] (I1) at (0, \yCenter) {};

% === Hidden layers ===
\foreach \layer in {1,...,\numLayers} {
	\foreach \i in {1,...,\numNeurons} {
		\pgfmathsetmacro{\y}{\totalHeight - (\i - 1)*\neuronSpacing}
		\pgfmathsetmacro{\x}{\hspaceneuron * \layer}
		\node[hiddenneuron] (H\layer\i) at (\x, \y) {};
	}
}
        
% === Output neuron aligned to center ===
\pgfmathsetmacro{\xOut}{\hspaceneuron * (\numLayers + 1)}
\node[outputneuron] (O1) at (\xOut, \yCenter) {};

% === Connections: input to hidden layer 1 ===
\foreach \i in {1,...,\numNeurons}
	\draw[->, thin] (I1) -- (H1\i);

% === Connections: hidden layer i to i+1 ===
\foreach \layer in {1,...,\numexpr\numLayers-1\relax} {
	\foreach \i in {1,...,\numNeurons} {
		\foreach \j in {1,...,\numNeurons} {
			\draw[->, thin] (H\layer\i) -- (H\the\numexpr\layer+1\relax\j);
		}
	}
}

% === Connections: last hidden layer to output ===
\foreach \i in {1,...,\numNeurons}
	\draw[->, thin] (H\numLayers\i) -- (O1);

\node at (0, -2) {\small\textbf{Input Layer}};
\node at (0, -1) {$\underbrace{\hspace{2cm}}$};

\pgfmathsetmacro{\xHiddenLabel}{\hspaceneuron * (\numLayers + 1)/2}
\node at (\xHiddenLabel, -2) {\small\textbf{Hidden Layers}};

\node at (\xHiddenLabel, -1) {$\underbrace{\hspace{6cm}}$};

\node at ({\hspaceneuron * (\numLayers + 1)}, -2) {\small\textbf{Output Layer}};

\node at ({\hspaceneuron * (\numLayers + 1)}, -1) {$\underbrace{\hspace{2cm}}$};

\end{tikzpicture}
\caption{Neural network architecture}\label{fig:nn_architecture}
\end{figure}
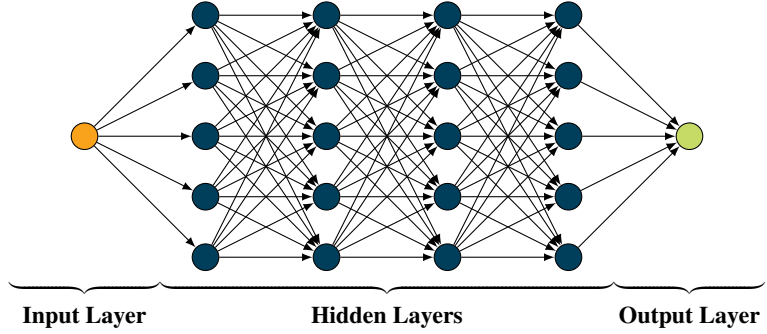

The architecture is shown in \figref{fig:nn_architecture}. The trained network is exported to the Open Neural Network Exchange (ONNX) format~\citep{onnxruntime}, a framework-independent representation of machine learning models. The exported model is loaded into the flow solver at runtime via the ONNX runtime: at each iteration, the solver passes the local shear rate to the model and receives the corresponding viscosity.

%%%%%%%%%%%%%%%%%%%%%%%%%%%%%%%%%%%%%%%%%%%%%%%%%%%%%%%%%%%%%%%%%%%%%%%%%%%%%%%%%%%%%

\section{Experimental Set Up}\label{sec:ExperimentalSetUp}
\subsection{Bubble Rise with Non-Newtonian Fluid Droplet}

Silicone oil (polydimethylsiloxane, PDMS, 1000 cSt, Gelest), fumed silicon dioxide, and perfluorodecalin was purchased from ThermoFisher Scientific and a pigment, Silc Blue Pig, was purchased from Smooth-on. Droplet inks were made by combining PDMS with 1.0 weight percent Silc Blue Pig to provide visual contrast within the high-speed motion videos. Either 1.0 weight percent fumed silicon dioxide (Material A) or 0.5 weight percent fumed silicon dioxide (Material B) was also incorporated to provide different viscous behavior. Each ink formulation was rigorously mixed using a Hauschild SpeedMixer at 3000 rpm for three minutes.

\begin{figure}[t!]
    \centering
    \includegraphics[width=0.7\linewidth,trim={0.0cm 0.0cm 0.0cm 1.8cm},clip]{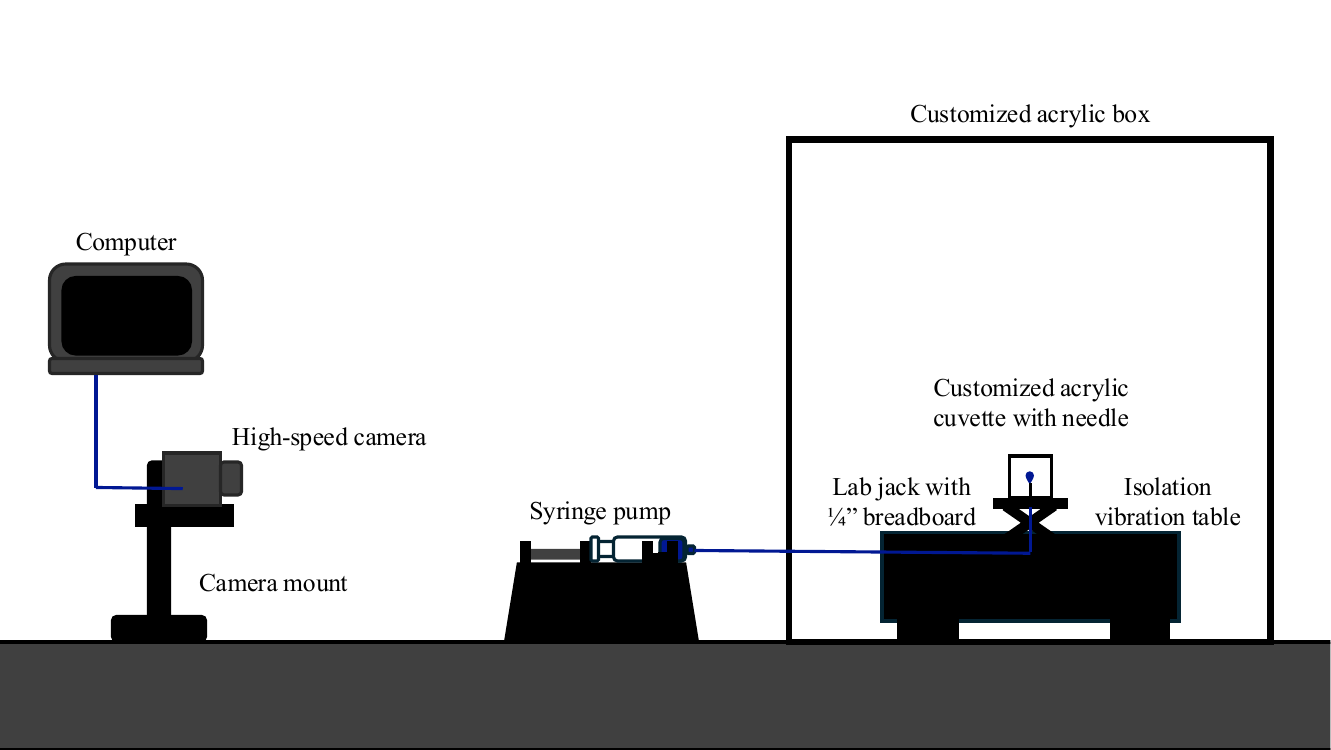}
    \caption{Schematic for rising droplet experimental setup.}
    \label{fig:bubble_rise_setup}
\end{figure}

The experimental setup for the bubble rise is shown in \figref{fig:bubble_rise_setup}. An acrylic box was made to protect the experiment from external air drafts or dust contamination. The needle to form droplets, either a 20G (outer diameter of 0.908 mm) or 25G (outer diameter of 0.515 mm), was attached to a customized acrylic cuvette. \RIII{The cuvette has a square internal cross-section of $25.7 \times 25.7$~mm and an internal height of $49.9$~mm, and was completely filled with perfluorodecalin. The needle tip was approximately centered in the cross-section, $12.4$~mm from the side wall, at a height of approximately $3.5$~mm (20G) or $10$~mm (25G) above the cuvette floor. The cuvette width is $18$--$21$ times the droplet diameter (\tableref{tab:dimensional_params}).} The cuvette was mounted on a lab jack with a 1/4-inch breadboard and placed on an isolation table to remove external vibrations.

Samples were loaded into a syringe and connected to the needle in the acrylic cuvette by appropriately sized tubing. The syringe was driven by a syringe pump, where minimum pressure was applied to allow droplets to form in the bath in the cuvette and to snap off with minimal initial velocity. The cuvette was filled with perfluorodecalin to maximize the travel height of each droplet within the imaging frame. Videos were captured using a Phantom Miro C211 with a 20 mm lens and a frame rate of 1000 fps. Videos were processed using PCC software (Phantom) and ImageJ.  

\subsection{Bubble Ink Characterization}

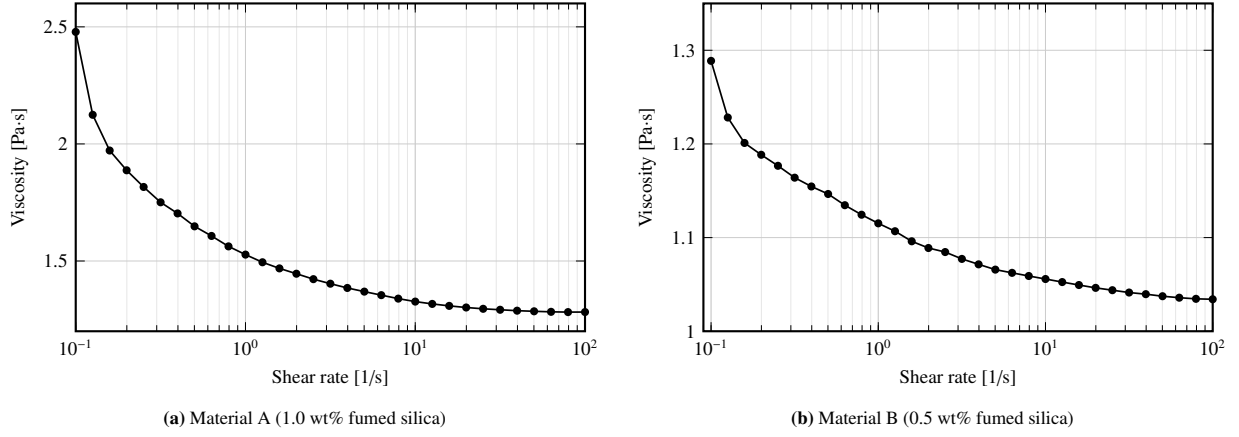
\begin{figure}[t!]
    \centering
    \begin{subfigure}[t]{0.49\textwidth}
\centering
\begin{tikzpicture}[scale=0.8, transform shape, >=Stealth]
\begin{axis}[
    width=10cm, height=7cm,
    xlabel={Shear rate [1/s]},
    ylabel={Viscosity [Pa·s]},
    xmode=log,
    ymode=normal,
    xmin=0.1, xmax=100,
    ymin=1.2, ymax=2.6,
    grid=both,
    major grid style={line width=.2pt,draw=gray!40},
    minor grid style={line width=.1pt,draw=gray!20},
    tick style={black},
    line width=1pt,
    mark size=1.5pt,
    every axis plot/.append style={thick},
    legend style={
        at={(0.97,0.97)},
        anchor=north east,
        draw=none,
        fill=none,
        font=\small
    },
    label style={font=\small},
    tick label style={font=\small}
]

\addplot[
    color=black,
    mark=*,
    mark options={fill=black},
]
table[row sep=crcr]{
0.100001 2.47845\\
0.125899 2.12438\\
0.158484 1.97183\\
0.199537 1.8874\\
0.251172 1.81617\\
0.316224 1.75073\\
0.398113 1.70321\\
0.501185 1.64804\\
0.630961 1.607\\
0.794335 1.56246\\
1 1.52733\\
1.2589 1.49463\\
1.58491 1.46844\\
1.99527 1.44568\\
2.51191 1.42247\\
3.16224 1.40308\\
3.98107 1.38489\\
5.01199 1.36923\\
6.30955 1.35419\\
7.94319 1.33934\\
9.99996 1.32664\\
12.5892 1.31674\\
15.849 1.30834\\
19.9526 1.30143\\
25.1188 1.29586\\
31.6228 1.29176\\
39.8108 1.28783\\
50.1189 1.28519\\
63.0957 1.28294\\
79.4329 1.28179\\
100 1.28198\\
};
\end{axis}
\end{tikzpicture}
\caption{Material A (1.0 wt\% fumed silica)}\label{fig:viscosity_mat_A}
\end{subfigure}
\hfill
\begin{subfigure}[t]{0.49\textwidth}
\centering
\begin{tikzpicture}[scale=0.8, transform shape, >=Stealth]
\begin{axis}[
    width=10cm, height=7cm,
    xlabel={Shear rate [1/s]},
    ylabel={Viscosity [Pa·s]},
    xmode=log,
    ymode=normal,
    xmin=0.09, xmax=100,
    ymin=1.0, ymax=1.35,
    grid=both,
    major grid style={line width=.2pt,draw=gray!40},
    minor grid style={line width=.1pt,draw=gray!20},
    tick style={black},
    line width=1pt,
    mark size=1.5pt,
    every axis plot/.append style={thick},
    legend style={
        at={(0.97,0.97)},
        anchor=north east,
        draw=none,
        fill=none,
        font=\small
    },
    label style={font=\small},
    tick label style={font=\small}
]

\addplot[
    color=black,
    mark=*,
    mark options={fill=black},
]
table[row sep=crcr]{
0.0999899 1.28869\\
0.125881 1.22821\\
0.158482 1.20104\\
0.199533 1.18842\\
0.251206 1.17665\\
0.316231 1.16397\\
0.398106 1.15457\\
0.501175 1.14655\\
0.630958 1.13458\\
0.794338 1.12441\\
0.999995 1.11522\\
1.25893 1.10679\\
1.58492 1.09605\\
1.99526 1.08888\\
2.51191 1.08456\\
3.1623 1.07724\\
3.98095 1.07143\\
5.01184 1.0658\\
6.30958 1.06236\\
7.9433 1.059\\
9.99997 1.05577\\
12.5892 1.05251\\
15.8489 1.04935\\
19.9527 1.04632\\
25.1189 1.04383\\
31.6227 1.04133\\
39.8108 1.03959\\
50.1186 1.0374\\
63.0957 1.03582\\
79.4329 1.03461\\
99.9999 1.03415\\
};
\end{axis}
\end{tikzpicture}
\caption{Material B (0.5 wt\% fumed silica)}\label{fig:viscosity_mat_B}
\end{subfigure}
    \caption{Steady-shear viscosity as a function of shear rate for the two silicone ink formulations used in the droplet rise experiments.}
    \label{fig:viscosity_curves}
\end{figure}

Shear rate vs. viscosity plots were generated using an AR-G2 Rheometer from TA Instruments using 40 mm parallel plates at a gap height of 0.5 mm. Samples were pre-sheared at 100 s$^{-1}$ for 60 seconds, followed by an equilibration time of 60 minutes to account for thixotropic behavior. After equilibration, logarithmic sweeps were performed from 0.1 s$^{-1}$ to 100 s$^{-1}$. This was repeated three times for each material and averaged to be used for simulations. The averaged shear-rate--viscosity curves for the two materials are shown in \figref{fig:viscosity_curves}.

The surface tension of the fluids was measured by performing reverse pendant drop tensiometry using a Theta Flex tensiometer from Biolin Scientific. A customized acrylic cuvette, similar to the cuvette in \figref{fig:bubble_rise_setup}, was filled with perfluorodecalin. A 1 mL precision syringe with a hooked needle attachment to form the pendant drop was filled with either Material A or Material B, which was then submerged into the cuvette. The surface tension was measured by taking the average of three separate 10-second videos.

Rise velocities and droplet trajectories were extracted from the high-speed video recordings using a custom OpenCV-based image processing algorithm that performs adaptive thresholding, morphological cleanup, and interface tracking across successive frames. A full description of the algorithm, together with a representative illustration of the detected interfaces, is provided in \ref{sec:ImageProcessingFramework}; the implementation is available at \texttt{\url{https://github.com/baskargroup/fluidFlowVideoAnalysis}}.

\section{Results and Discussion}\label{sec:Results}

\subsection{Validation with Benchmark Problems}\label{sec:ValidationwithBenchmarkCases}

We validate the framework against the benchmark of a single bubble rising in a quiescent shear-thinning fluid reported by Pang and Lu~\citep{pang2018numerical}, in which the background fluid follows the Carreau--Yasuda law. We compare bubble shapes at a quasi-steady time across four cases at varying power-law index $n$ and Weber number $We$ (Cases 1--4, \tableref{tab:nn_sim_params}). For each case we compare three shapes: the one reported by Pang and Lu, the one from our solver with the analytical Carreau--Yasuda closure, and the one from our solver with a Lipschitz-regularized neural network closure trained on synthetic shear-rate/viscosity data generated from the same analytical formula and exported to ONNX. For an additional case at $Re=3$, $We=5$, $n=0.4$ (Case~5), we also compare the simulated rise velocity against the terminal velocity reported by Pang and Lu. Agreement between the analytical-CY and NN runs isolates the neural-constitutive pipeline itself, encompassing the trained network, its ONNX export, and its integration with the solver; agreement of both with Pang and Lu confirms that the underlying solver is consistent with prior work.

\begin{figure}[b!]
\centering
\begin{tikzpicture}[scale=0.58,>=Stealth] % Adjust the scale to control size
	% Draw the rectangular domain with custom thickness
	\fill[lightgray] (0,0) rectangle (10,12);
	\draw[line width=2pt] (0,0) rectangle (10,12);

	\draw[<->, thick] (0,-0.5) -- (10,-0.5); 
	\node[below] at (5,-0.75) {\(l\)}; % Label for length of the box

	\draw[<->, thick] (11,0) -- (11,12); 
    \node[right] at (10.3,5) {\(h\)}; % Label for height of the box

	% Label the domain boundaries
	\node[above] at (5,11) {Top Boundary};
	\node[below] at (5,1) {Bottom Boundary};
	\node[rotate=90] at (0.5,6) {Left Boundary};  
	% Rotate text 90 degrees for left boundary
	\node[rotate=90] at (9.5,6) {Right Boundary};  
	% Rotate text 90 degrees for right boundary
	
	% Draw the bubble (circle)
	\draw[fill=white] (5,3) circle (1.5);
	
	\fill (5,3) circle (2pt); % Dot at the center
	\node[below right] at (4.5,3.0) {\(x_o, y_o\)}; 

	\draw[->, thick] (5,3) -- (5.5,4.4); % Arrow for radius
	\node[right] at (4.2,3.75) {\(r_o\)}; % Label for radius

	% \node at (3,6.5) {Fluid 2};  % Label for Fluid 1
	\node at (5,5.5) {(+)};  % Label for Fluid 1
	
	% \node at (3,3.25) {Fluid 1 };  % Label for Fluid 2
	\node at (5,2.0) {(--)};  % Label for Fluid 2		
	
	% Gravitational force
	\draw[->, thick] (5,9.5) -- (5,8);
	\node[right] at (5.25,9.00) {$\vec{g}$}; 
	
	% Draw the coordinate axis
	\draw[->, line width=1pt] (0.5,0.5) -- ++(1.0,0) node[right] {\(x\)};
	\draw[->, line width=1pt] (0.5,0.5) -- ++(0,1.0) node[above] {\(y\)};
\end{tikzpicture}
\caption{Schematic for droplet rising problem.}\label{fig:bubble_rise_schematic}
\end{figure}
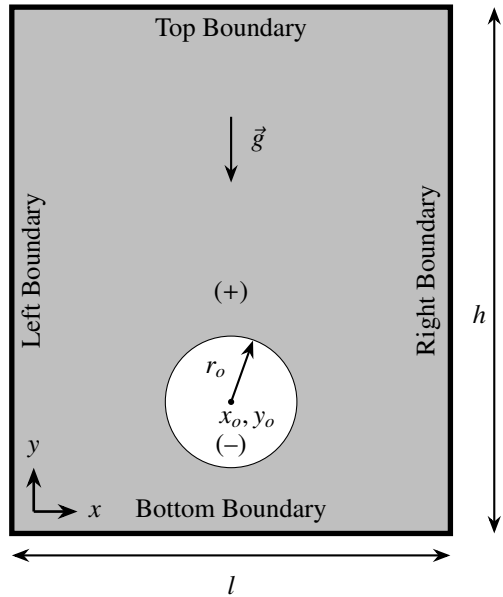

The non-dimensional Carreau--Yasuda law used by Pang and Lu~\citep{pang2018numerical} is
\begin{equation}\label{eq:carreau_yasuda_nd}
    \eta^{+}(\dot{\gamma}) = \eta_{\infty}^{+} + \left(\eta_{o}^{+} - \eta_{\infty}^{+}\right)\left[1 + \left(\lambda \dot{\gamma}\right)^{2}\right]^{(n-1)/2},
\end{equation}
where, in non-dimensional form, $\eta_o^{+}$ and $\eta_{\infty}^{+}$ are the zero-shear and infinite-shear viscosities, $\lambda$ is the relaxation time constant, $n$ is the power-law index, and $\dot{\gamma}$ is the shear rate defined in \eqnref{eq:6}. We use this law in two ways: as the analytical reference closure inside our solver, and as the generator of synthetic training data for the neural-network closure. Parameters for the four shape-comparison cases are given in \tableref{tab:nn_sim_params}. Case~5 shares the structural settings of Case~4, with the power-law index changed to $n=0.4$.

\figref{fig:bubble_rise_schematic} shows the problem setup. The simulations in this section are two-dimensional, matching the setup of Pang and Lu~\citep{pang2018numerical}. The computational domain has length $l = 4.0$ and height $h = 8.0$, with a circular bubble of radius $r_o = 0.5$ placed initially at $(x_o,\, y_o) = (2.0,\, 2.0)$. With no characteristic velocity available, the reference velocity is chosen so that the Froude number is unity, $u_r=\sqrt{g_r\,l_r}$. The Peclet number follows from the scaling relation $Pe=1/(3\,Cn^2)$ for the chosen Cahn number $Cn$~\citep{magaletti2013sharp}. The bubble diameter is the length scale ($l_r=2r_0$), and the heavier fluid sets the density and viscosity scales ($\rho_r=\rho^+$, $\eta_r=\eta^+$). The top and bottom walls are no-slip ($u_i=0$). The side walls use free-slip ($u_1=0$, $\partial u_2/\partial x_1=0$). The phase field is given zero normal gradient on all walls ($n_i \, \partial \phi/\partial x_i=0$, $n_i \, \partial \mu/\partial x_i=0$).

\begin{figure}[t!]
    \centering
    \begin{tikzpicture}[scale=0.9, transform shape, >=Stealth]
\begin{axis}[
    width=12cm, height=8cm,
    xlabel={Non-dimensional time $t$},
    ylabel={Normalized rise velocity $U/U_{n=0.2}$},
    xmin=0, xmax=4,
    ymin=0, ymax=1.2,
    grid=both,
    major grid style={line width=.2pt,draw=gray!40},
    minor grid style={line width=.1pt,draw=gray!20},
    tick style={black},
    line width=1pt,
    every axis plot/.append style={line width=2pt},
    legend style={
        at={(0.97,0.03)},
        anchor=south east,
        draw=black,
        fill=white,
        font=\small
    },
    label style={font=\small},
    tick label style={font=\small}
]

% --- Pang & Lu reference (terminal value at n=0.4, normalized by n=0.2 terminal) ---
\addplot[
    color=black,
    dashed,
    line width=1.2pt,
    domain=0:15,
    forget plot,
]{0.9533};
\addlegendimage{black, dashed, line width=1.2pt}
% \addlegendentry{Pang \& Lu terminal velocity (Fig.~10, $Ga=3$, $Eo=5$)}
\addlegendentry{Pang \& Lu }

% --- Analytical Carreau-Yasuda trajectory ---
\addplot[
    color=blue,
    mark=none,
]
table[row sep=crcr]{
0.0000 0.4003\\
0.1000 0.4618\\
0.2000 0.5179\\
0.3000 0.5654\\
0.4000 0.6057\\
0.5000 0.6438\\
0.6000 0.7228\\
0.7000 0.7957\\
0.8000 0.8557\\
0.9000 0.8978\\
1.0000 0.9321\\
1.1000 0.9551\\
1.2000 0.9684\\
1.3000 0.9746\\
1.4000 0.9770\\
1.5000 0.9778\\
1.6000 0.9733\\
1.7000 0.9719\\
1.8000 0.9703\\
1.9000 0.9698\\
2.0000 0.9720\\
2.1000 0.9705\\
2.2000 0.9681\\
2.3000 0.9657\\
2.4000 0.9654\\
2.5000 0.9637\\
2.6000 0.9619\\
2.7000 0.9600\\
2.8000 0.9591\\
2.9000 0.9611\\
3.0000 0.9590\\
3.1000 0.9587\\
3.2000 0.9585\\
3.3000 0.9574\\
3.4000 0.9594\\
3.5000 0.9593\\
3.6000 0.9606\\
3.7000 0.9612\\
3.8000 0.9631\\
3.9000 0.9652\\
4.0000 0.9645\\
4.1000 0.9675\\
4.2000 0.9689\\
4.3000 0.9707\\
4.4000 0.9744\\
4.5000 0.9753\\
4.6000 0.9768\\
4.7000 0.9795\\
4.8000 0.9822\\
4.9000 0.9836\\
5.0000 0.9851\\
5.1000 0.9887\\
5.2000 0.9907\\
5.3000 0.9928\\
5.4000 0.9953\\
5.5000 0.9973\\
5.6000 0.9999\\
5.7000 1.0004\\
5.8000 1.0015\\
5.9000 1.0041\\
6.0000 1.0063\\
6.1000 1.0082\\
6.2000 1.0094\\
6.3000 1.0099\\
6.4000 1.0100\\
6.5000 1.0105\\
6.6000 1.0121\\
6.7000 1.0141\\
6.8000 1.0156\\
6.9000 1.0154\\
7.0000 1.0149\\
7.1000 1.0154\\
7.2000 1.0165\\
7.3000 1.0186\\
7.4000 1.0187\\
7.5000 1.0180\\
7.6000 1.0190\\
7.7000 1.0189\\
7.8000 1.0186\\
7.9000 1.0184\\
8.0000 1.0191\\
};
\addlegendentry{Carreau--Yasuda (present)}

% --- Neural network trajectory ---
\addplot[
    color=red,
    mark=none,
    dash pattern=on 8pt off 8pt,
]
table[row sep=crcr]{
0.0000 0.4016\\
0.1000 0.4628\\
0.2000 0.5191\\
0.3000 0.5667\\
0.4000 0.6062\\
0.5000 0.6443\\
0.6000 0.7221\\
0.7000 0.7950\\
0.8000 0.8561\\
0.9000 0.8988\\
1.0000 0.9326\\
1.1000 0.9545\\
1.2000 0.9678\\
1.3000 0.9733\\
1.4000 0.9755\\
1.5000 0.9771\\
1.6000 0.9730\\
1.7000 0.9734\\
1.8000 0.9715\\
1.9000 0.9694\\
2.0000 0.9716\\
2.1000 0.9705\\
2.2000 0.9686\\
2.3000 0.9659\\
2.4000 0.9653\\
2.5000 0.9648\\
2.6000 0.9626\\
2.7000 0.9597\\
2.8000 0.9581\\
2.9000 0.9601\\
3.0000 0.9587\\
3.1000 0.9590\\
3.2000 0.9586\\
3.3000 0.9569\\
3.4000 0.9593\\
3.5000 0.9597\\
3.6000 0.9602\\
3.7000 0.9606\\
3.8000 0.9633\\
3.9000 0.9658\\
4.0000 0.9653\\
4.1000 0.9681\\
4.2000 0.9692\\
4.3000 0.9715\\
4.4000 0.9752\\
4.5000 0.9757\\
4.6000 0.9779\\
4.7000 0.9802\\
4.8000 0.9830\\
4.9000 0.9856\\
5.0000 0.9876\\
5.1000 0.9908\\
5.2000 0.9924\\
5.3000 0.9947\\
5.4000 0.9968\\
5.5000 0.9986\\
5.6000 1.0018\\
5.7000 1.0028\\
5.8000 1.0042\\
5.9000 1.0055\\
6.0000 1.0060\\
6.1000 1.0076\\
6.2000 1.0093\\
6.3000 1.0106\\
6.4000 1.0106\\
6.5000 1.0114\\
6.6000 1.0128\\
6.7000 1.0137\\
6.8000 1.0143\\
6.9000 1.0144\\
7.0000 1.0154\\
7.1000 1.0166\\
7.2000 1.0173\\
7.3000 1.0187\\
7.4000 1.0187\\
7.5000 1.0179\\
7.6000 1.0181\\
7.7000 1.0183\\
7.8000 1.0191\\
7.9000 1.0191\\
8.0000 1.0188\\
};
\addlegendentry{Neural network (present)}

\end{axis}
\end{tikzpicture}
    \caption{Normalized rise velocity versus non-dimensional time for Case~5 ($Re=3$, $We=5$, $n=0.4$). \RIII{Velocities are normalized by $U_{n=0.2}$, the terminal velocity from an additional simulation at $n=0.2$, matching the convention of Figure~10 in Pang and Lu~\citep{pang2018numerical}.} The solid blue curve uses the analytical Carreau--Yasuda closure and the dashed red curve uses the trained neural-network closure; the black dashed horizontal line is the terminal velocity reported by Pang and Lu for the $Ga=3$, $Eo=5$, $n=0.4$ case. The two simulated trajectories are visually indistinguishable and both plateau close to the Pang and Lu reference.}
    \label{fig:rise_velocity_benchmark}
\end{figure}
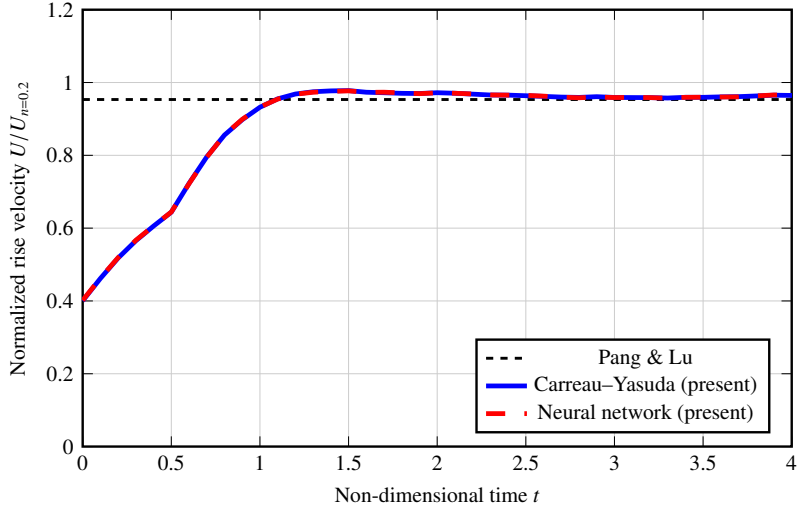

For Case~5, \citet{pang2018numerical} report a terminal velocity, which lets us compare it against both of our simulated trajectories. \figref{fig:rise_velocity_benchmark} shows the normalized rise velocities from our solver with the analytical Carreau--Yasuda closure and with the trained neural-network closure, along with the Pang and Lu terminal velocity as a horizontal dashed line. \RIII{All values are normalized by $U_{n=0.2}$, the terminal velocity obtained from an additional simulation in which the power-law index is set to $n=0.2$ and all other parameters are left unchanged. Each trajectory is normalized using the $n=0.2$ simulation performed with the same closure. This matches the normalization convention of Pang and Lu's Figure~10.} Both simulated trajectories rise monotonically, reach a peak, and settle onto a quasi-steady plateau. The NN plateau is $0.9608$ and the analytical-CY plateau is $0.9609$, a difference of about $0.01\%$. Both differ from the Pang and Lu reference value of $0.9533$ by approximately $0.8\%$.

\begin{table}[t!]
\centering
\caption{Benchmark simulation parameters for the four bubble-shape comparison cases. For each case, synthetic shear-rate/viscosity training data are generated from the Carreau--Yasuda formula with the listed parameters, and a dedicated neural network is trained on that dataset and used as the viscosity closure in the simulation.}
\label{tab:nn_sim_params}
\begin{tabular}{llllll}
	\toprule
	Parameter & & Case 1 & Case 2 & Case 3 & Case 4\\
	\midrule
	Reynolds number  & $Re$   & 3 &3 & 3 & 3\\
	Weber Number & $We$   & 20 &20 & 20 & 5\\
	Froude Number & $Fr$ & 1.0 &1.0 & 1.0 & 1.0\\
	Cahn Number & $Cn$ & 0.01 &0.01 & 0.01 & 0.01\\
	Peclet Number & $Pe$ & 3333.33 &3333.33 & 3333.33 & 3333.33\\
	Density of heavy fluid & $\rho^+$ & 1.0 &1.0 & 1.0 & 1.0\\
	Density of light fluid & $\rho^-$ & 0.001 &0.001 & 0.001 & 0.001\\
    Zero-shear viscosity of heavy fluid & $\eta_o^{+}$ & 1.0 &1.0 & 1.0 & 1.0\\
	Infinite-shear viscosity of heavy fluid & $\eta_{\infty}^{+}$ & 1/50 &1/50 & 1/50 & 1/50\\
	Viscosity of light fluid & $\eta^-$ & 0.01 &0.01 & 0.01 & 0.01\\
	Power-law index & $n$ & 0.8 &0.4 & 0.6 & 0.6\\
	Relaxation time constant & $\lambda$ & 48.5 &48.5 & 48.5 & 48.5\\
	\bottomrule
\end{tabular}
\end{table}

\begin{figure}[t!]
\centering
\begin{subfigure}[b]{0.99\textwidth}
\begin{subfigure}[b]{0.3\textwidth}
\centering
\includegraphics[width=0.9\linewidth,trim={2.0cm 2.6cm 1.6cm 2.6cm},clip]{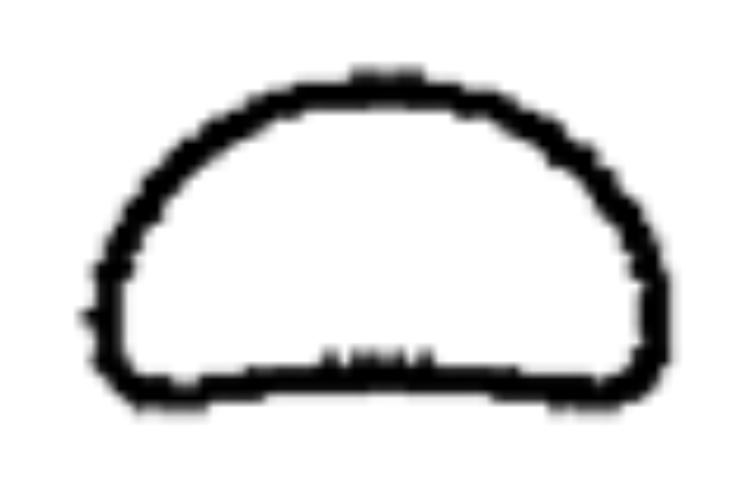}
\caption*{Pang and Lu~\citep{pang2018numerical}}
\end{subfigure}
\hfill
\begin{subfigure}[b]{0.3\textwidth}
\centering
\includegraphics[width=0.8\linewidth,trim={81.6cm 76.9cm 91.1cm 37.9cm},clip]{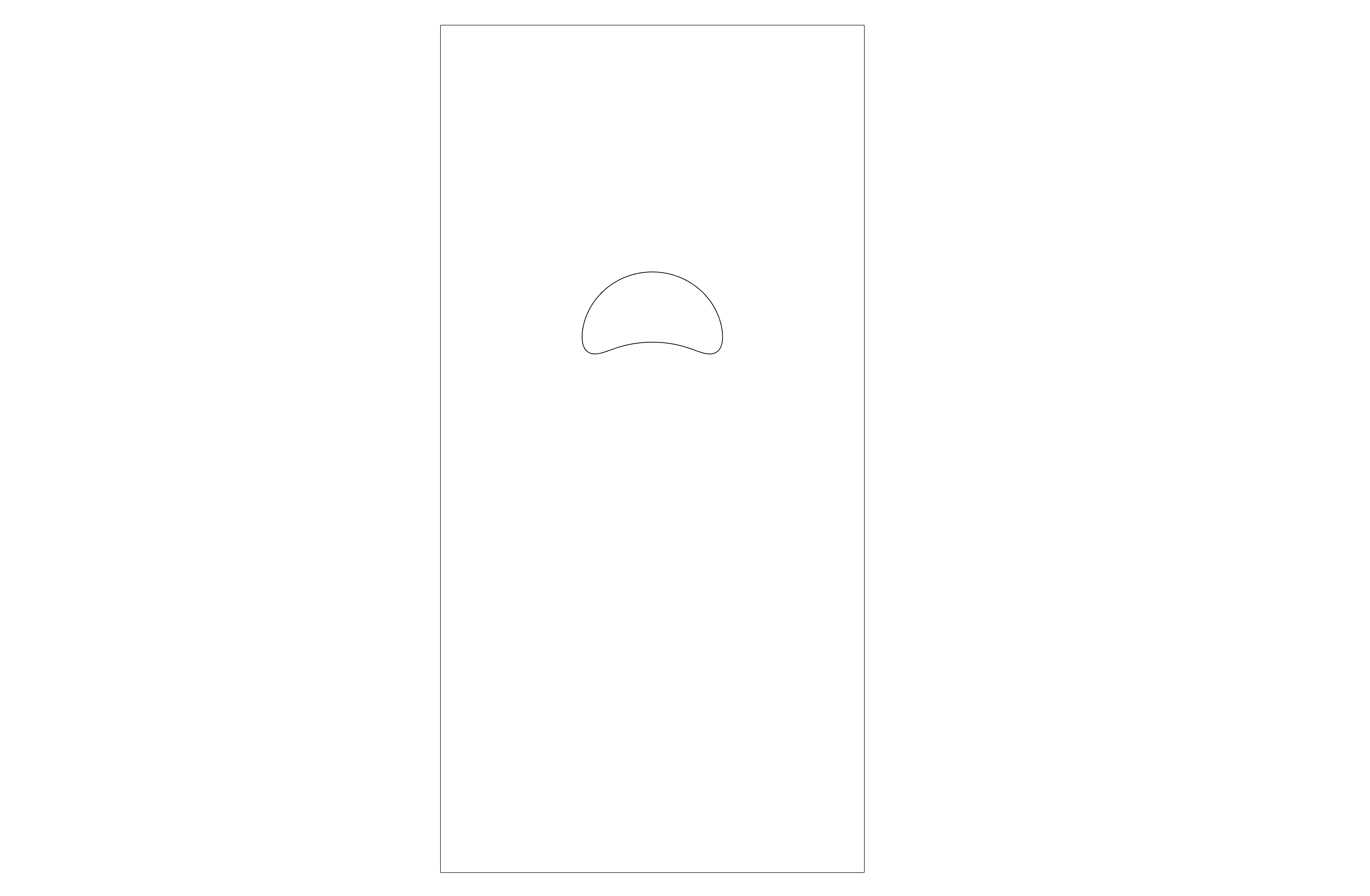}
\caption*{Analytical Carreau--Yasuda}
\end{subfigure}
\hfill
\begin{subfigure}[b]{0.3\textwidth}
\centering
\includegraphics[width=0.8\linewidth,trim={86.2cm 77.3cm 86.2cm 37.2cm},clip]{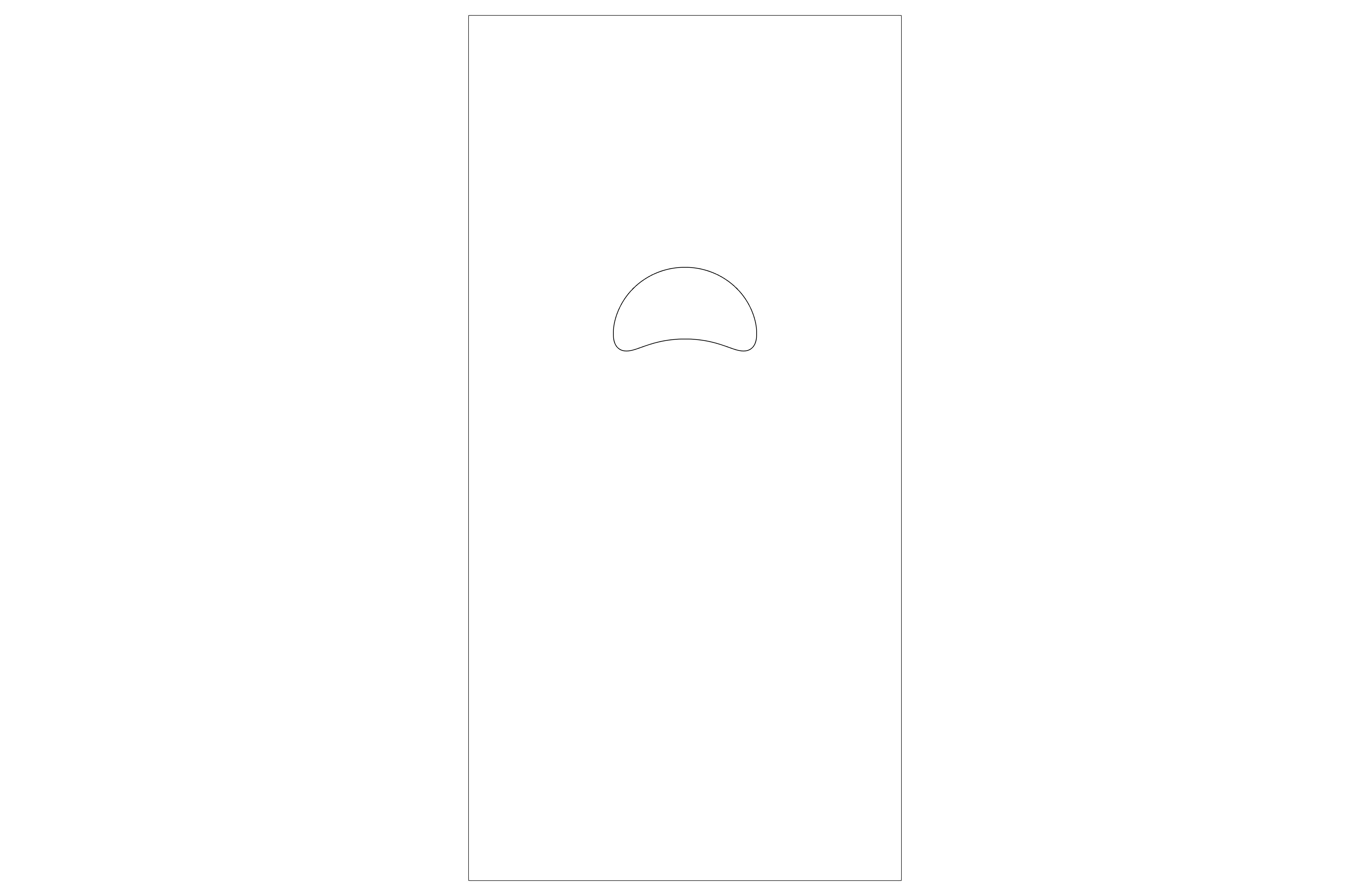}
\caption*{Neural network}
\end{subfigure}
\caption{Case 1 ($n=0.8$, $We=20$).}\label{fig:case1_shape}
\end{subfigure}
%%%%%%%%%%%%%%%%%%%%%%%%%%%%%%%%%%%%%%%%%%%%%%%%%%%%%%%%%%%%%%%%%%%%%%%%%%%%%%%%%%%%%
% CASE 2 
\begin{subfigure}[b]{0.99\textwidth}
\begin{subfigure}[b]{0.3\textwidth}
\centering
\includegraphics[width=1.0\linewidth,trim={2.0cm 4cm 1.6cm 3.0cm},clip]{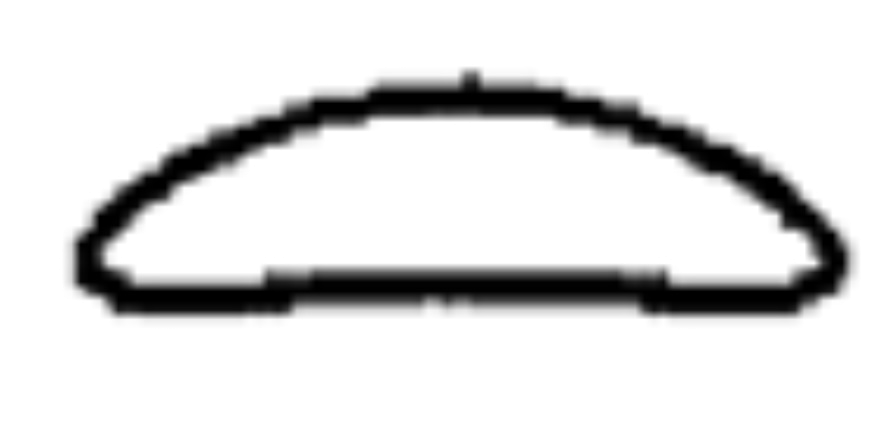}
\caption*{Pang and Lu~\citep{pang2018numerical}}
\end{subfigure}
\hfill
\begin{subfigure}[b]{0.3\textwidth}
\centering
\includegraphics[width=0.9\linewidth,trim={82.0cm 92.3cm 82.0cm 24.9cm},clip]{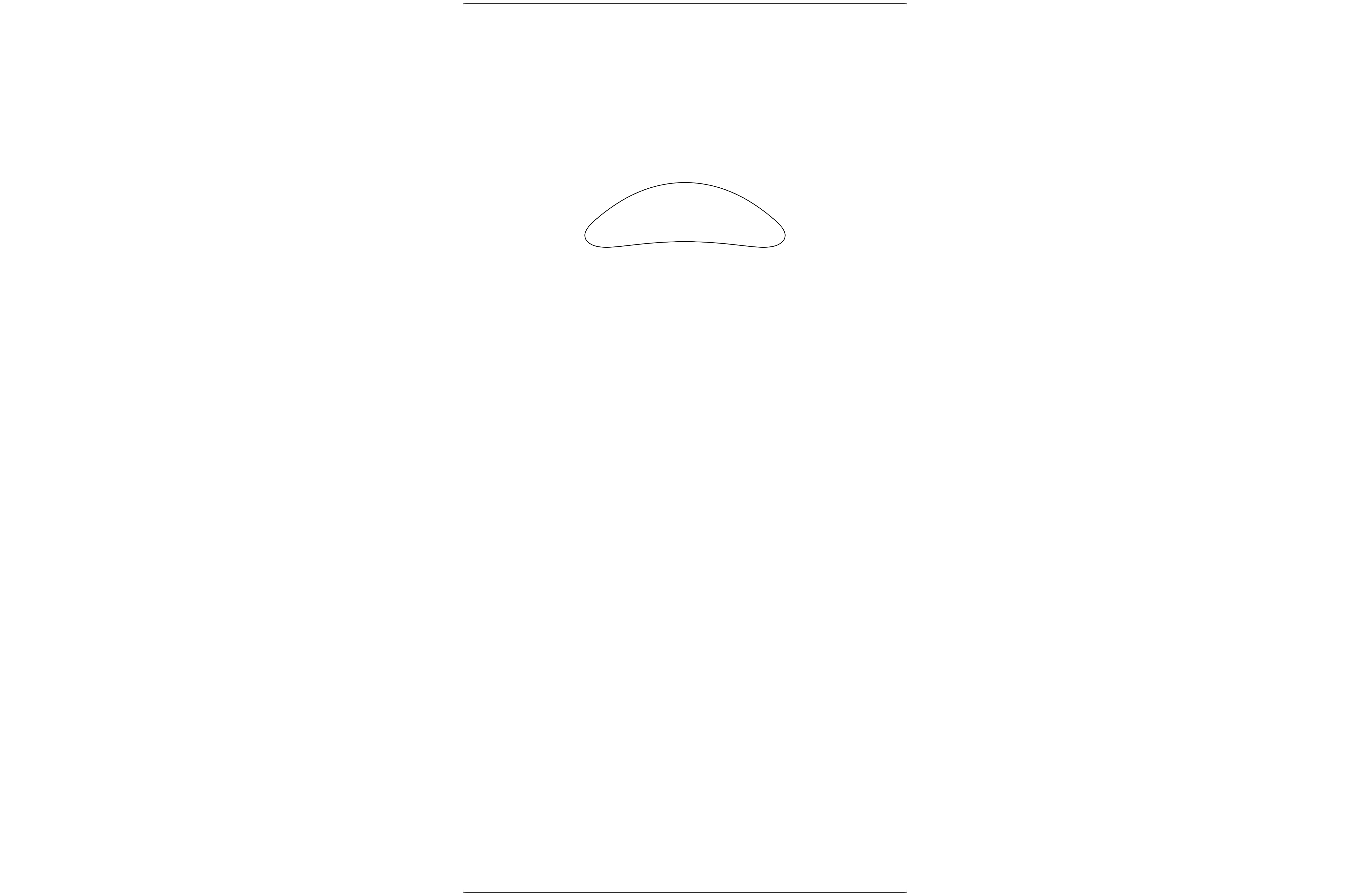}
\caption*{Analytical Carreau--Yasuda}
\end{subfigure}
\hfill
\begin{subfigure}[b]{0.3\textwidth}
\centering
\includegraphics[width=0.93\linewidth,trim={82.2cm 91.9cm 82.2cm 25.4cm},clip]{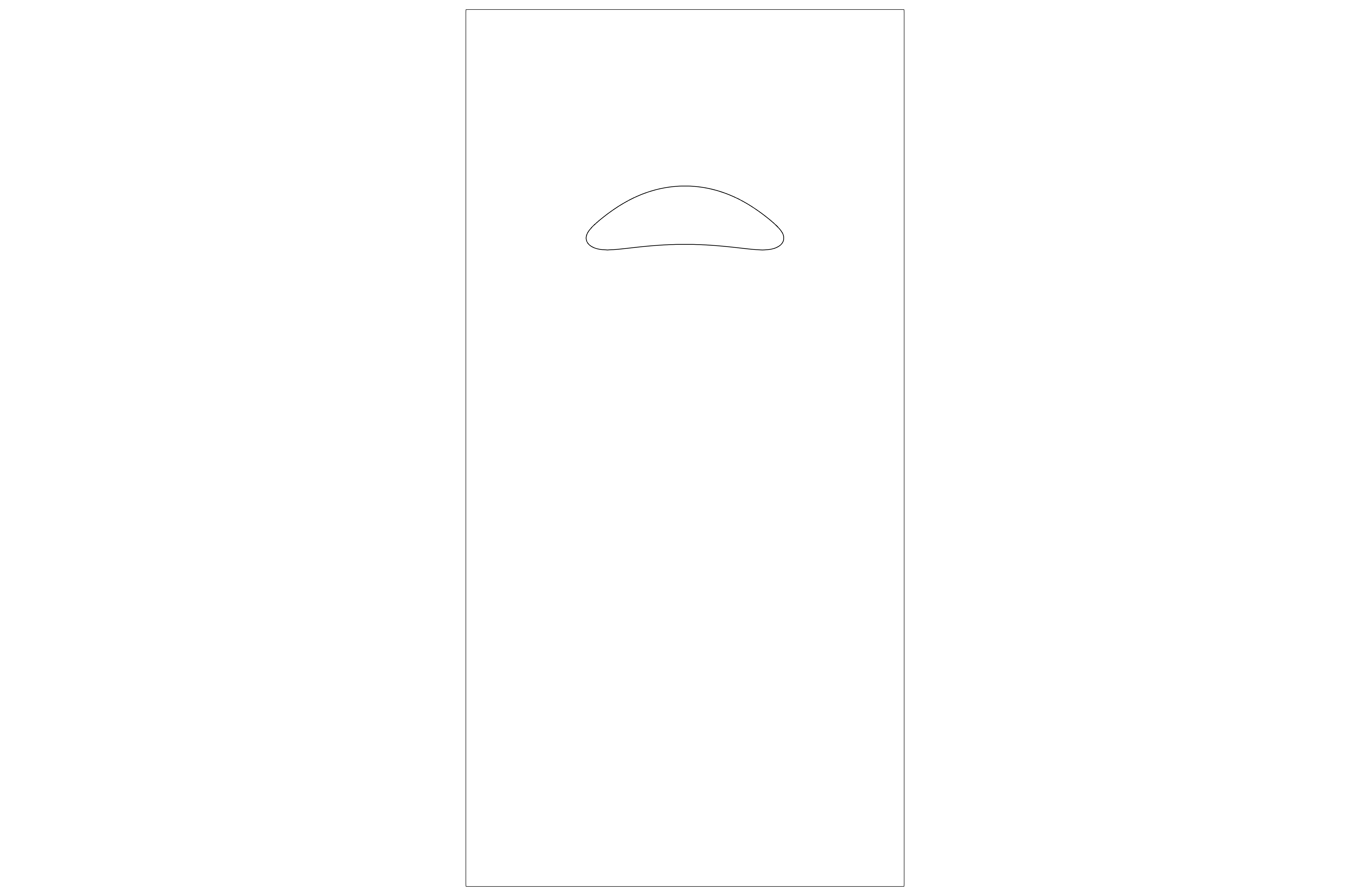}
\caption*{Neural network}
\end{subfigure}
\caption{Case 2 ($n=0.4$, $We=20$).}\label{fig:case2_shape}
\end{subfigure}
%%%%%%%%%%%%%%%%%%%%%%%%%%%%%%%%%%%%%%%%%%%%%%%%%%%%%%%%%%%%%%%%%%%%%%%%%%%%%%%%%%%%
% CASE 3
\begin{subfigure}[b]{0.99\textwidth}
\begin{subfigure}[b]{0.3\textwidth}
\centering
\includegraphics[width=0.9\linewidth,trim={2.3cm 3cm 2.3cm 3.4cm},clip]{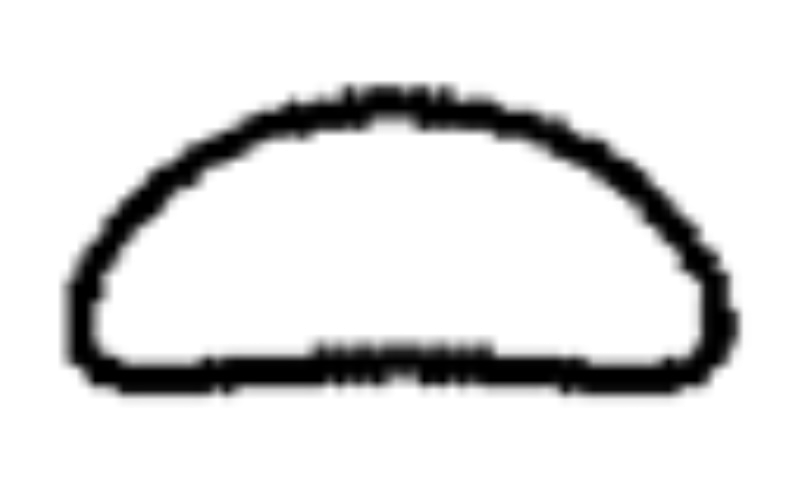}
\caption*{Pang and Lu~\citep{pang2018numerical}}
\end{subfigure}
\hfill
\begin{subfigure}[b]{0.3\textwidth}
\centering
\includegraphics[width=\linewidth,trim={83.8cm 88.8cm 83.8cm 27.3cm},clip]{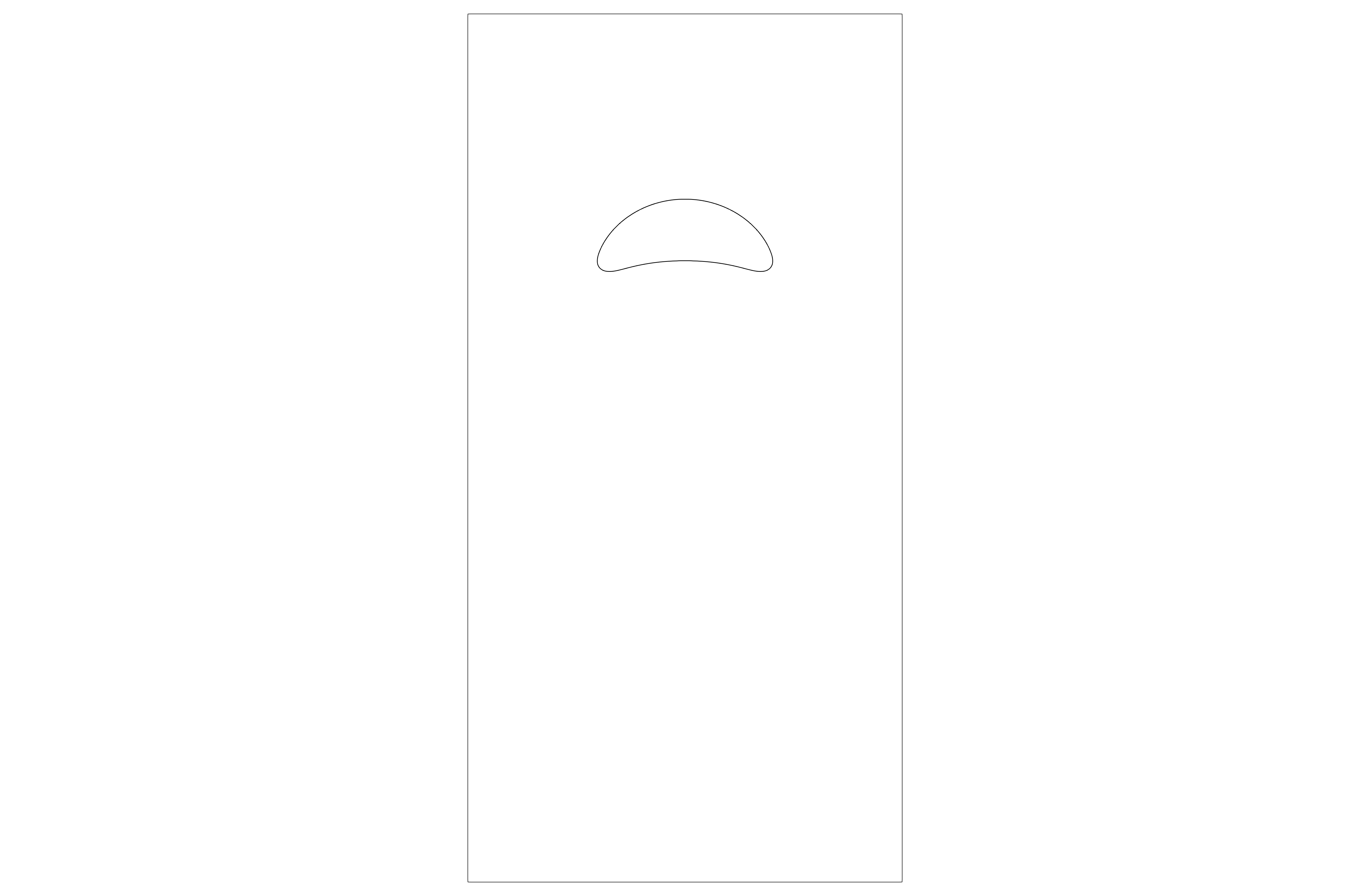}
\caption*{Analytical Carreau--Yasuda}
\end{subfigure}
\hfill
\begin{subfigure}[b]{0.3\textwidth}
\centering
\includegraphics[width=\linewidth,trim={83.6cm 89.2cm 83.6cm 26.8cm},clip]{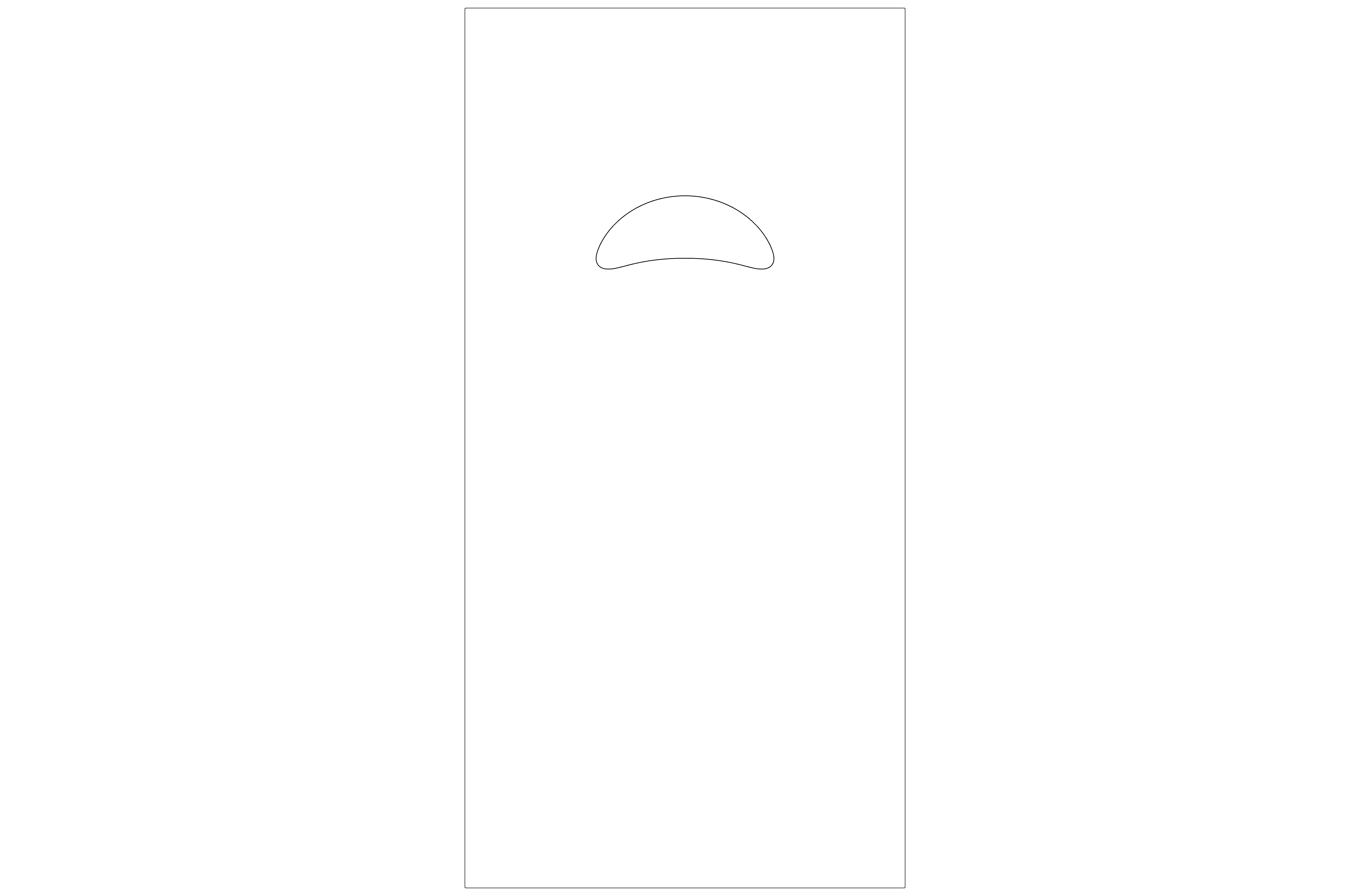}
\caption*{Neural network}
\end{subfigure}
\caption{Case 3 ($n=0.6$, $We=20$).}\label{fig:case3_shape}
\end{subfigure}
%%%%%%%%%%%%%%%%%%%%%%%%%%%%%%%%%%%%%%%%%%%%%%%%%%%%%%%%%%%%%%%%%%%%%%%%%%%%%%%%%%%%%
\begin{subfigure}[b]{0.99\textwidth}
\centering
\begin{subfigure}[b]{0.3\textwidth}
\centering
\includegraphics[width=\linewidth,trim={1cm 1cm 1cm 2cm},clip]{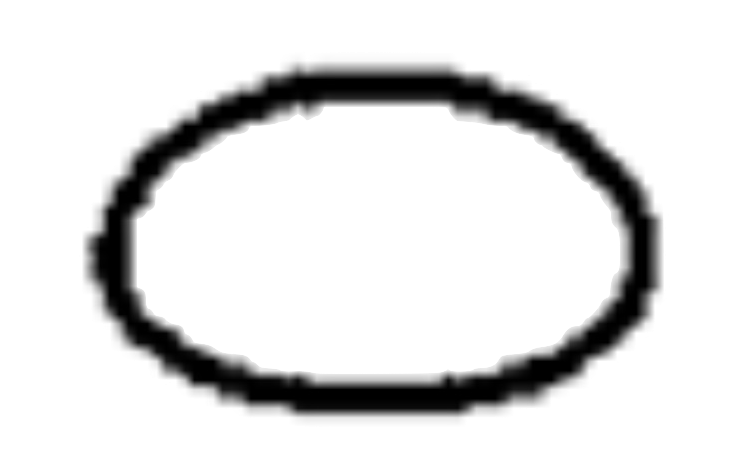}
\caption*{Pang and Lu~\citep{pang2018numerical}}
\end{subfigure}
\hfill
\begin{subfigure}[b]{0.3\textwidth}
\centering
\includegraphics[width=\linewidth,trim={85.9cm 96.2cm 85.9cm 17.5cm},clip]{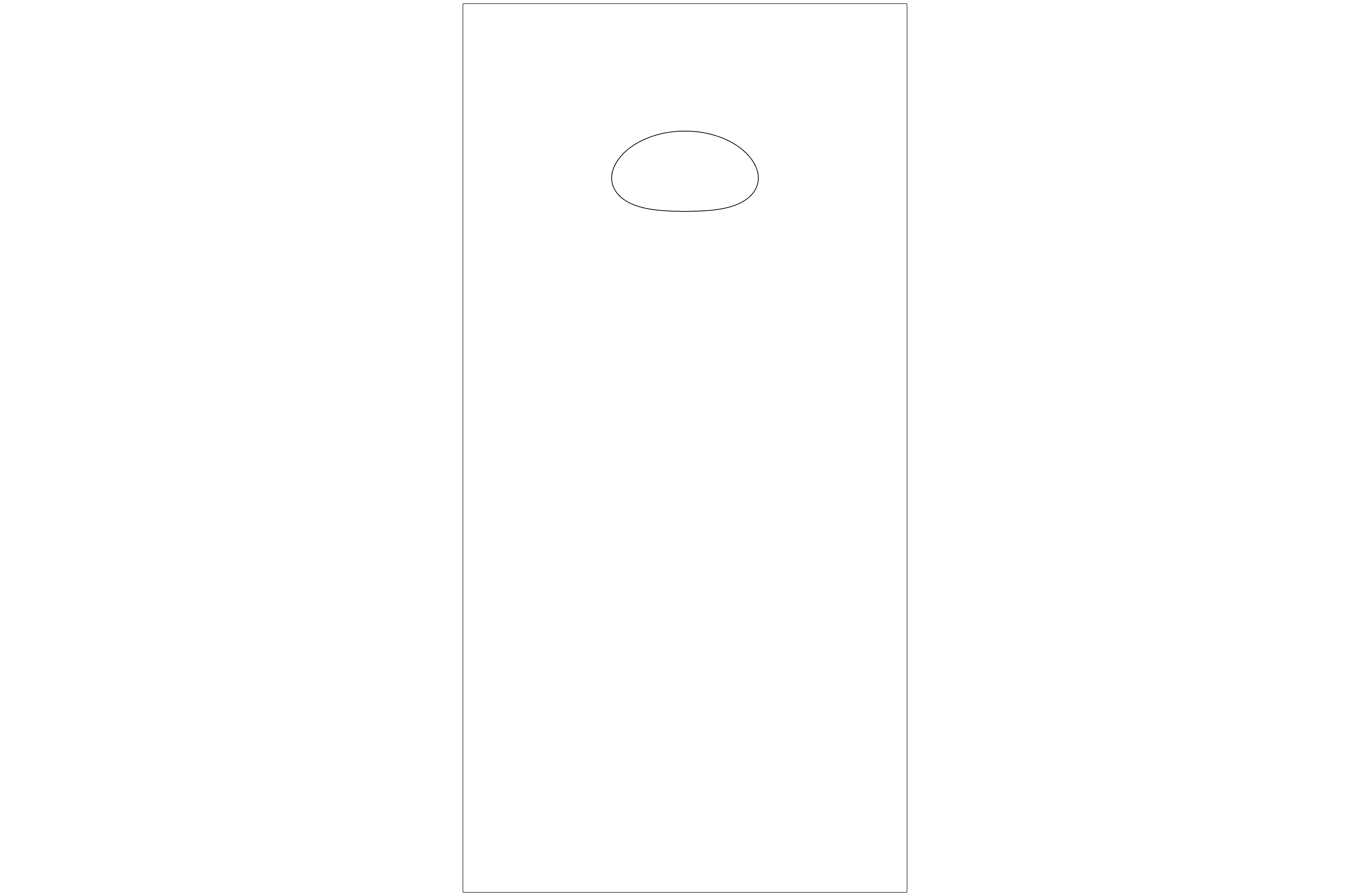}
\caption*{Analytical Carreau--Yasuda}
\end{subfigure}
\hfill
\begin{subfigure}[b]{0.3\textwidth}
\centering
\includegraphics[width=\linewidth,trim={86.2cm 96.6cm 86.2cm 18.7cm},clip]{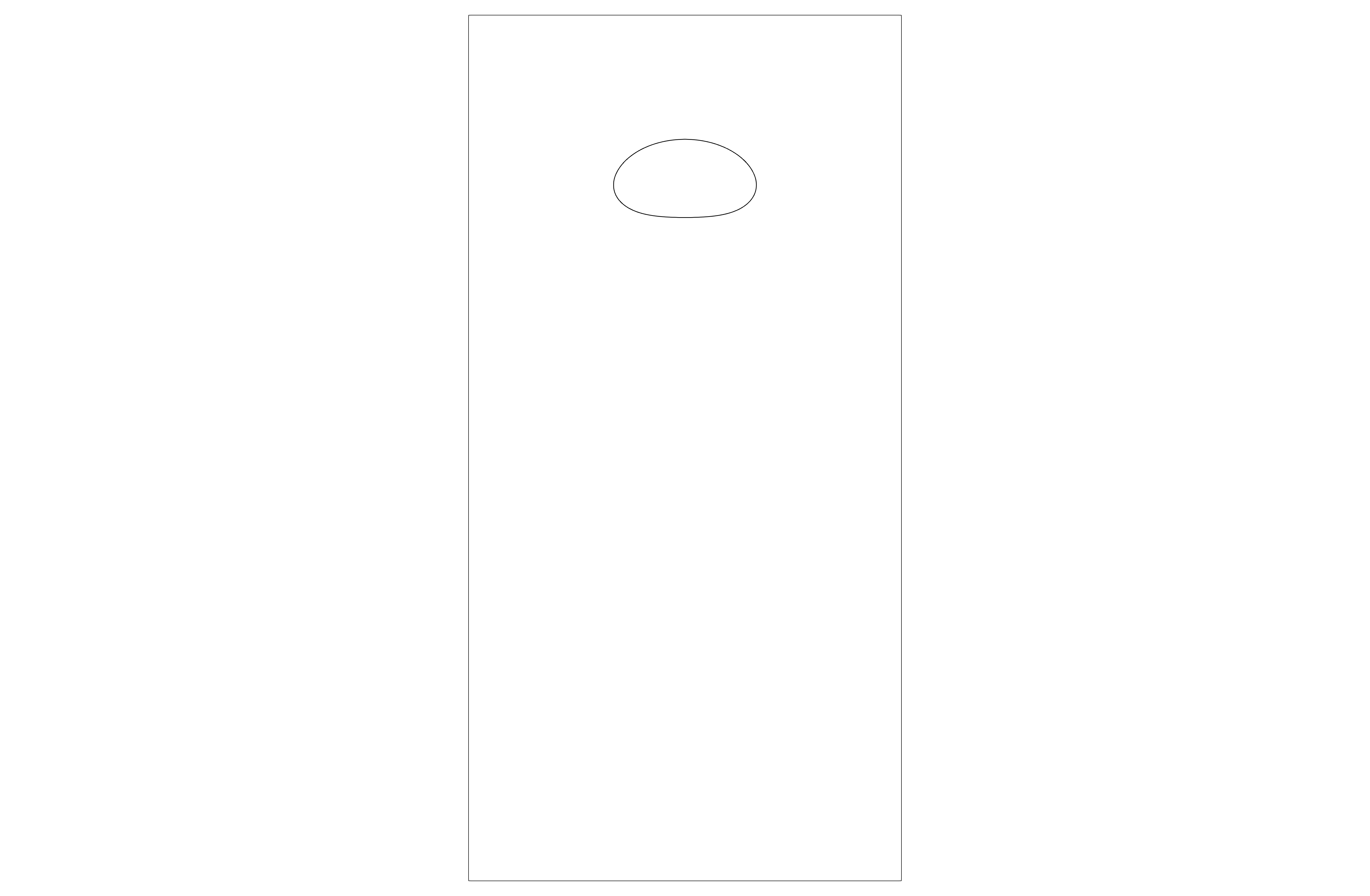}
\caption*{Neural network}
\end{subfigure}
\caption{Case 4 ($n=0.6$, $We=5$).}\label{fig:case4_shape}
\end{subfigure}
\caption{Comparison of the bubble shape between experimental, analytical, and our neural network-based approach.}
\label{fig:bubble_shape}
\end{figure}

\figref{fig:bubble_shape} shows the three-way shape comparisons for the four cases. In each case, the analytical-CY and neural-network shapes are visually indistinguishable, confirming that the neural closure reproduces the analytical target inside the solver. Both also agree with the shapes reported by \citet{pang2018numerical}, confirming solver consistency with prior work.

The shapes vary with the governing dimensionless parameters in the expected way. Across Cases 1--3 (all at $We=20$, Figures~\ref{fig:case1_shape}--\ref{fig:case3_shape}), reducing the power-law index $n$ from $0.8$ to $0.4$ produces progressively greater deformation and a more pronounced skirted shape: lower $n$ means stronger shear-thinning, which lowers the effective viscosity around the bubble and lets it deform more easily under buoyancy. Comparing Case 3 ($n=0.6$, $We=20$, \figref{fig:case3_shape}) with Case 4 ($n=0.6$, $We=5$, \figref{fig:case4_shape}) shows the effect of Weber number: at $We=5$, surface tension dominates inertia and the bubble retains a near-elliptical shape. The agreement among Pang and Lu, the analytical-CY runs, and the NN runs across this range of $n$ and $We$ supports both the solver and the neural-constitutive pipeline.

%%%%%%%%%%%%%%%%%%%%%%%%%%%%%%%%%%%%%%%%%%%%%%%%%%%%%%%%%%%%%%%%%%%%%%%%%%%%%%%%%%%%%
\subsection{Synthetic Non-Monotonic Case}\label{sec:SyntheticNonMonotonic}

The benchmark cases in \secref{sec:ValidationwithBenchmarkCases} use a Carreau--Yasuda constitutive law, which is monotone in shear rate. To test the pipeline on a viscosity shape outside this family, we construct a synthetic constitutive law in which $\eta(\dot{\gamma})$ first decreases with shear rate and then increases at high shear:
\begin{equation}\label{eq:nonmonotonic_law}
    \eta^{+}(\dot{\gamma}) = \eta_{\infty}^{+} + \left(\eta_{o}^{+} - \eta_{\infty}^{+}\right)\left[1 + (\lambda\dot{\gamma})^{2}\right]^{(n_1 - 1)/2} + K\,\dot{\gamma}^{\,n_2}.
\end{equation}
The first two terms reproduce a Carreau--Yasuda-type shear-thinning branch with zero-shear viscosity $\eta_o^{+}$, infinite-shear floor $\eta_\infty^{+}$, relaxation time $\lambda$, and exponent $n_1 < 1$. The third term adds a power-law thickening branch with coefficient $K$ and exponent $n_2 > 1$, so the viscosity curve is non-monotonic in $\dot{\gamma}$. Parameters are listed in \tableref{tab:nonmonotonic_params}. Shear-thinning--to--thickening crossovers of this kind are reported for associative polymer solutions and concentrated particulate suspensions~\citep{wagner2009shear}.

The flow setup matches Case~4 of \secref{sec:ValidationwithBenchmarkCases}: $Re = 3$, $We = 5$, $Fr = 1$, $Cn = 0.01$, $Pe = 3333.33$, with the same domain, initial bubble position, and boundary conditions. Only the constitutive law changes. We run two simulations: one in which the solver evaluates \eqnref{eq:nonmonotonic_law} directly at each iteration, and one in which the solver evaluates a Lipschitz-regularized neural network trained on shear-rate/viscosity pairs generated from \eqnref{eq:nonmonotonic_law} and exported to ONNX. The training procedure is the same as in \secref{sec:ValidationwithBenchmarkCases}.

\figref{fig:nonmonotonic_shape} compares the bubble shapes at the quasi-steady time, and \figref{fig:nonmonotonic_rise} compares the rise velocity trajectories. The analytical-law and network simulations produce matching results in both. This shows that the workflow handles a constitutive shape outside the standard shear-thinning class without modification to the network architecture, training procedure, or solver. Compared with the purely shear-thinning benchmark Case~4 at the same flow conditions, the bubble here remains nearly spherical. This is consistent with the non-monotonic constitutive law: the shear-thickening branch increases viscosity in regions of high strain rate, which resists deformation. 

\begin{table}[t!]
\centering
\caption{Constitutive parameters for the synthetic non-monotonic case.}
\label{tab:nonmonotonic_params}
\begin{tabular}{lll}
    \toprule
    Parameter & Symbol & Value \\
    \midrule
    Zero-shear viscosity     & $\eta_o^{+}$       & 20.0 \\
    Infinite-shear viscosity & $\eta_\infty^{+}$  & 0.2 \\
    Relaxation time          & $\lambda$          & 8.0 \\
    Shear-thinning exponent  & $n_1$              & 0.35 \\
    Thickening coefficient   & $K$                & 0.015 \\
    Thickening exponent      & $n_2$              & 1.3 \\
    \bottomrule
\end{tabular}
\end{table}
\begin{figure}[t!]
\centering
\begin{tikzpicture}[scale=0.9, transform shape, >=Stealth]
\begin{axis}[
    width=10cm, height=7cm,
    xlabel={Shear rate $\dot{\gamma}$},
    ylabel={Viscosity $\eta^{+}$},
    xmode=log,
    ymode=normal,
    xmin=0.01, xmax=100,
    ymin=0, ymax=22,
    grid=both,
    major grid style={line width=.2pt,draw=gray!40},
    minor grid style={line width=.1pt,draw=gray!20},
    tick style={black},
    line width=1pt,
    legend style={
        at={(0.97,0.97)},
        anchor=north east,
        draw=black,
        fill=white,
        font=\small
    },
    label style={font=\small},
    tick label style={font=\small}
]

\addplot[
    color=blue,
    line width=1.5pt,
    mark=none,
    samples=400,
    domain=0.01:100,
]
{0.2 + (20.0 - 0.2)*(1 + (8.0*x)^2)^((0.35 - 1)/2) + 0.015*x^1.3};
\addlegendentry{Analytical law}

\end{axis}
\end{tikzpicture}
\caption{Synthetic non-monotonic constitutive law.}
\label{fig:nonmonotonic_law}
\end{figure}
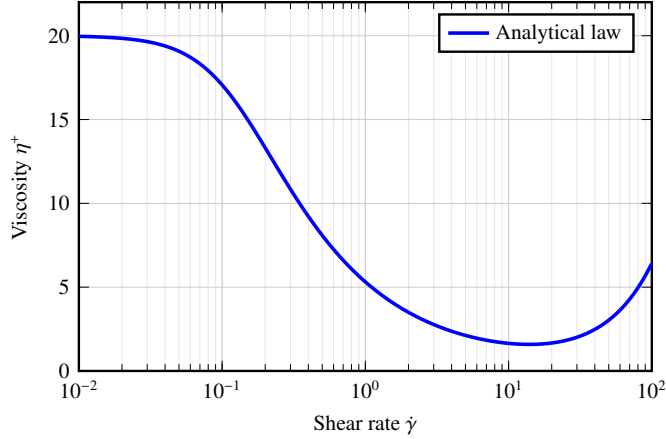
\begin{figure}[t!]
    \centering
    \begin{subfigure}[b]{0.48\linewidth}
        \centering
        \begin{tikzpicture}
        \begin{axis}[
            width=0.95\linewidth,
            xlabel={$x$},
            ylabel={$y$},
            xmin=1.3, xmax=2.7,
            ymin=1.7, ymax=2.8,
            axis equal image,
            grid=major,
            grid style={dashed,gray!30},
            legend style={at={(0.5,0.5)}, anchor=center, font=\small},
            tick label style={font=\small},
            label style={font=\small},
        ]
        \addplot[only marks, mark=*, mark size=0.5pt, color=blue]
            table[x index=1, y index=2, col sep=comma, header=true]
            {SyntheticAN_Phi_150.tex};
        \addlegendentry{Analytical}
        \addplot[only marks, mark=o, mark size=2pt, color=red, each nth point=20, filter discard warning=false, unbounded coords=discard]
            table[x index=1, y index=2, col sep=comma, header=true]
            {SyntheticNN_Phi_150.tex};
        \addlegendentry{Neural network}
        \end{axis}
        \end{tikzpicture}
        \caption{Bubble shape at quasi-steady time.}
        \label{fig:nonmonotonic_shape}
    \end{subfigure}
    \hfill
    \begin{subfigure}[b]{0.48\linewidth}
        \centering
        \begin{tikzpicture}
        \begin{axis}[
            width=0.95\linewidth,
            xlabel={$t$},
            ylabel={$U$},
            xmin=0, xmax=12,
            scaled y ticks=false,
            y tick label style={/pgf/number format/fixed, /pgf/number format/precision=4},
            grid=major,
            grid style={dashed,gray!30},
            legend pos=south east,
            legend style={font=\small},
            tick label style={font=\small},
            label style={font=\small},
        ]
        \addplot[blue, solid, line width=1.0pt]
            table[x=time, y=velocity, col sep=comma]
            {syntheticAN.tex};
        \addlegendentry{Analytical}
        \addplot[red, dashed, line width=1.0pt]
            table[x=time, y=velocity, col sep=comma]
            {syntheticNN.tex};
        \addlegendentry{Neural network}
        \end{axis}
        \end{tikzpicture}
        \caption{Rise velocity versus non-dimensional time.}
        \label{fig:nonmonotonic_rise}
    \end{subfigure}
    \caption{Synthetic non-monotonic case. (a) Bubble shape at quasi-steady time. (b) Rise velocity versus non-dimensional time. In both panels: solid blue is the solver with the analytical constitutive law as closure; red is the solver with the neural-network closure trained on data from the analytical law.}
    \label{fig:nonmonotonic_combined}
\end{figure}
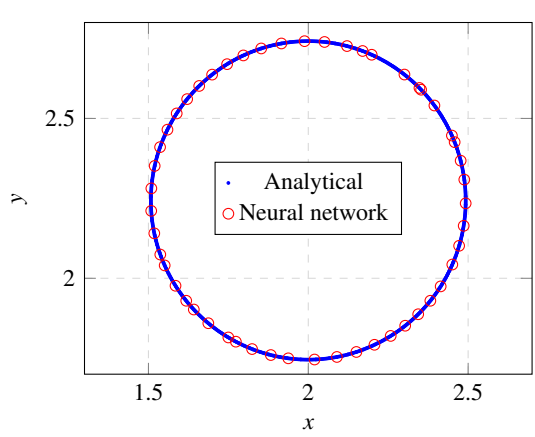
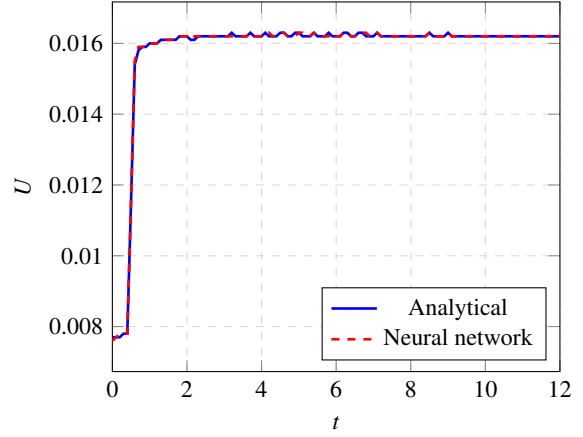
%%%%%%%%%%%%%%%%%%%%%%%%%%%%%%%%%%%%%%%%%%%%%%%%%%%%%%%%%%%%%%%%%%%%%%%%%%%%%%%%%%%%%

\subsection{Droplet Rise Simulations}\label{sec:DropletRiseSimulations}

The experiments use two silicone-based materials, Material A and Material B, containing $1.0$ and $0.5$ weight percent silica respectively. Their measured viscosity-shear-rate curves are shown in \figref{fig:viscosity_curves}; Material A has the higher zero-shear viscosity. For each material, the viscosity at each shear rate was averaged over three independent rheometer trials, and the averaged data were used as input to the simulations. Droplets were generated using two needle gauges, 20G (outer diameter $0.908$~mm) and 25G (outer diameter $0.515$~mm). Cases 1 and 3 use the 20G needle and Cases 2 and 4 use the 25G needle. In our setup the droplet size at detachment is set by the combined effects of needle geometry, ink viscosity, and pinch-off dynamics, and is only weakly sensitive to needle outer diameter in the range tested; the four cases therefore span two ink formulations and two release conditions, giving an independent test of the framework for each material.

The simulations in this section are three-dimensional, matching the experiments. We non-dimensionalize using the following reference scales. The reference length $l_r$ is the maximum horizontal extent of the initial droplet shape extracted from the experiment. The reference density and viscosity are those of the Newtonian background fluid (perfluorodecalin): $\rho_r = \rho_{bg} = 1920~\text{kg/m}^3$ and $\eta_r = \eta_{bg} = 0.0051~\text{Pa}\cdot\text{s}$. The reference gravity is $g_r = 9.81~\text{m/s}^2$. With no characteristic velocity available from the experiments, the reference velocity is set so that the Froude number is unity, $u_r = \sqrt{g_r\,l_r}$, with corresponding time scale $t_r = l_r / u_r$. The Peclet number follows from the scaling relation $Pe = 1/(3\,Cn^2)$ for the chosen Cahn number $Cn$~\citep{magaletti2013sharp}.

\RIII{The computational domain is $4 \times 8 \times 4$, with the rise direction along the second coordinate; the droplet is initially centered in the cross-section at $(2,\, 2,\, 2)$, two units above the bottom wall. The top and bottom walls are no-slip ($u_i = 0$) and the four side walls are free-slip; the phase field and chemical potential satisfy zero normal gradient on all walls ($n_i\,\partial\phi/\partial x_i = 0$, $n_i\,\partial\mu/\partial x_i = 0$), as in the benchmark cases.}

\RII{To assess grid sensitivity of the three-dimensional simulations, Case~1 was run on three meshes, M1--M3, all with base refinement level 6 and interface refinement levels 7, 8, and 9 respectively. The steady-state rise velocity, averaged over $6 \le t \le 8$, is 0.473, 0.359, and 0.347: successive refinement changes the velocity by 32\% and then 3.6\%. Richardson extrapolation of the sequence gives 0.345, within 0.5\% of the M3 value. All three-dimensional results reported in this work use mesh M3.}

\begin{figure}[t!]
    \centering
    \input{Rheology}
    \caption{Comparison of the trained Lipschitz-regularized neural network closure with the averaged rheometer measurements for the two silicone ink formulations. Open circles: experimental rheometer data. Solid blue line: prediction from the trained neural network used as the viscosity closure in the droplet rise simulations. Both axes are non-dimensional, scaled by the reference quantities defined in \secref{sec:DropletRiseSimulations}.}
    \label{fig:nn_viscosity_comparison}
\end{figure}

\figref{fig:nn_viscosity_comparison} shows the trained network alongside the rheometer measurements in non-dimensional form. Figures~\ref{fig:case1_phi}--\ref{fig:case4_phi} show the phase-field evolution for the four cases. Each simulation is initialized with the droplet shape extracted directly from the corresponding experimental frame, so the initial geometries differ between cases. In every case, the droplet evolves toward a near-spherical shape as it rises, regardless of its initial shape.

The Material B curve shows a mild change in slope around $\dot{\gamma} \approx 5\times10^{-3}$, suggesting more than one shear-thinning regime. The neural closure learns this shape directly from the data, without committing to a functional form.

\begin{table}[t!]
\centering
\caption{Simulation parameters for the droplet rise cases --- dimensional.}
\label{tab:dimensional_params}
\begin{tabular}{llllll}
	\toprule
	Parameter & & Case 1 & Case 2 & Case 3 & Case 4\\
	\midrule
	Material  & $\,$   & A &A & B & B\\
	Needle gauge & $\,$ & 20G & 25G & 20G & 25G \\
	Density of background fluid ($kg/m^3$) & $\rho_{bg}$ & 1920 &1920 & 1920 & 1920\\
	Density of droplet ($kg/m^3$) & $\rho_{d}$ & 970 &970 & 970 & 970\\
	Viscosity of background fluid ($Pa\cdot s$) & $\eta_{bg}$ & 0.0051 &0.0051 & 0.0051 & 0.0051\\
	Horizontal Extent of droplet ($mm$) & $d$ & 1.2771 & 1.2800 & 1.4401 & 1.3812\\
	Surface Tension ($N/m$) & $\sigma$ & 0.0047 &0.0047 & 0.0047 & 0.0047\\
	\bottomrule
\end{tabular}
\end{table}

\begin{table}[t!]
\centering
\caption{Simulation parameters for the droplet rise cases --- non-dimensional.}
\begin{tabular}{llllll}
	\toprule
	Parameter & & Case 1 & Case 2 & Case 3 & Case 4\\
	\midrule
	Material  & $\,$   & A &A & B & B\\
	Reynolds number  & $Re$  & 53.8149 &53.9984 & 64.4399 & 60.5272\\
	Froude Number & $Fr$ & 1.0 &1.0 & 1.0 & 1.0\\
	Cahn Number & $Cn$ & 0.01 &0.01 & 0.01 & 0.01\\
	Peclet Number & $Pe$ & 3333.33 &3333.33 & 3333.33 & 3333.33\\
	Weber Number & $We$ & 6.536 & 6.566 & 8.311 & 7.645 \\
	Density of background fluid & $\rho_{bg}$ & 1.0 &1.0 & 1.0 & 1.0\\
	Density of droplet & $\rho_{d}$ & 0.51 &0.51 & 0.51 & 0.51\\
	Viscosity of background fluid & $\eta_{bg}$ & 1.0 &1.0 & 1.0 & 1.0\\
	\bottomrule
\end{tabular}
\end{table}

To quantify shape agreement, we compute the Hausdorff~\citep{huttenlocher2002comparing} and Chamfer~\citep{borgefors1988hierarchical} distances between the simulated droplet contour and the experimental contour extracted from the high-speed video, for each of the four cases. For two finite point sets $A$ and $B$, these are defined as
\begin{equation}\label{eq:hausdorff}
    d_H(A, B) = \max\!\left\{ \max_{a \in A} \min_{b \in B} \|a - b\|,\; \max_{b \in B} \min_{a \in A} \|a - b\| \right\},
\end{equation}
\begin{equation}\label{eq:chamfer}
    d_C(A, B) = \frac{1}{|A|} \sum_{a \in A} \min_{b \in B} \|a - b\| \;+\; \frac{1}{|B|} \sum_{b \in B} \min_{a \in A} \|a - b\|,
\end{equation}
where $\|\cdot\|$ is the Euclidean norm. The Hausdorff distance is the worst-case nearest-neighbor deviation; the Chamfer distance is the symmetric mean nearest-neighbor deviation. \tableref{tab:shape_metrics_droplet} reports the values for all four cases.

\begin{table}[t!]
\centering
\caption{Hausdorff and Chamfer distances between the simulated and experimentally observed droplet contours for the four droplet rise cases. Each contour is normalized by the average of its horizontal and vertical extents ($l_{r,\mathrm{avg}}$) before comparison, accounting for the changing aspect ratio of deforming droplets; the reported values are dimensionless fractions of the average droplet size. Smaller values indicate closer agreement.}
\label{tab:shape_metrics_droplet}
\begin{tabular}{lllll}
	\toprule
	Metric & Case 1 & Case 2 & Case 3 & Case 4 \\
	\midrule
	Hausdorff distance & 0.044282 & 0.078821 & 0.050732 & 0.066279 \\
	Chamfer distance   & 0.040280 & 0.062229 & 0.051514 & 0.056702 \\
	\bottomrule
\end{tabular}
\end{table}

\begin{figure}[t!]
    \centering
    \input{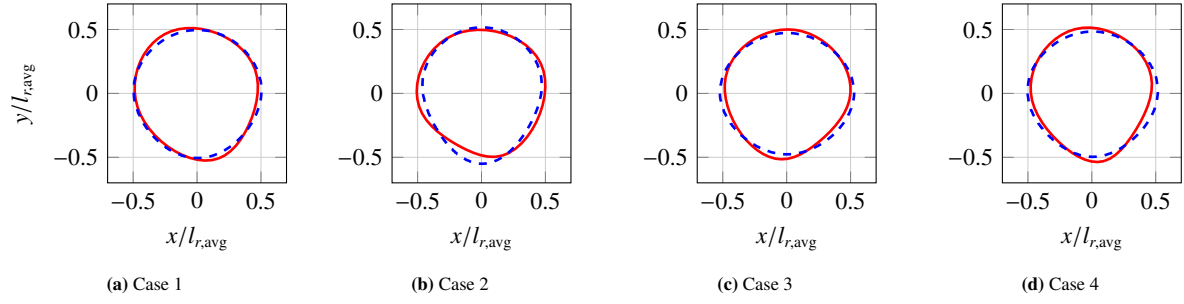}
    \caption{Overlay of simulated and experimental droplet contours for the four cases. Solid red line: experimental contour. Dashed blue line: simulated contour.}
    \label{fig:contour_overlays}
\end{figure}

%%%%%%%%%%%%%%%%%%%%%%%%%%%%%%%%%%%%%%%%%%%%%%%%%%%%%%%%%%%%%%%%%%%%%%%%%%%%%%%%%%%%%

\begin{figure}[H]
\centering
\begin{subfigure}{0.18\textwidth}
\centering
\includegraphics[width=\linewidth,
        trim={65.0cm 0.8cm 65.0cm 1.3cm},clip]{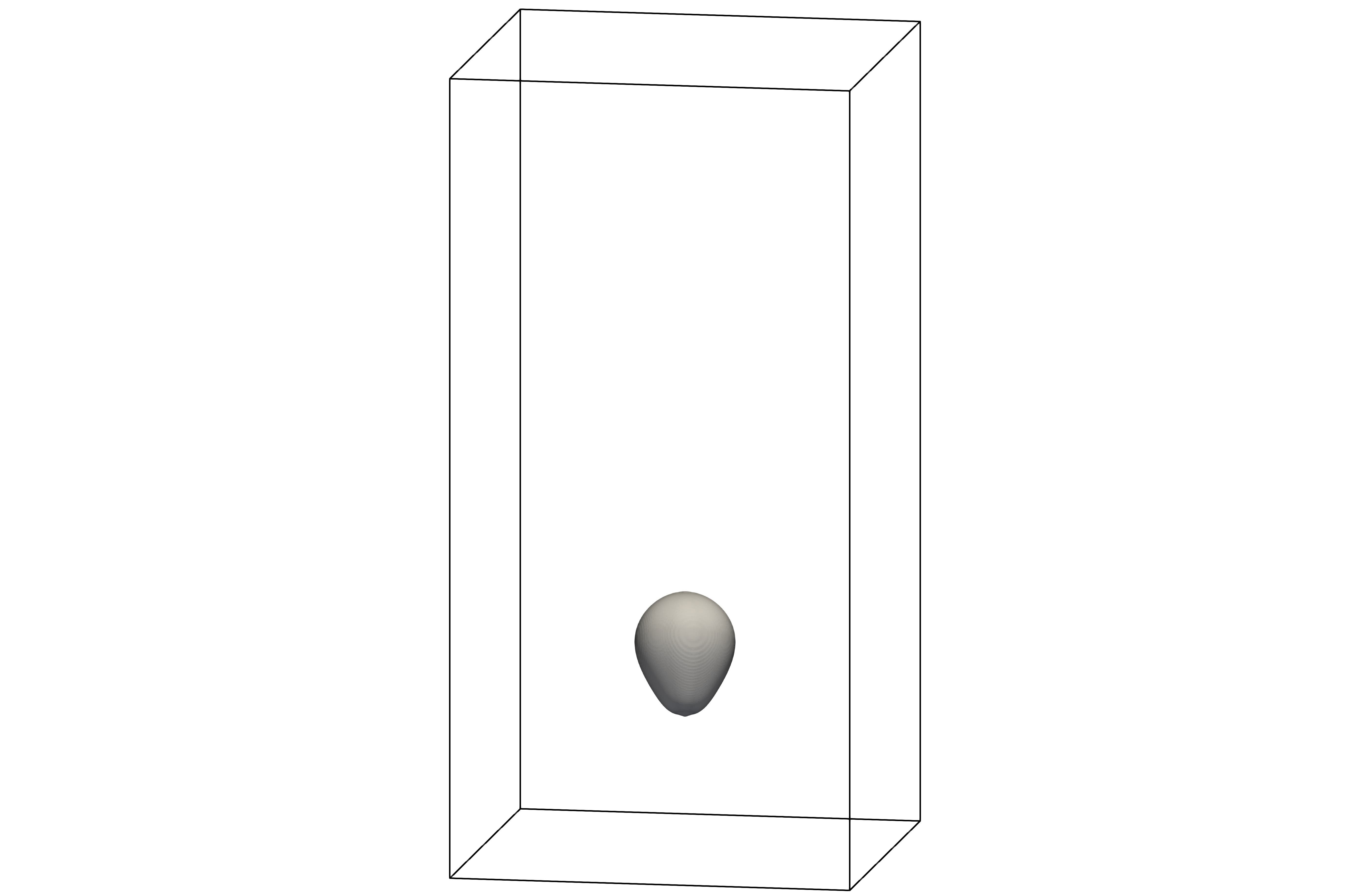}
\caption{t = 0}
\end{subfigure}
\begin{subfigure}{0.18\textwidth}
\centering
\includegraphics[width=\linewidth,
        trim={64.9cm 0.7cm 64.9cm 1.2cm},clip]{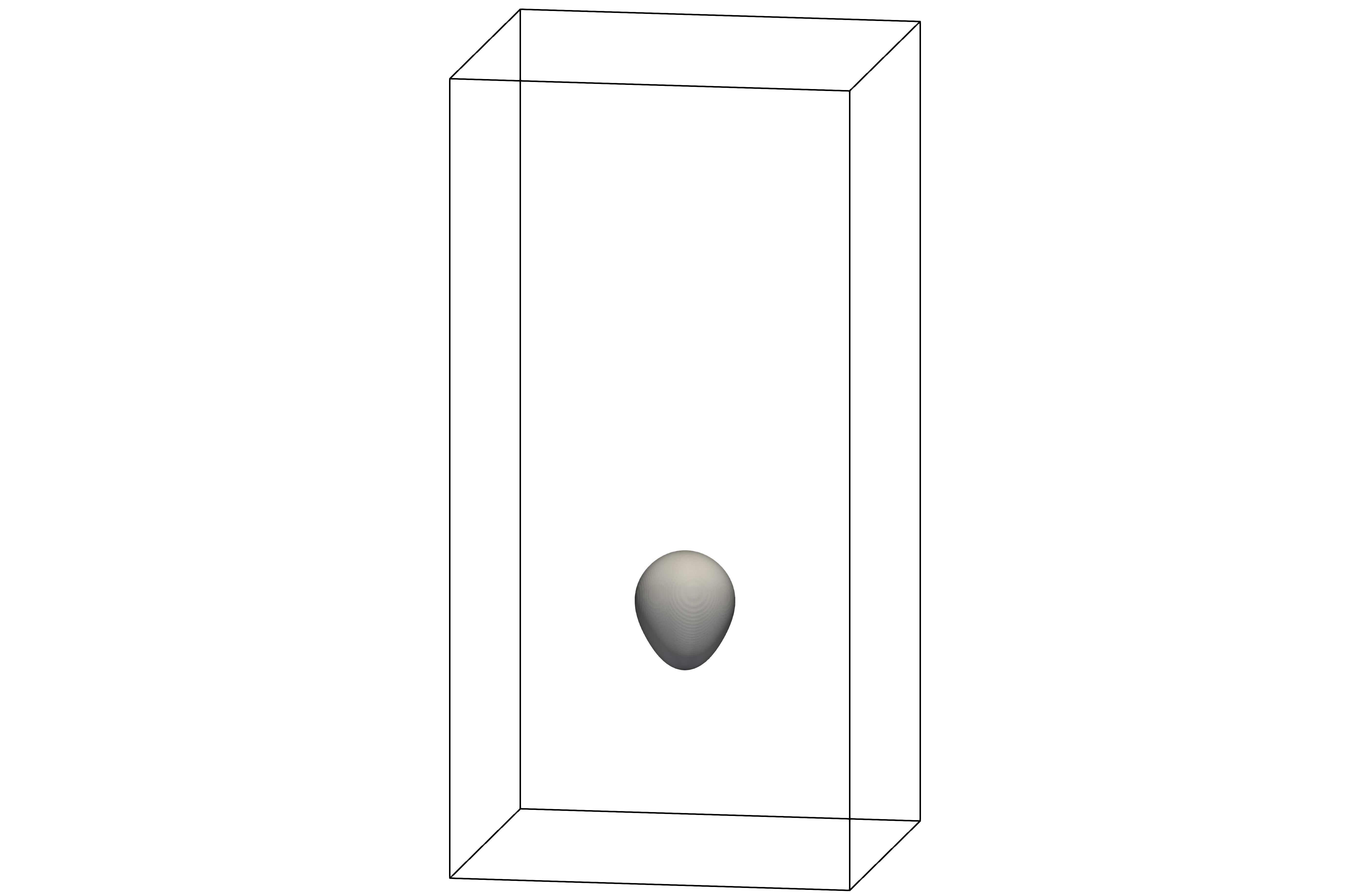}
\caption{t = 2.5}
\end{subfigure}
\begin{subfigure}{0.18\textwidth}
\centering
\includegraphics[width=\linewidth,
        trim={64.9cm 0.7cm 64.9cm 1.2cm},clip]{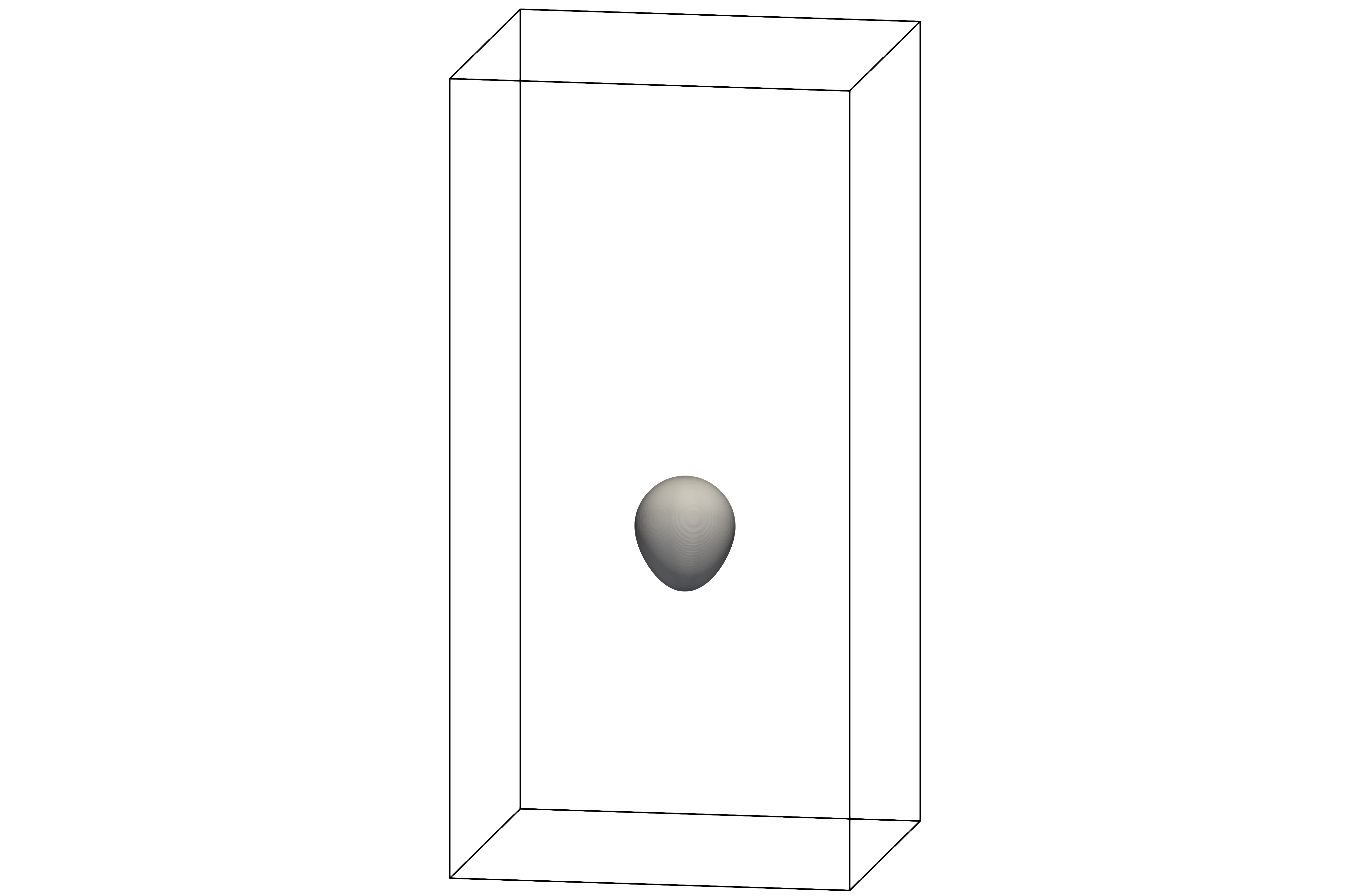}
\caption{t = 5}
\end{subfigure}
\begin{subfigure}{0.18\textwidth}
\centering
\includegraphics[width=\linewidth,
        trim={64.9cm 0.7cm 64.9cm 1.2cm},clip]{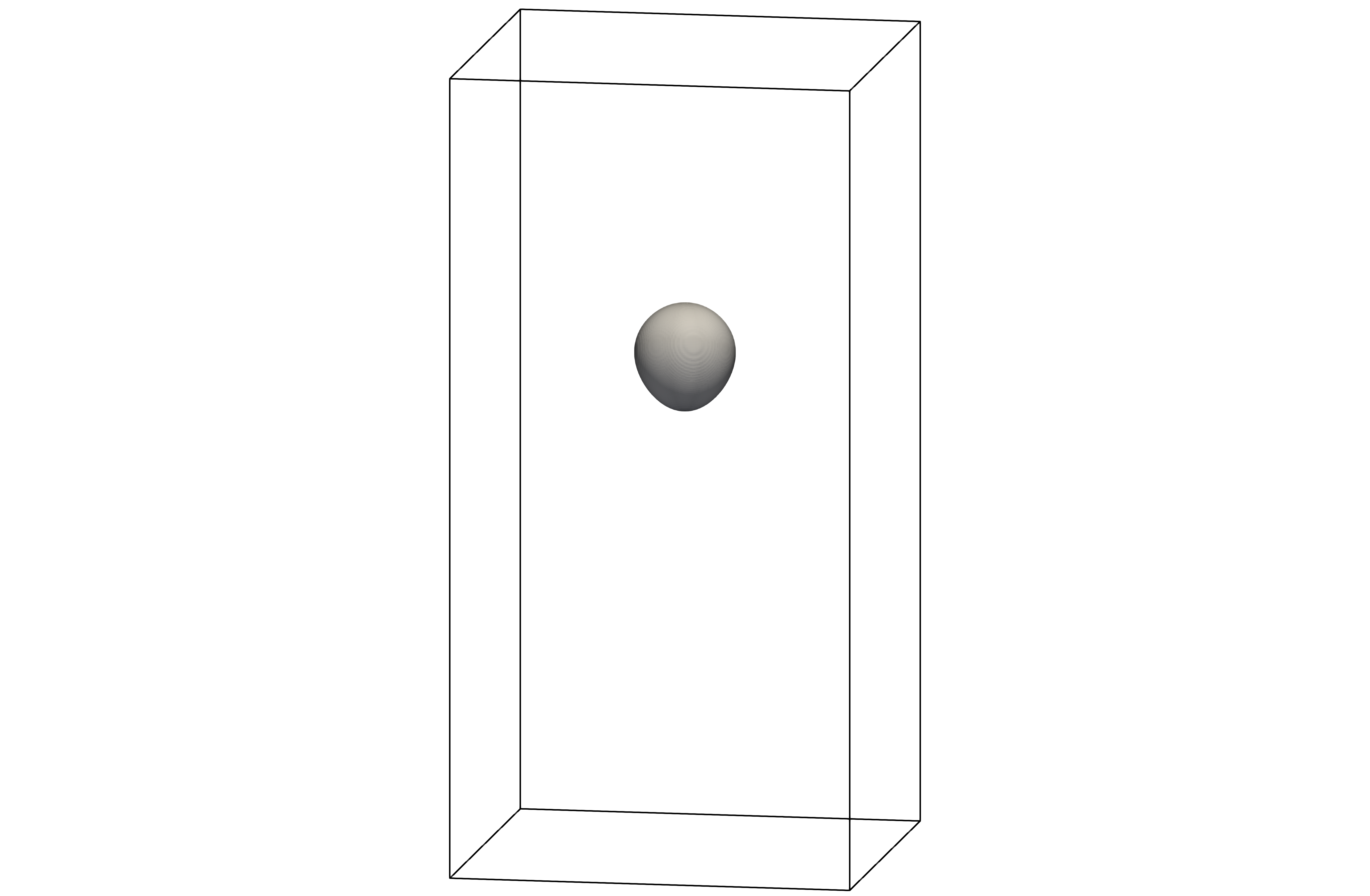}
\caption{t = 10}
\end{subfigure}
\caption{Case 1 (Material A, 20G needle): phase-field evolution.}\label{fig:case1_phi}
\end{figure}
%%%%%%%%%%%%%%%%%%%%%%%%%%%%%%%%%%%%%%%%%%%%%%%%%%%%%%%%%%%%%%%%%%%%%%%%%%%%%%%%%%%%%

\begin{figure}[H]
\centering
\begin{subfigure}{0.18\textwidth}
\centering
\includegraphics[width=\linewidth,
        trim={64.8cm 1.2cm 65.2cm 1.0cm},clip]{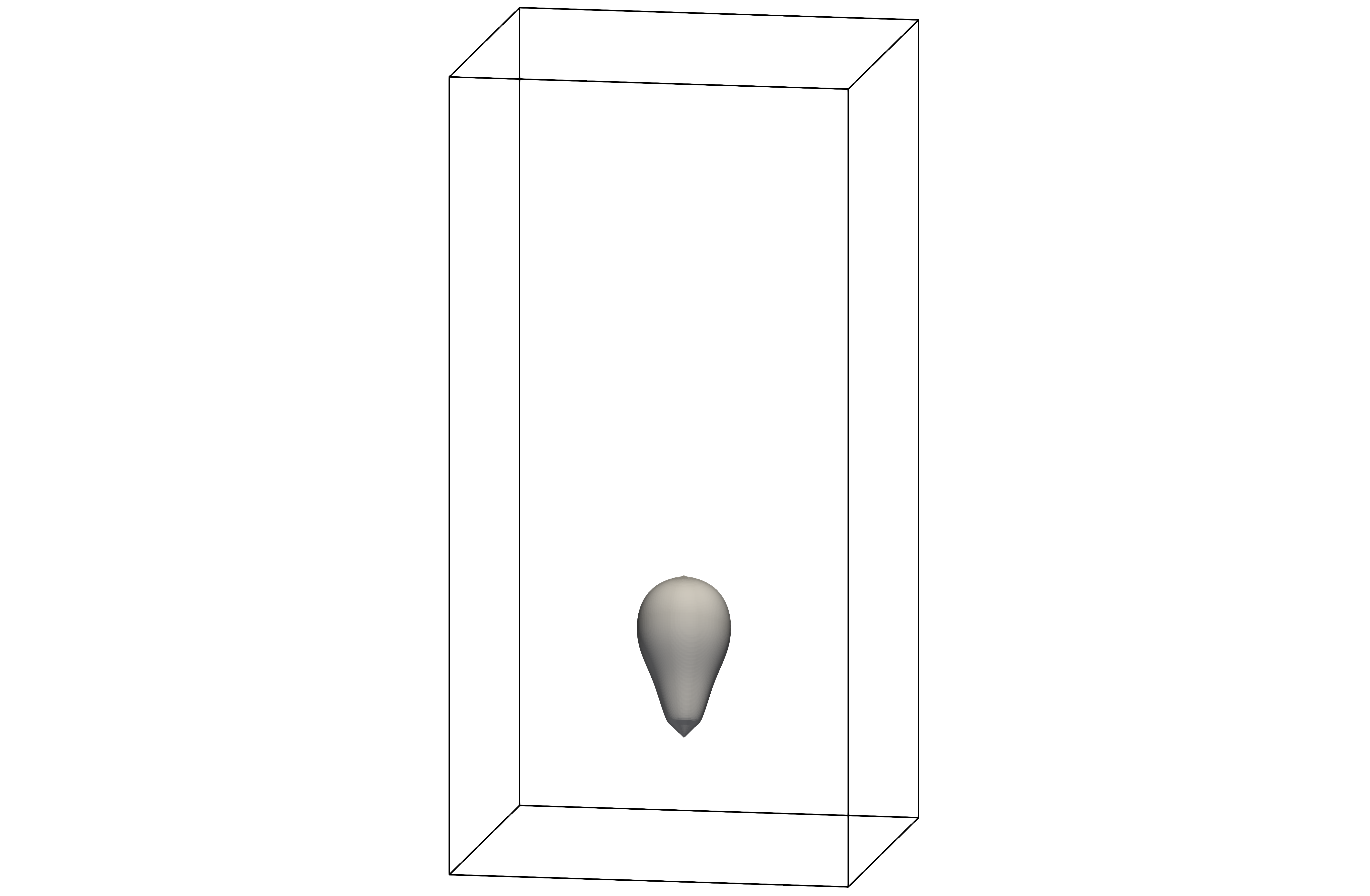}
\caption{t = 0}
\end{subfigure}
\begin{subfigure}{0.18\textwidth}
\centering
\includegraphics[width=\linewidth,
        trim={64.8cm 1.2cm 65.2cm 1.0cm},clip]{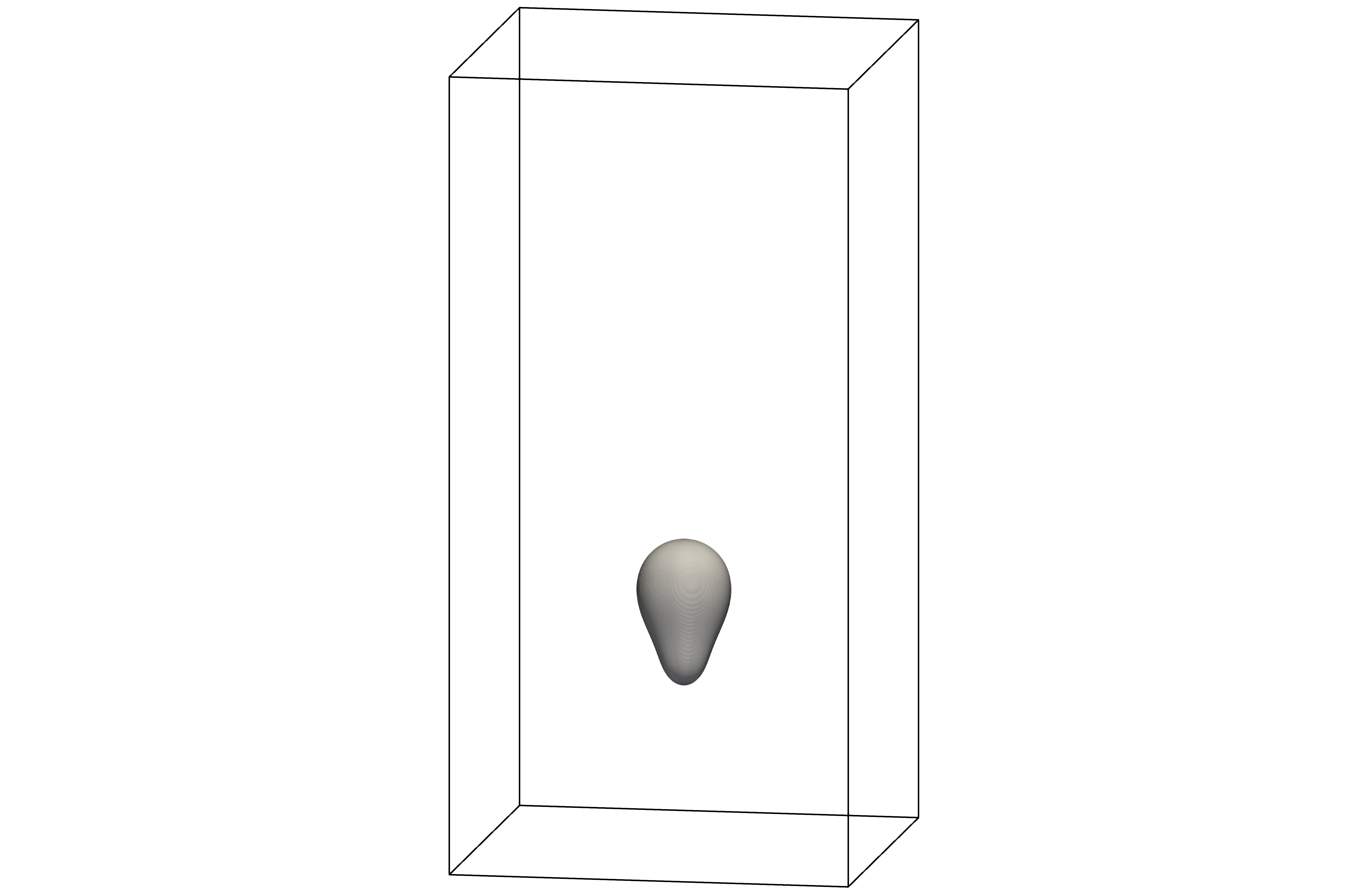}
\caption{t = 2.5}
\end{subfigure}
\begin{subfigure}{0.18\textwidth}
\centering
\includegraphics[width=\linewidth,
        trim={64.8cm 1.2cm 65.2cm 1.0cm},clip]{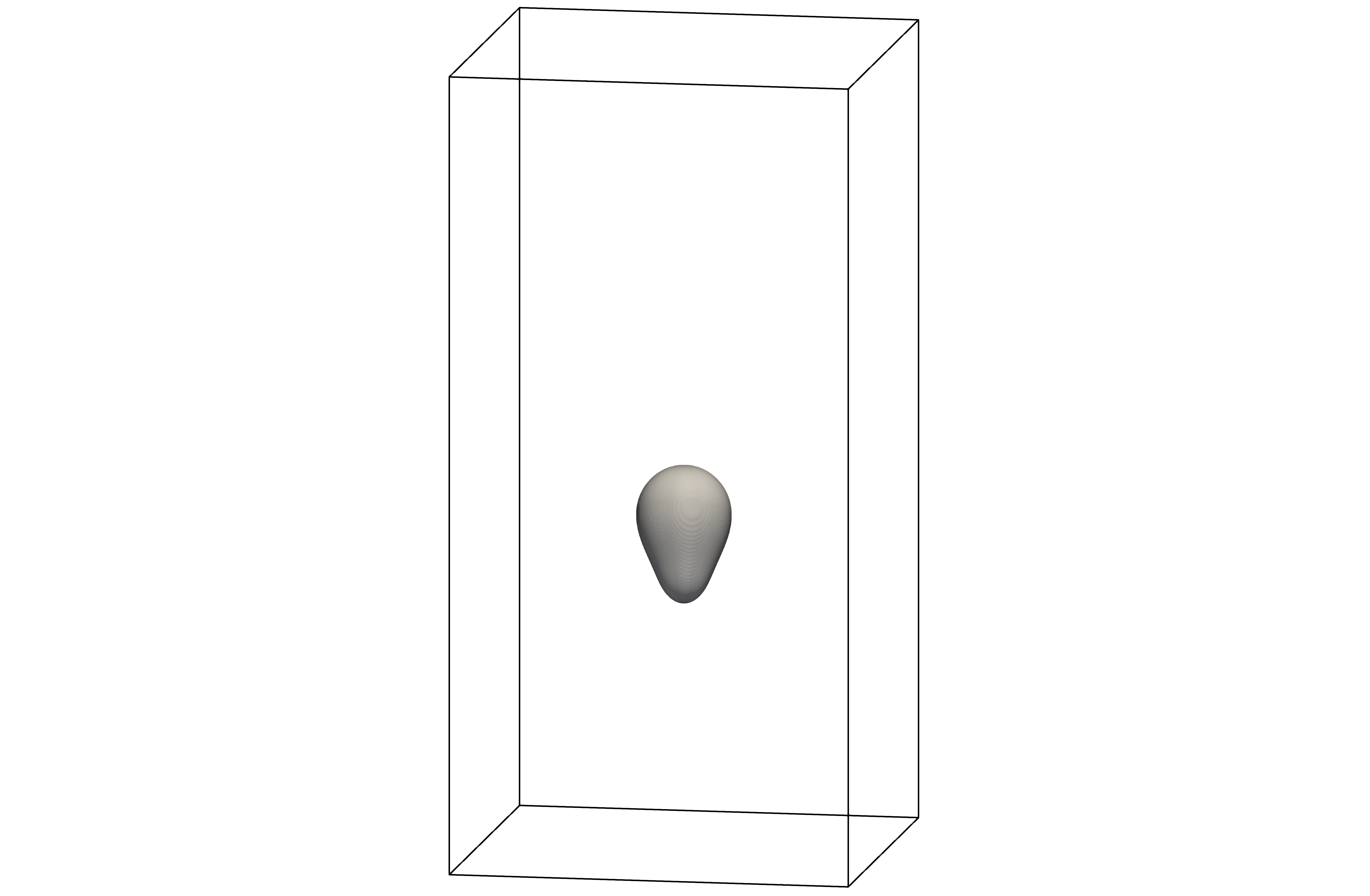}
\caption{t = 5}
\end{subfigure}
\begin{subfigure}{0.18\textwidth}
\centering
\includegraphics[width=\linewidth,
        trim={64.8cm 1.2cm 65.2cm 1.0cm},clip]{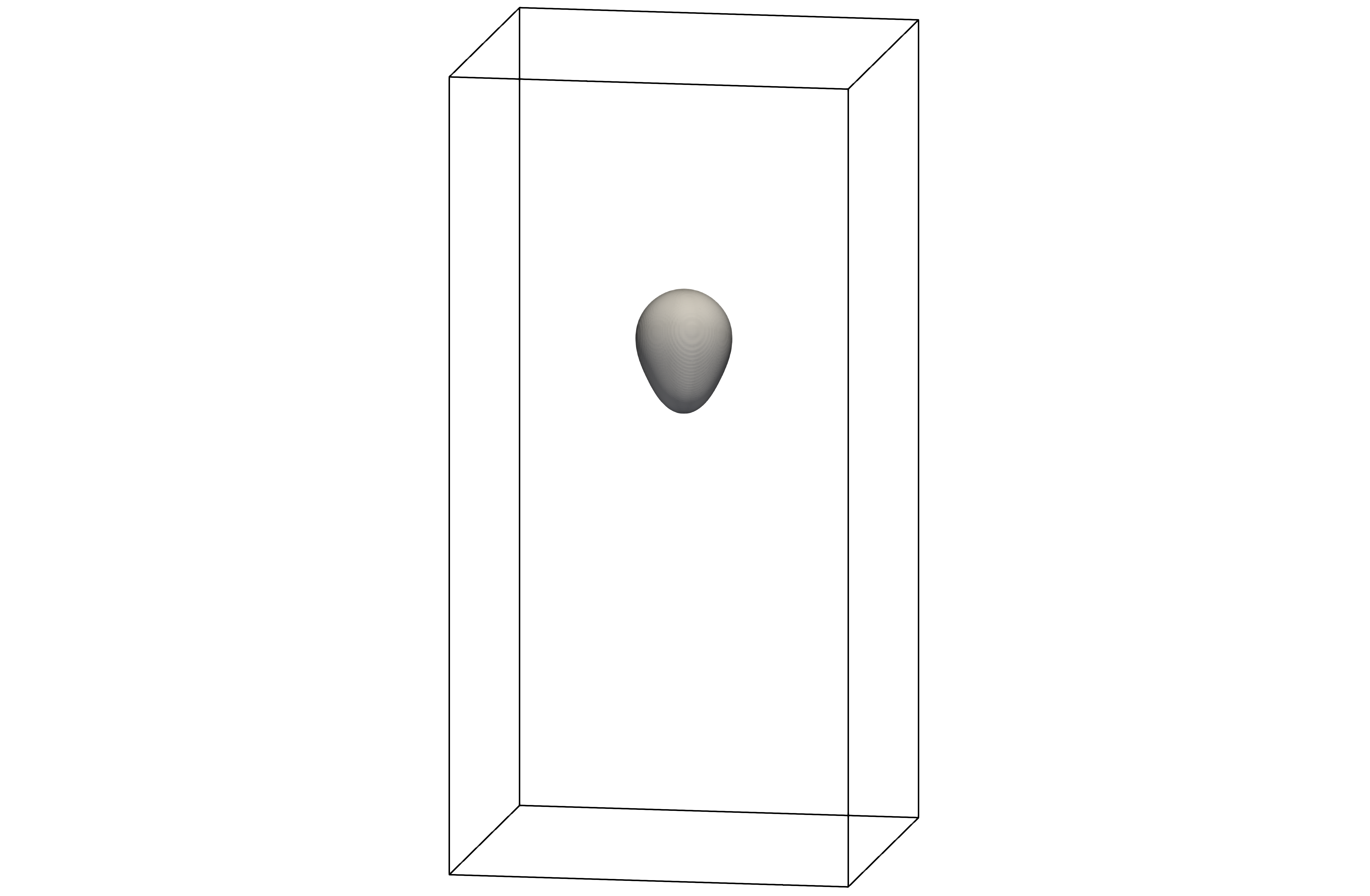}
\caption{t = 10}
\end{subfigure}
\caption{Case 2 (Material A, 25G needle): phase-field evolution.}\label{fig:case2_phi}
\end{figure}

%%%%%%%%%%%%%%%%%%%%%%%%%%%%%%%%%%%%%%%%%%%%%%%%%%%%%%%%%%%%%%%%%%%%%%%%%%%%%%%%%%%%%
%%%%%%%%%%%%%%%%%%%%%%%%%%%%%%%%%%%%%%%%%%%%%%%%%%%%%%%%%%%%%%%%%%%%%%%%%%%%%%%%%%%%%

\begin{figure}[H]
\centering
\begin{subfigure}{0.18\textwidth}
\centering
\includegraphics[width=\linewidth,
        trim={64.8cm 1.1cm 65.1cm 1.1cm},clip]{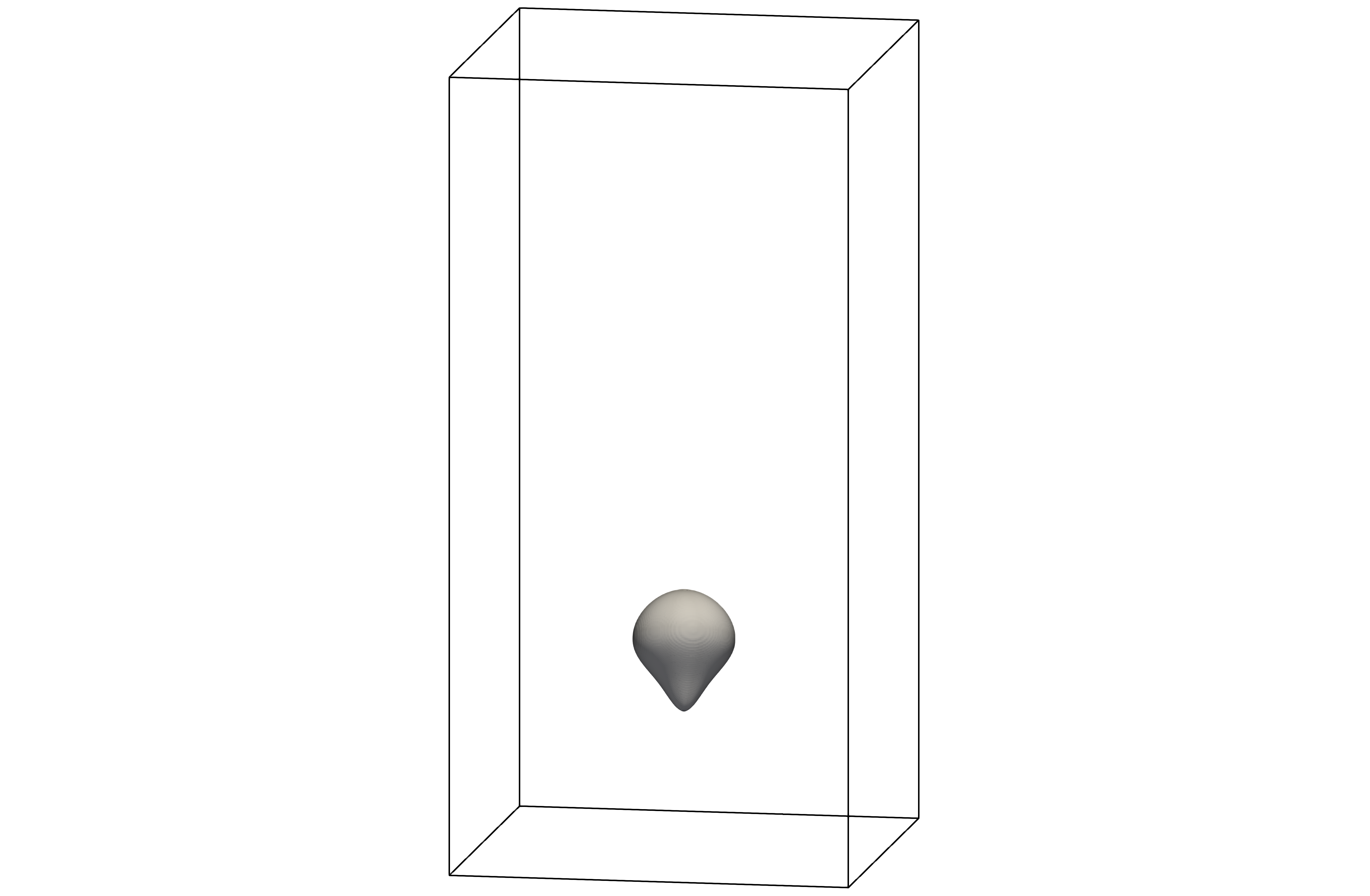}
\caption{t = 0}
\end{subfigure}
\begin{subfigure}{0.18\textwidth}
\centering
\includegraphics[width=\linewidth,
        trim={64.8cm 1.1cm 65.1cm 1.1cm},clip]{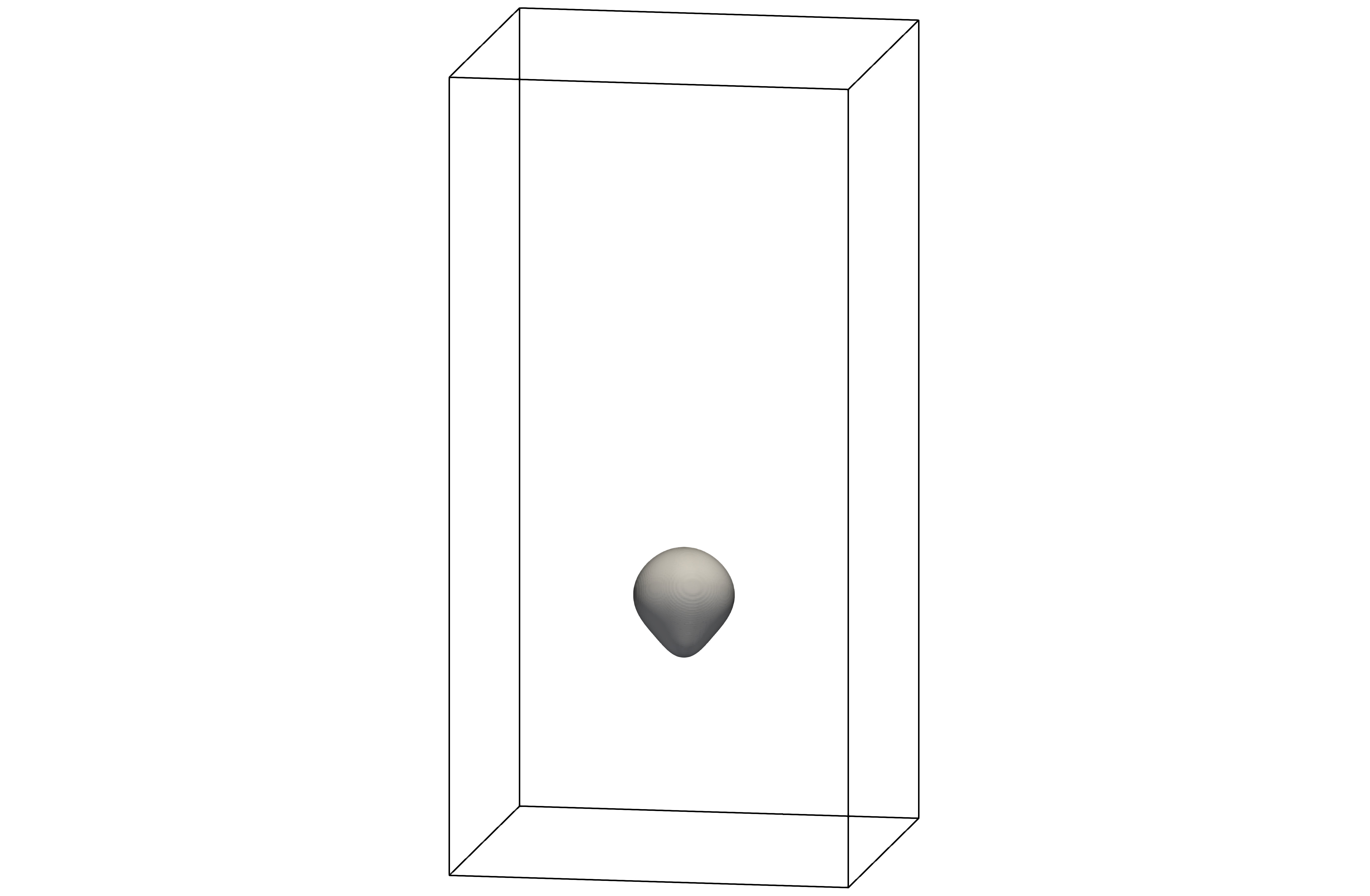}
\caption{t = 2.5}
\end{subfigure}
\begin{subfigure}{0.18\textwidth}
\centering
\includegraphics[width=\linewidth,
        trim={64.8cm 1.1cm 65.1cm 1.1cm},clip]{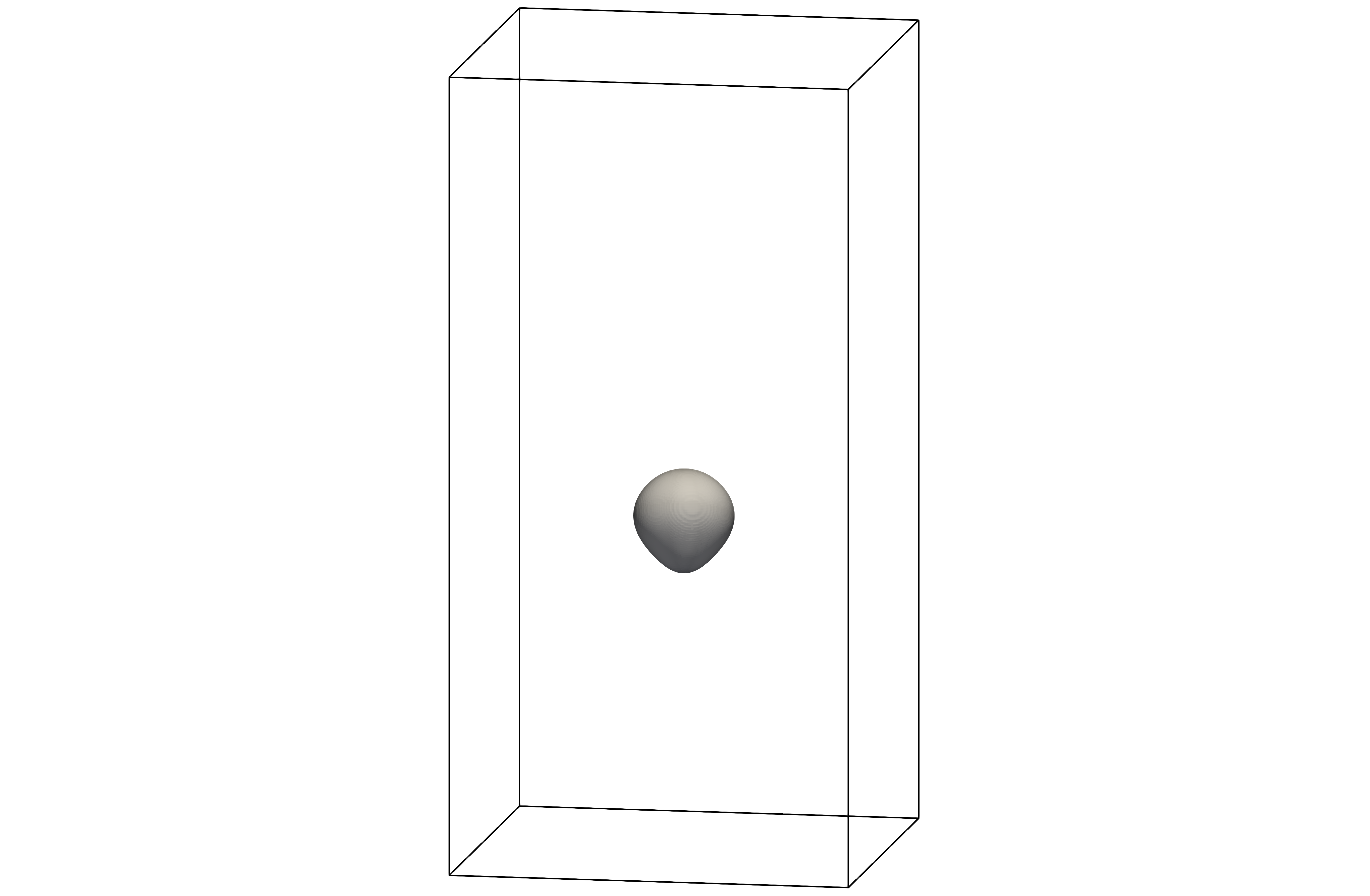}
\caption{t = 5}
\end{subfigure}
\begin{subfigure}{0.18\textwidth}
\centering
\includegraphics[width=\linewidth,
        trim={64.8cm 1.1cm 65.1cm 1.1cm},clip]{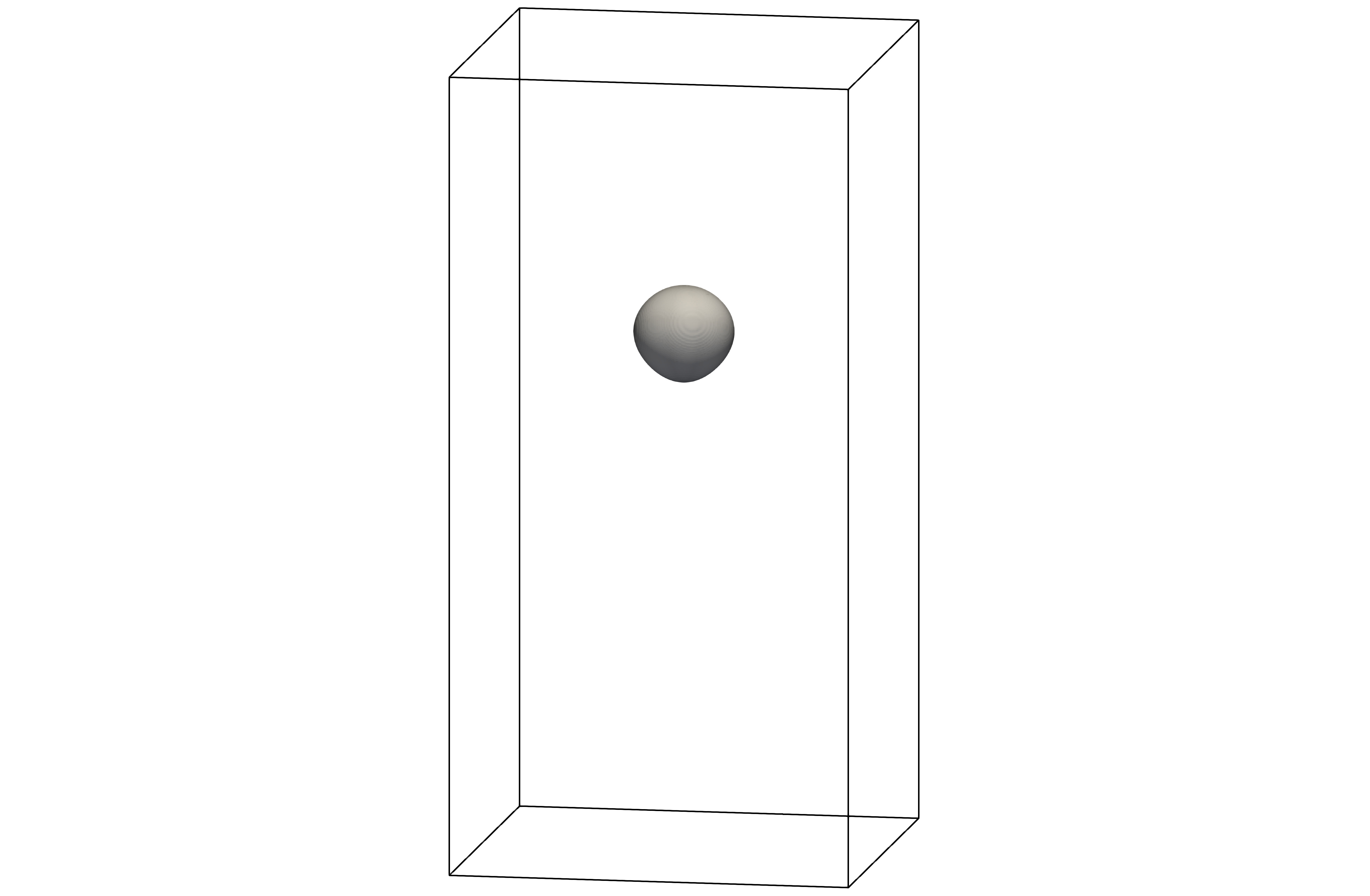}
\caption{t = 10}
\end{subfigure}
\caption{Case 3 (Material B, 20G needle): phase-field evolution.}\label{fig:case3_phi}
\end{figure}
%%%%%%%%%%%%%%%%%%%%%%%%%%%%%%%%%%%%%%%%%%%%%%%%%%%%%%%%%%%%%%%%%%%%%%%%%%%%%%%%%%%%%

\begin{figure}[H]
\centering
\begin{subfigure}{0.18\textwidth}
\centering
\includegraphics[width=\linewidth,
        trim={64.7cm 1.2cm 65.0cm 0.5cm},clip]{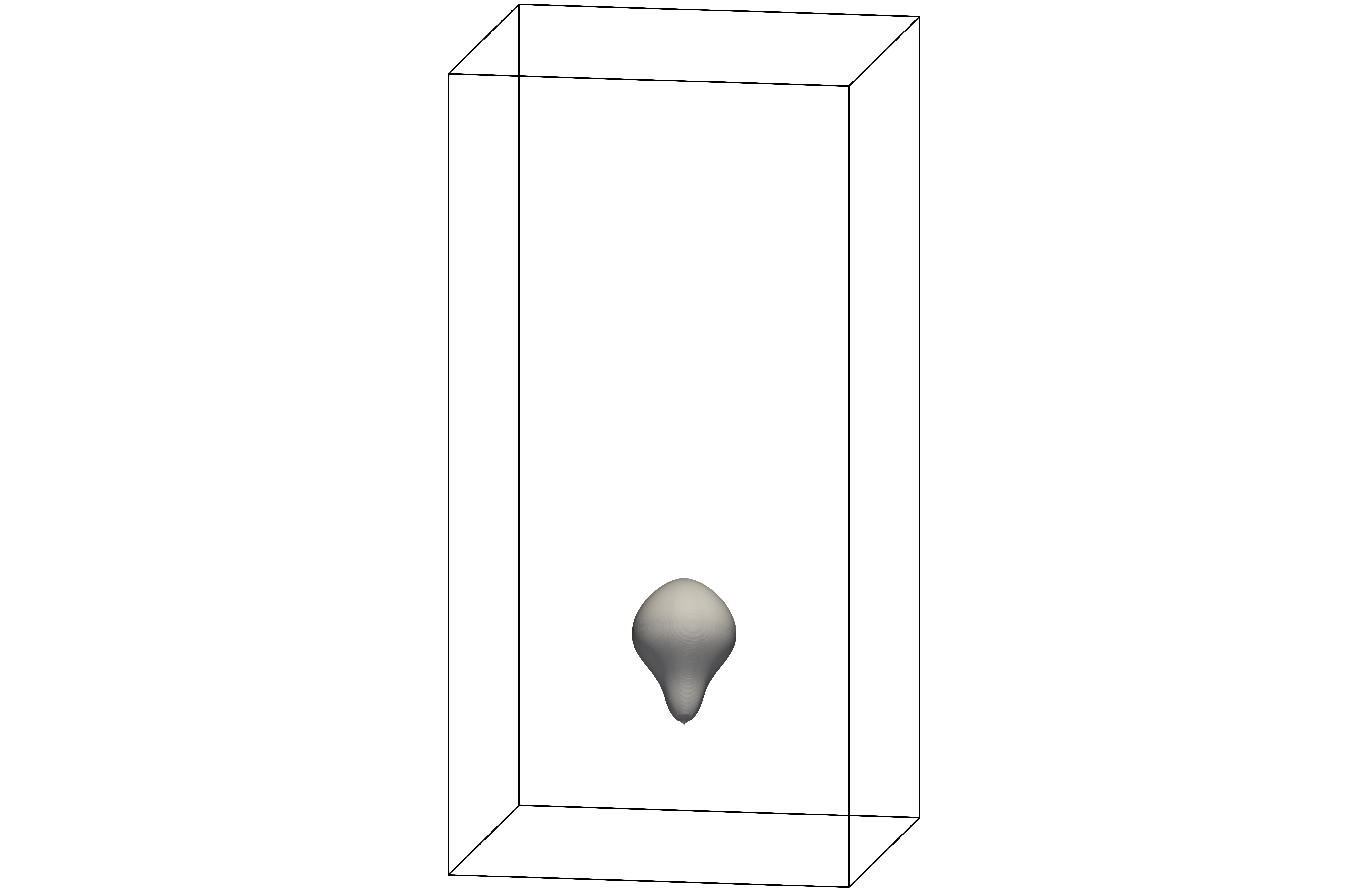}
\caption{t = 0}
\end{subfigure}
\begin{subfigure}{0.18\textwidth}
\centering
\includegraphics[width=\linewidth,
        trim={64.7cm 1.2cm 65.0cm 0.5cm},clip]{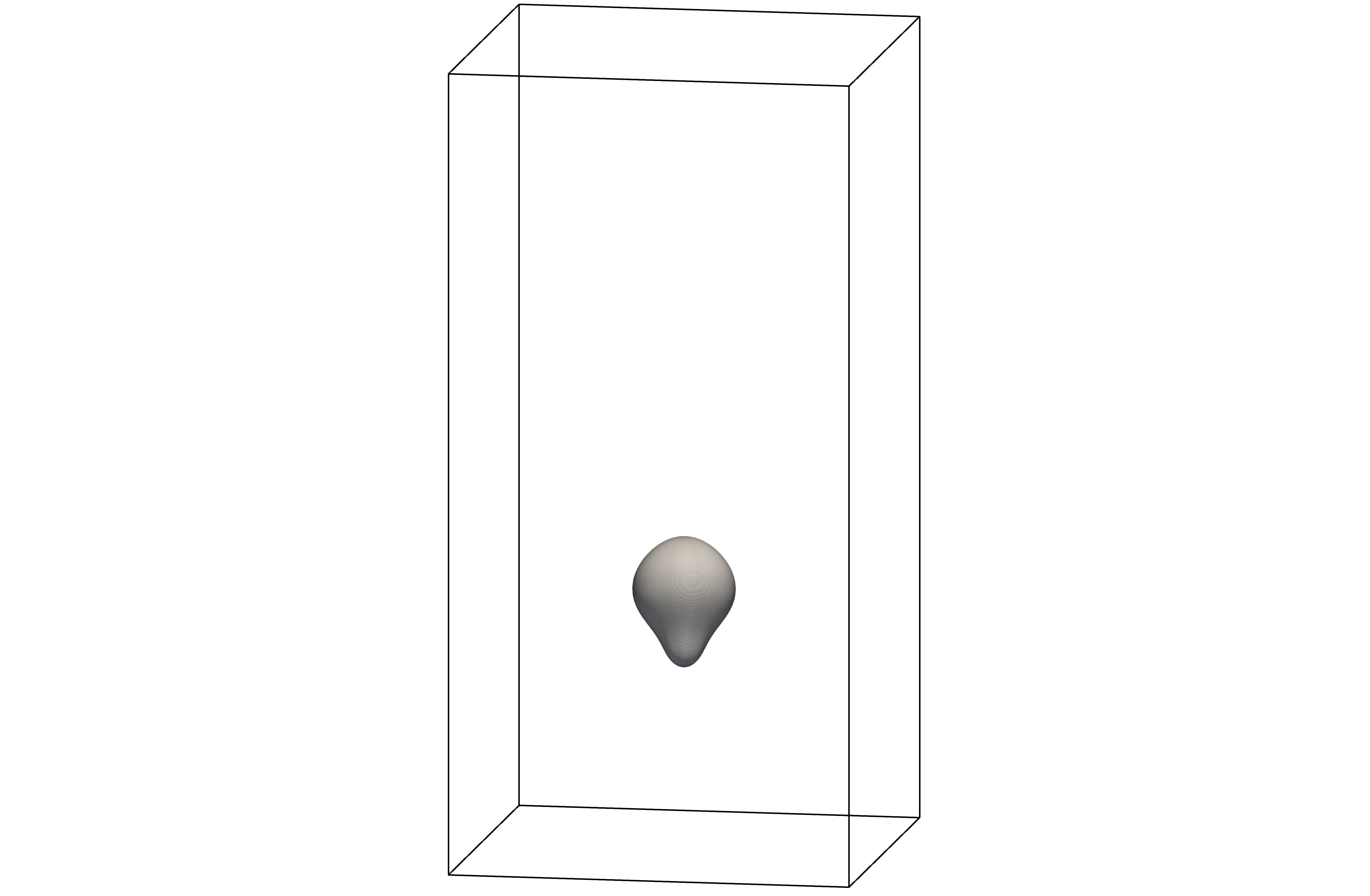}
\caption{t = 2.5}
\end{subfigure}
\begin{subfigure}{0.18\textwidth}
\centering
\includegraphics[width=\linewidth,
        trim={64.7cm 1.2cm 65.0cm 0.5cm},clip]{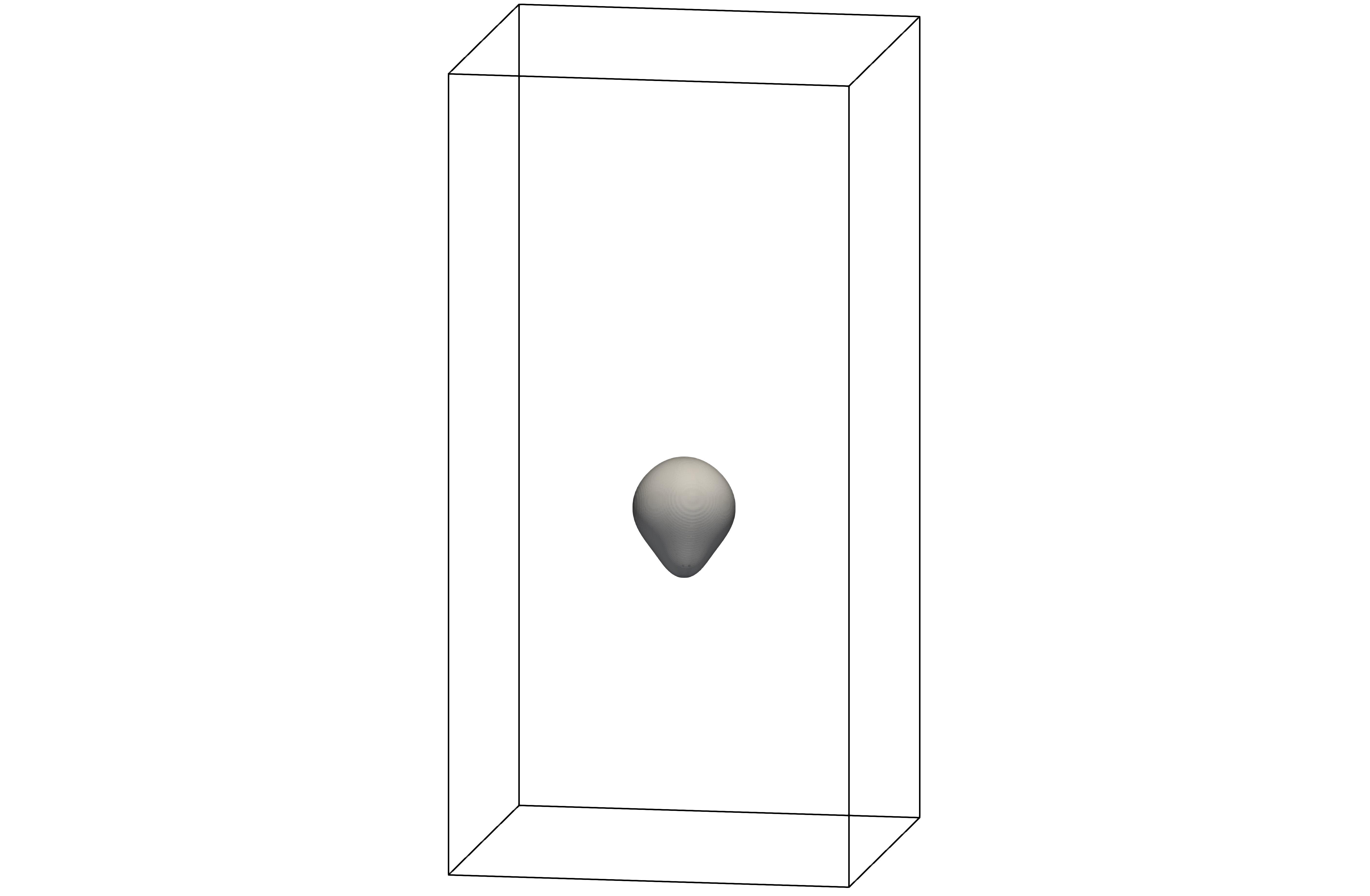}
\caption{t = 5}
\end{subfigure}
\begin{subfigure}{0.18\textwidth}
\centering
\includegraphics[width=\linewidth,
        trim={64.7cm 1.2cm 65.0cm 0.5cm},clip]{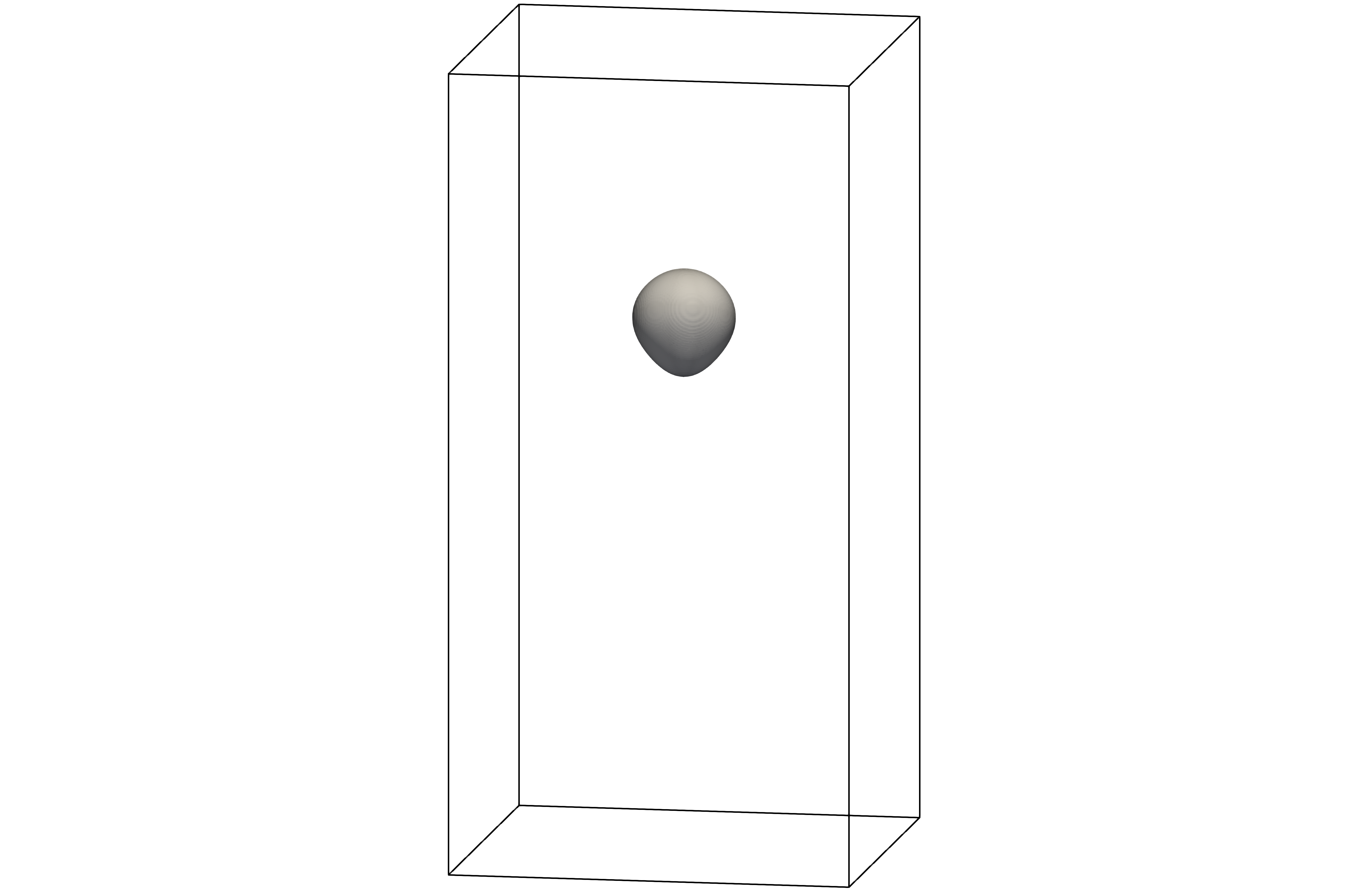}
\caption{t = 10}
\end{subfigure}
\caption{Case 4 (Material B, 25G needle): phase-field evolution.}\label{fig:case4_phi}
\end{figure}
%%%%%%%%%%%%%%%%%%%%%%%%%%%%%%%%%%%%%%%%%%%%%%%%%%%%%%%%%%%%%%%%%%%%%%%%%%%%%%%%%%%%%

The Hausdorff distances span $0.044$--$0.079$ and the Chamfer distances span $0.040$--$0.062$, with the smallest values for Case~1 and the largest for Case~2. Each contour is normalized by the average of its horizontal and vertical extents before comparison, so these values are fractions of the average droplet size: simulated and experimental contour points are within $4$--$6\%$ of the average droplet size (Chamfer), and the worst-case deviation is bounded by about $8\%$ (Hausdorff). \figref{fig:contour_overlays} shows the simulated and experimental contours overlaid for each case, with the simulated shape closely tracking the experimental one across all four. The neural-constitutive workflow therefore reproduces the experimentally observed droplet shape to within a few percent across all four cases.

\begin{figure}[t!]
    \centering
    % Row 1
    \begin{subfigure}[b]{0.49\textwidth}
        \centering
        \includegraphics[width=0.9\textwidth]{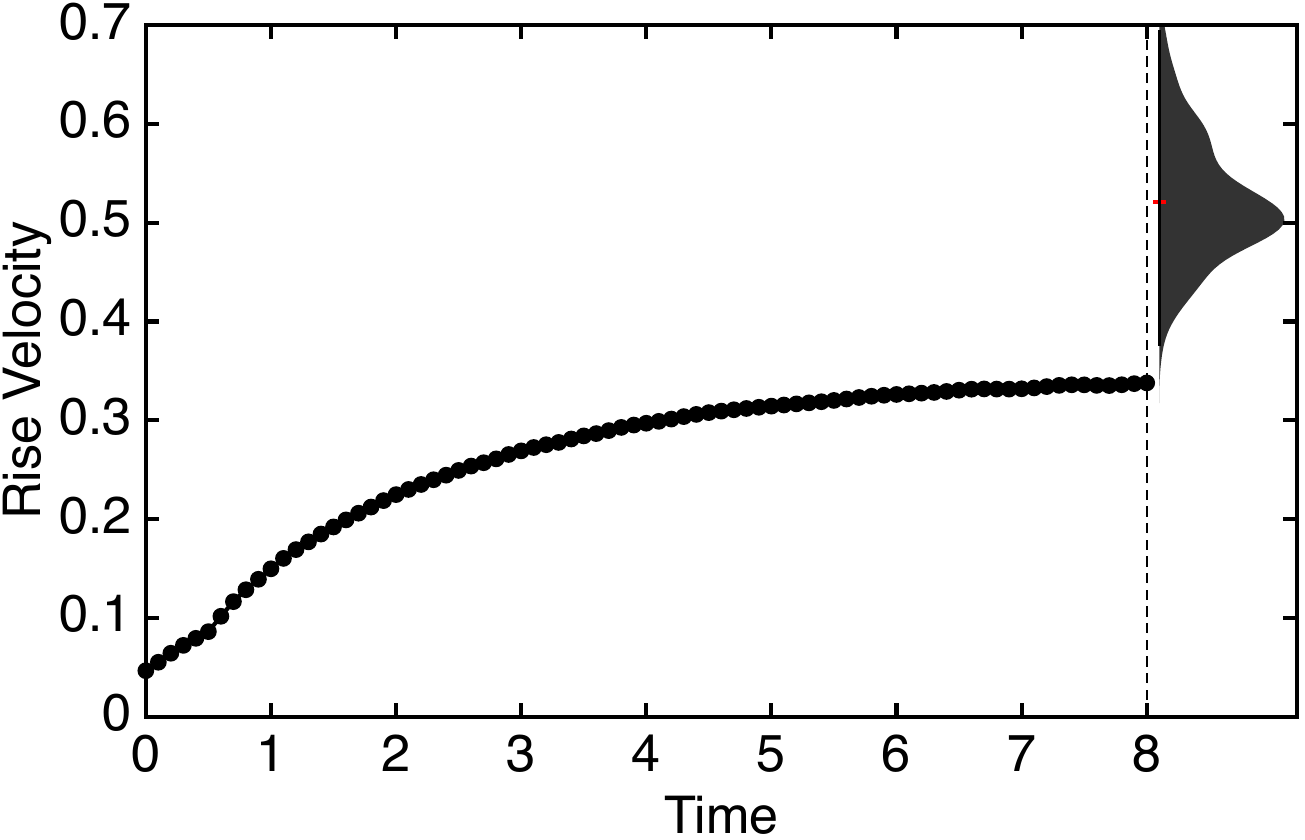}
        \caption{Case 1}
    \end{subfigure}
    \begin{subfigure}[b]{0.49\textwidth}
        \centering
        \includegraphics[width=0.9\textwidth]{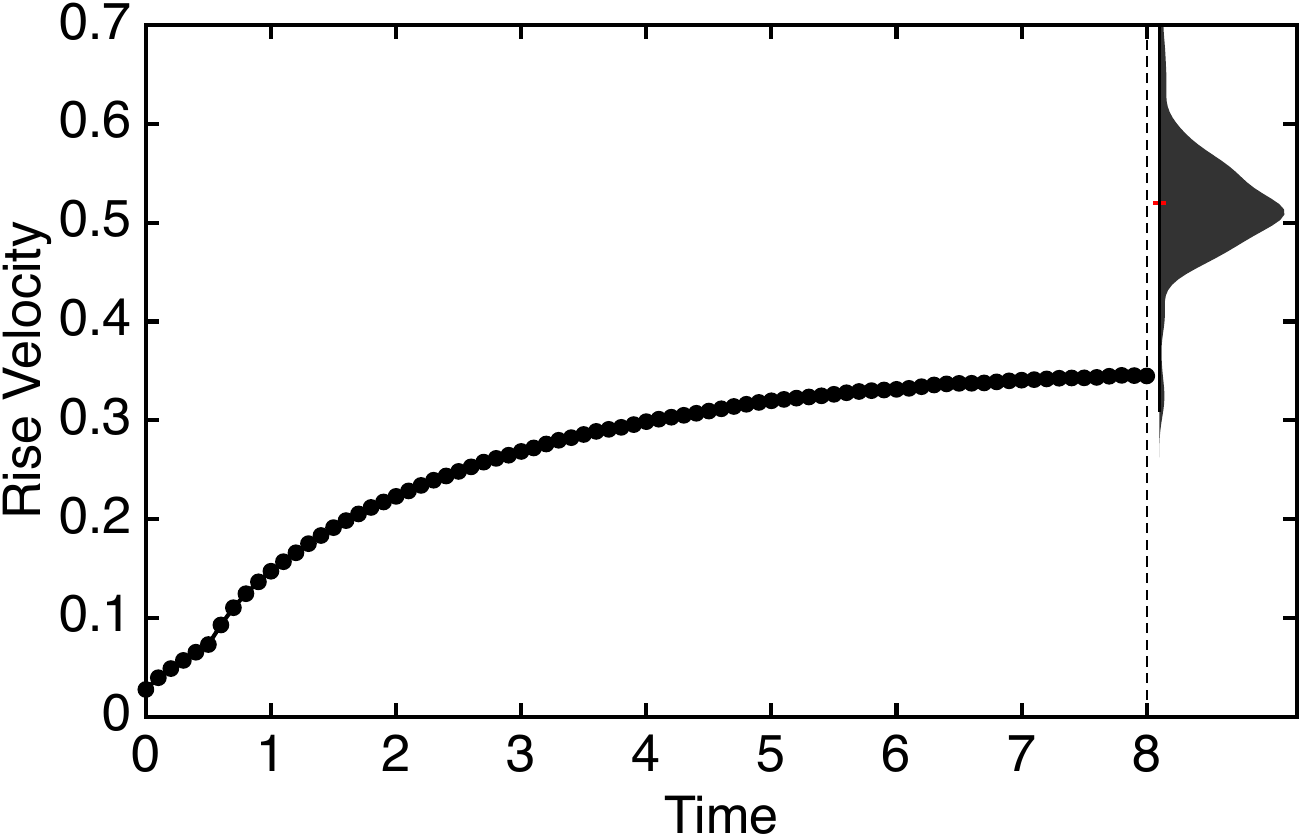}
        \caption{Case 2}
    \end{subfigure}
    % Row 2
    \begin{subfigure}[b]{0.49\textwidth}
        \centering
        \includegraphics[width=0.9\textwidth]{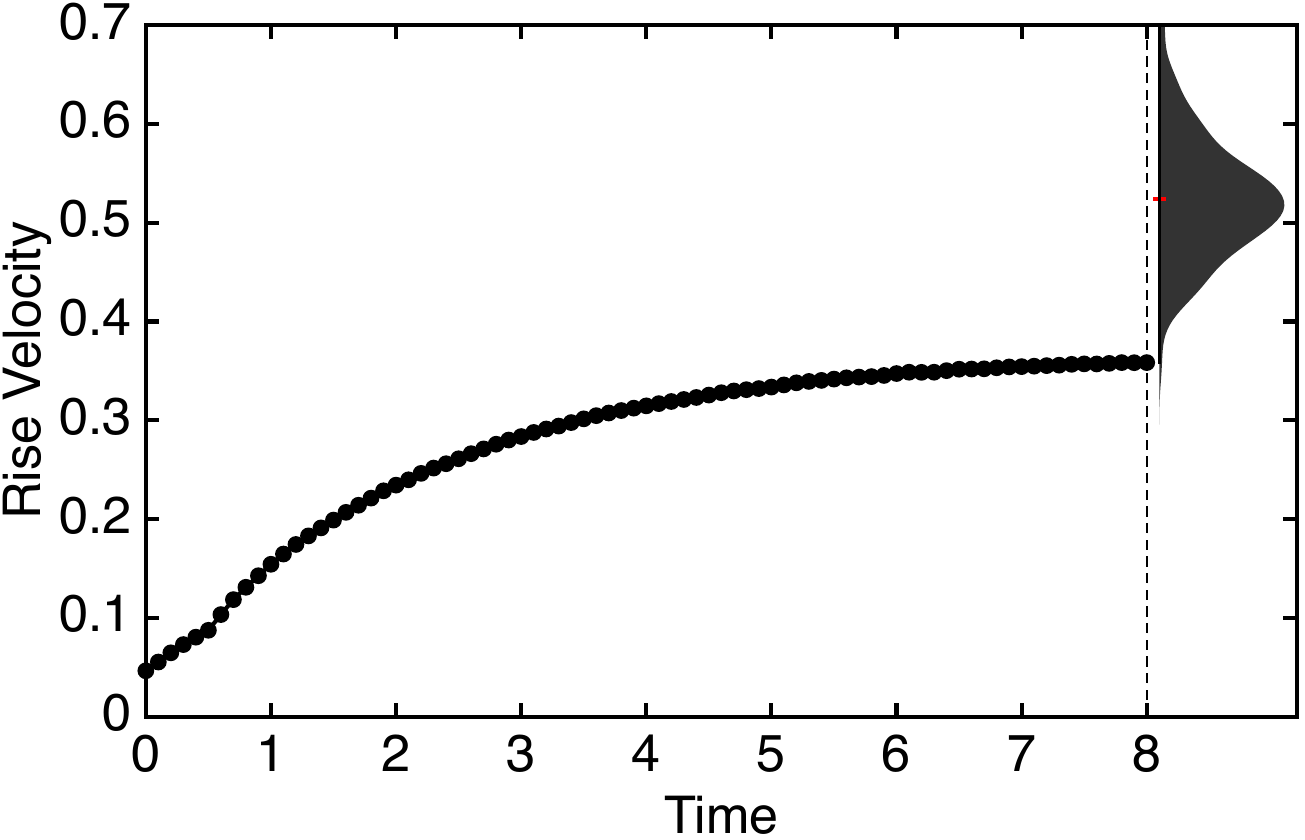}
        \caption{Case 3}
    \end{subfigure}
    \begin{subfigure}[b]{0.49\textwidth}
        \centering
        \includegraphics[width=0.9\textwidth]{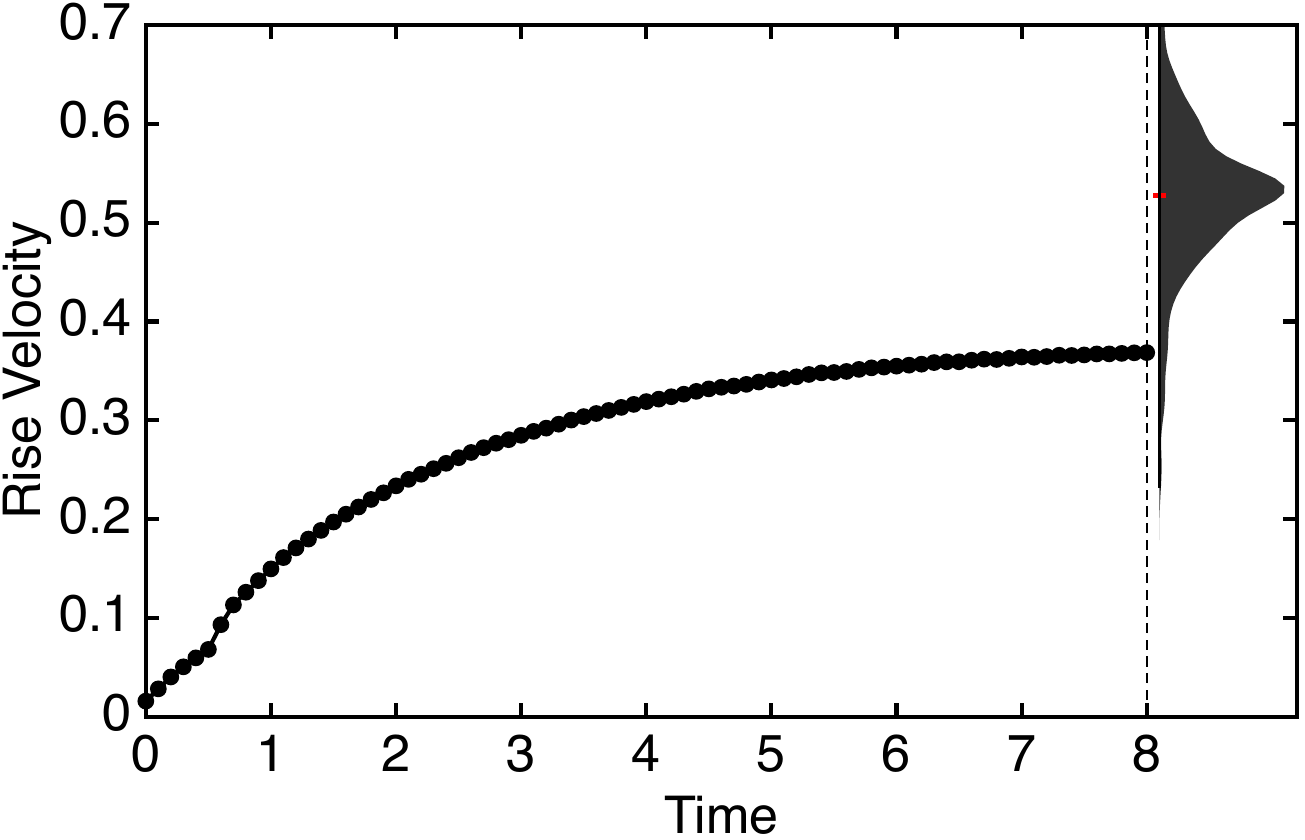}
        \caption{Case 4}
    \end{subfigure}

    \caption{Computed and experimentally measured rise velocities for the four droplet cases (Cases 1--2: Material A; Cases 3--4: Material B). The violin plots show the distribution of measured rise velocities over the duration of each droplet's rise, and the curve shows the simulated rise velocity from the neural-constitutive CHNS solver.}

    % \caption{Rise velocity for all four cases. The violin plots represent the distribution of experimentally measured rise velocities, and the black curve represents the simulated rise velocity.}
    \label{fig:rise_velocity_all}
\end{figure}

\figref{fig:rise_velocity_all} shows the rise velocity for all four cases. The simulations use one trained Lipschitz-regularized neural network per material as the non-Newtonian viscosity closure, trained on the rheometer data in \figref{fig:viscosity_curves} and exported to ONNX. The rise velocity is computed by tracking the topmost point of the droplet interface in both simulation and experiment. The violin plots show the distribution of measured rise velocities across each experimental trial. The simulated trajectory increases, reaches a peak, and settles onto a quasi-steady value, falling within the experimental spread for all four cases.

\subsection{Scope of the Validation and Practical Considerations}\label{sec:ScopeOfValidation}

The two validation studies test different aspects of the framework. In the benchmark cases (\secref{sec:ValidationwithBenchmarkCases}), the neural network is trained on synthetic data generated from an analytical Carreau--Yasuda law, and the analytical law is also available as a reference inside our solver, so the comparison checks the neural-constitutive pipeline against a known target. In the droplet cases (\secref{sec:DropletRiseSimulations}), the network is trained on experimental rheometry and the ``true'' constitutive law is unknown, so the comparison checks the full end-to-end workflow on real data. Agreement in both is needed before drawing conclusions from the pipeline. In \figref{fig:rise_velocity_all}, the violin plots show the distribution of measured rise velocities over the duration of each droplet's rise. We compare the simulated quasi-steady rise velocity against this distribution; in all four cases it falls within the measured range. ONNX inference for a scalar-input, scalar-output network of this size is cheap compared to the linear-algebra cost of the Navier--Stokes and Cahn--Hilliard solves at each time step, so replacing the analytical viscosity evaluation with a neural one adds negligible runtime overhead.

The rheometer sweeps cover shear rates from $0.1$ to $100~\mathrm{s}^{-1}$, while local shear rates near the droplet interface can briefly exceed this range. \RIII{In this regime the Lipschitz regularization discourages rapid changes of the predicted viscosity with respect to shear rate.}

\section{Conclusions}\label{sec:ConclusionAndFutureWork}

We presented a deployment workflow for data-driven constitutive modeling in non-Newtonian multiphase flows, with a Lipschitz-regularized neural network as the viscosity closure inside a Cahn--Hilliard--Navier--Stokes finite element solver. The network is trained on experimental rheological data and exported in the Open Neural Network Exchange (ONNX) format, so it can be loaded into the solver at runtime without reimplementation or solver modification. We validated the underlying CHNS solver against benchmark bubble rise cases in shear-thinning fluids, with agreement on bubble shape across varying power-law indices and Weber numbers. We then applied the workflow to silicone ink droplets rising in perfluorodecalin. Simulated rise velocities fell within the experimentally measured spread for all four cases, and simulated and experimental droplet contours agreed in Hausdorff distance to within approximately $8\%$ and in Chamfer distance to within approximately $6\%$ of the average droplet size. \RIII{The closure used throughout is generalized Newtonian and instantaneous.}

\section*{Code and Data Availability}
The trained ONNX viscosity models, training scripts, averaged rheometer data, and benchmark input data supporting this study are available at \texttt{\url{https://github.com/baskargroup/neuralViscosityCHNS}}. The image processing pipeline used to extract rise velocities from the high-speed video is available at \texttt{\url{https://github.com/baskargroup/fluidFlowVideoAnalysis}}.

\section*{Acknowledgments}
This work was funded by the National Science Foundation under award number DMREF 2323715/2323716.

\section*{Author contributions}
All authors have accepted responsibility for the entire content of this
manuscript and approved its submission.

\section*{Competing interests}
The authors state no conflict of interest.

\bibliographystyle{IEEEtranN}
\bibliography{references} 

@IEEEtranBSTCTL{BSTcontrol,
  CTLdash_repeated_names = "no"
}

@article{yasuda1981shear,
  title={Shear flow properties of concentrated solutions of linear and star branched polystyrenes},
  author={Yasuda, KY and Armstrong, RC and Cohen, RE},
  journal={Rheologica Acta},
  volume={20},
  number={2},
  pages={163--178},
  year={1981},
  publisher={Springer},
  doi={10.1007/BF01513059},
  note={(DOI: 10.1007/BF01513059)}
}

@article{wagner2009shear,
  title={Shear thickening in colloidal dispersions},
  author={Wagner, Norman J and Brady, John F},
  journal={Physics Today},
  volume={62},
  number={10},
  pages={27--32},
  year={2009},
  publisher={AIP Publishing},
  doi={10.1063/1.3248476},
  note={(DOI: 10.1063/1.3248476)}
}

@article{khara2025semi,
  title={A Semi-Implicit Variational Multiscale Formulation for the Incompressible Navier-Stokes Equations via Exact Adjoint Linearization},
  author={Khara, Biswajit and Murugaiyan, Suresh and Dhakshinamoorthy, Suriya and Khanwale, Makrand and Hsu, Ming-Chen and Ganapathysubramanian, Baskar},
  journal={arXiv preprint arXiv:2512.21773},
  year={2025},
  doi={10.48550/arXiv.2512.21773},
  note={(DOI: 10.48550/arXiv.2512.21773)}
}

@article{huttenlocher2002comparing,
  title={Comparing images using the Hausdorff distance},
  author={Huttenlocher, Daniel P and Klanderman, Gregory A. and Rucklidge, William J},
  journal={IEEE Transactions on pattern analysis and machine intelligence},
  volume={15},
  number={9},
  pages={850--863},
  year={1993},
  publisher={IEEE},
  doi={10.1109/34.232073},
  note={(DOI: 10.1109/34.232073)}
}

@article{borgefors1988hierarchical,
  title={Hierarchical chamfer matching: A parametric edge matching algorithm},
  author={Borgefors, Gunilla},
  journal={IEEE Transactions on pattern analysis and machine intelligence},
  volume={10},
  number={6},
  pages={849--865},
  year={1988},
  publisher={IEEE},
  doi={10.1109/34.9107},
  note={(DOI: 10.1109/34.9107)}
}

@article{freund2015quantitative,
  title={Quantitative rheological model selection: good fits versus 
         credible models using Bayesian inference},
  author={Freund, Jonathan B. and Ewoldt, Randy H.},
  journal={Journal of Rheology},
  volume={59},
  number={3},
  pages={667--701},
  year={2015},
  doi={10.1122/1.4915299},
  note={(DOI: 10.1122/1.4915299)}
}

@article{hysing2009quantitative,
  title={Quantitative benchmark computations of two-dimensional bubble dynamics},
  author={Hysing, S and Turek, Stefan and Kuzmin, Dmitri and Parolini, Nicola and Burman, Erik and Ganesan, Sashikumaar and Tobiska, Lutz},
  journal={International Journal for Numerical Methods in Fluids},
  volume={60},
  number={11},
  pages={1259--1288},
  year={2009},
  publisher={Wiley Online Library},
  doi={10.1002/fld.1934},
  note={(DOI: 10.1002/fld.1934)}
}

@article{tripathi2015dynamics,
  title={Dynamics of an initially spherical bubble rising in quiescent liquid},
  author={Tripathi, Manoj Kumar and Sahu, Kirti Chandra and Govindarajan, Rama},
  journal={Nature communications},
  volume={6},
  number={1},
  pages={6268},
  year={2015},
  publisher={Nature Publishing Group UK London},
  doi={10.1038/ncomms7268},
  note={(DOI: 10.1038/ncomms7268)}
}

@phdthesis{khanwale2021energy,
  title={Energy stable and conservative numerical schemes for simulating two-phase flows using Cahn-Hilliard Navier Stokes equations},
  author={Khanwale, Makrand Ajay},
  year={2021},
  school={Iowa State University},
  doi={10.31274/td-20240329-278},
  note={(DOI: 10.31274/td-20240329-278)}
}

@article{tripathi2015non,
  title={Non-isothermal bubble rise: non-monotonic dependence of surface tension on temperature},
  author={Tripathi, Manoj Kumar and Sahu, KC and Karapetsas, G and Sefiane, K and Matar, OK},
  journal={Journal of Fluid Mechanics},
  volume={763},
  pages={82--108},
  year={2015},
  publisher={Cambridge University Press},
  doi={10.1017/jfm.2014.659},
  note={(DOI: 10.1017/jfm.2014.659)}
}

@article{bunner1999direct,
  title={Direct numerical simulations of three-dimensional bubbly flows},
  author={Bunner, Bernard and Tryggvason, Gr{\'e}tar},
  journal={Physics of Fluids},
  volume={11},
  number={8},
  pages={1967--1969},
  year={1999},
  publisher={American Institute of Physics},
  doi={10.1063/1.870105},
  note={(DOI: 10.1063/1.870105)}
}

@article{sussman2000coupled,
  title={A coupled level set and volume-of-fluid method for computing 3D and axisymmetric incompressible two-phase flows},
  author={Sussman, Mark and Puckett, Elbridge Gerry},
  journal={Journal of computational physics},
  volume={162},
  number={2},
  pages={301--337},
  year={2000},
  publisher={Elsevier},
  doi={10.1006/jcph.2000.6537},
  note={(DOI: 10.1006/jcph.2000.6537)}
}

@article{hua2008numerical,
  title={Numerical simulation of 3D bubbles rising in viscous liquids using a front tracking method},
  author={Hua, Jinsong and Stene, Jan F and Lin, Ping},
  journal={Journal of Computational Physics},
  volume={227},
  number={6},
  pages={3358--3382},
  year={2008},
  publisher={Elsevier},
  doi={10.1016/j.jcp.2007.12.002},
  note={(DOI: 10.1016/j.jcp.2007.12.002)}
}

@article{pivello2014fully,
  title={A fully adaptive front tracking method for the simulation of two phase flows},
  author={Pivello, M{\'a}rcio Ricardo and Villar, Millena Martins and Serfaty, Ricardo and Roma, Alexandre Megiorin and Silveira-Neto, Aristeu da},
  journal={International Journal of Multiphase Flow},
  volume={58},
  pages={72--82},
  year={2014},
  publisher={Elsevier},
  doi={10.1016/j.ijmultiphaseflow.2013.08.009},
  note={(DOI: 10.1016/j.ijmultiphaseflow.2013.08.009)}
}

@book{chhabra2023bubbles,
  title={Bubbles, drops, and particles in non-Newtonian fluids},
  author={Chhabra, Raj P and Patel, Swati A},
  year={2023},
  publisher={CRC press},
  isbn={978-0-367-20302-3},
  note={(ISBN: 978-0-367-20302-3)}
}

@article{tsamopoulos2008steady,
  title={Steady bubble rise and deformation in Newtonian and viscoplastic fluids and conditions for bubble entrapment},
  author={Tsamopoulos, John and Dimakopoulos, Y and Chatzidai, N and Karapetsas, G and Pavlidis, M},
  journal={Journal of Fluid Mechanics},
  volume={601},
  pages={123--164},
  year={2008},
  publisher={Cambridge University Press},
  doi={10.1017/s0022112008000517},
  note={(DOI: 10.1017/s0022112008000517)}
}

@misc{clift1978drops,
  title={Bubbles, Drops, and Particles},
  author={Clift, R and Grace, JR and Weber, ME},
  year={1978},
  publisher={Academic Press New York},
  isbn={978-0-12-176950-5},
  note={(ISBN: 978-0-12-176950-5)}
}

@article{zhang2010numerical,
  title={Numerical simulation of a bubble rising in shear-thinning fluids},
  author={Zhang, Li and Yang, Chao and Mao, Zai-Sha},
  journal={Journal of Non-Newtonian Fluid Mechanics},
  volume={165},
  number={11-12},
  pages={555--567},
  year={2010},
  publisher={Elsevier},
  doi={10.1016/j.jnnfm.2010.02.012},
  note={(DOI: 10.1016/j.jnnfm.2010.02.012)}
}

@misc{onnxruntime,
  title={ONNX Runtime},
  author={ONNX Runtime developers},
  year={2021},
  howpublished={\url{https://onnxruntime.ai/}},
  note={Version: 1.15.1}
}

@article{rabeh2024modeling,
  title={Modeling and simulations of high-density two-phase flows using projection-based Cahn-Hilliard Navier-Stokes equations},
  author={Rabeh, Ali and Khanwale, Makrand A and Lee, John J and Ganapathysubramanian, Baskar},
  journal={arXiv preprint arXiv:2406.17933},
  year={2024},
  doi={10.48550/arXiv.2406.17933},
  note={(DOI: 10.48550/arXiv.2406.17933)}
}

@article{khanwale2023projection,
  title={A projection-based, semi-implicit time-stepping approach for the Cahn-Hilliard Navier-Stokes equations on adaptive octree meshes},
  author={Khanwale, Makrand A and Saurabh, Kumar and Ishii, Masado and Sundar, Hari and Rossmanith, James A and Ganapathysubramanian, Baskar},
  journal={Journal of Computational Physics},
  volume={475},
  pages={111874},
  year={2023},
  publisher={Elsevier},
  doi={10.1016/j.jcp.2022.111874},
  note={(DOI: 10.1016/j.jcp.2022.111874)}
}

@article{premlata2017dynamics,
  title={Dynamics of an air bubble rising in a non-Newtonian liquid in the axisymmetric regime},
  author={Premlata, AR and Tripathi, Manoj Kumar and Karri, Badarinath and Sahu, Kirti Chandra},
  journal={Journal of non-Newtonian fluid mechanics},
  volume={239},
  pages={53--61},
  year={2017},
  publisher={Elsevier},
  doi={10.1016/j.jnnfm.2016.12.003},
  note={(DOI: 10.1016/j.jnnfm.2016.12.003)}
}

@article{hu2022numerical,
  title={Numerical study on the influence of liquid viscosity ratio on the hydrodynamics of a single bubble in shear-thinning liquid},
  author={Hu, Bo and Pang, Mingjun},
  journal={Applied Mathematical Modelling},
  volume={103},
  pages={122--140},
  year={2022},
  publisher={Elsevier},
  doi={10.1016/j.apm.2021.10.009},
  note={(DOI: 10.1016/j.apm.2021.10.009)}
}

@article{reyes2021learning,
  title={Learning unknown physics of non-Newtonian fluids},
  author={Reyes, Brandon and Howard, Amanda A and Perdikaris, Paris and Tartakovsky, Alexandre M},
  journal={Physical Review Fluids},
  volume={6},
  number={7},
  pages={073301},
  year={2021},
  publisher={APS},
  doi={10.1103/physrevfluids.6.073301},
  note={(DOI: 10.1103/physrevfluids.6.073301)}
}

@article{tartakovsky2018learning,
  title={Learning parameters and constitutive relationships with physics informed deep neural networks},
  author={Tartakovsky, Alexandre M and Marrero, Carlos Ortiz and Perdikaris, Paris and Tartakovsky, Guzel D and Barajas-Solano, David},
  journal={arXiv preprint arXiv:1808.03398},
  year={2018},
  doi={10.48550/arXiv.1808.03398},
  note={(DOI: 10.48550/arXiv.1808.03398)}
}

@article{thakur2024viscoelasticnet,
  title={Viscoelasticnet: A physics informed neural network framework for stress discovery and model selection},
  author={Thakur, Sukirt and Raissi, Maziar and Ardekani, Arezoo M},
  journal={Journal of Non-Newtonian Fluid Mechanics},
  volume={330},
  pages={105265},
  year={2024},
  publisher={Elsevier},
  doi={10.1016/j.jnnfm.2024.105265},
  note={(DOI: 10.1016/j.jnnfm.2024.105265)}
}

@article{mahmoudabadbozchelou2021rheology,
  title={Rheology-informed neural networks ({RhINNs}) for forward and inverse metamodelling of complex fluids},
  author={Mahmoudabadbozchelou, Mohammadamin and Jamali, Safa},
  journal={Scientific reports},
  volume={11},
  number={1},
  pages={12015},
  year={2021},
  publisher={Nature Publishing Group UK London},
  doi={10.1038/s41598-021-91518-3},
  note={(DOI: 10.1038/s41598-021-91518-3)}
}

@article{mahmoudabadbozchelou2024unbiased,
  title={Unbiased construction of constitutive relations for soft materials from experiments via rheology-informed neural networks},
  author={Mahmoudabadbozchelou, Mohammadamin and Kamani, Krutarth M and Rogers, Simon A and Jamali, Safa},
  journal={Proceedings of the National Academy of Sciences},
  volume={121},
  number={2},
  pages={e2313658121},
  year={2024},
  publisher={National Academy of Sciences},
  doi={10.1073/pnas.2313658121},
  note={(DOI: 10.1073/pnas.2313658121)}
}

@article{tucny2024learning,
  title={Learning of viscosity functions in rarefied gas flows with physics-informed neural networks},
  author={Tucny, Jean-Michel and Durve, Mihir and Montessori, Andrea and Succi, Sauro},
  journal={Computers \& Fluids},
  volume={269},
  pages={106114},
  year={2024},
  publisher={Elsevier},
  doi={10.1016/j.compfluid.2023.106114},
  note={(DOI: 10.1016/j.compfluid.2023.106114)}
}

@article{lardy2025inferring,
  title={Inferring viscoplastic models from velocity fields: a physics-informed neural network approach},
  author={Lardy, Martin and Tlili, Sham and Gsell, Simon},
  journal={arXiv preprint arXiv:2506.17681},
  year={2025},
  doi={10.48550/arXiv.2506.17681},
  note={(DOI: 10.48550/arXiv.2506.17681)}
}

@article{negrini2021system,
  title={System identification through Lipschitz regularized deep neural networks},
  author={Negrini, Elisa and Citti, Giovanna and Capogna, Luca},
  journal={Journal of Computational Physics},
  volume={444},
  pages={110549},
  year={2021},
  publisher={Elsevier},
  doi={10.1016/j.jcp.2021.110549},
  note={(DOI: 10.1016/j.jcp.2021.110549)}
}

@inproceedings{liu2022learning,
  title={Learning smooth neural functions via lipschitz regularization},
  author={Liu, Hsueh-Ti Derek and Williams, Francis and Jacobson, Alec and Fidler, Sanja and Litany, Or},
  booktitle={ACM SIGGRAPH 2022 Conference Proceedings},
  pages={1--13},
  year={2022},
  doi={10.1145/3528233.3530713},
  note={(DOI: 10.1145/3528233.3530713)}
}

@article{magaletti2013sharp,
  title={The sharp-interface limit of the Cahn--Hilliard/Navier--Stokes model for binary fluids},
  author={Magaletti, Francesco and Picano, Francesco and Chinappi, Mauro and Marino, Luca and Casciola, Carlo Massimo},
  journal={Journal of Fluid Mechanics},
  volume={714},
  pages={95--126},
  year={2013},
  publisher={Cambridge University Press},
  doi={10.1017/jfm.2012.461},
  note={(DOI: 10.1017/jfm.2012.461)}
}

@article{pang2018numerical,
  title={Numerical study on dynamics of single bubble rising in shear-thinning power-law fluid in different gravity environment},
  author={Pang, Mingjun and Lu, Minjie},
  journal={Vacuum},
  volume={153},
  pages={101--111},
  year={2018},
  publisher={Elsevier},
  doi={10.1016/j.vacuum.2018.04.011},
  note={(DOI: 10.1016/j.vacuum.2018.04.011)}
}

@article{schmidmayer2019adaptive,
  title={Adaptive mesh refinement algorithm based on dual trees for cells and faces for multiphase compressible flows},
  author={Schmidmayer, Kevin and Petitpas, Fabien and Daniel, Eric},
  journal={Journal of Computational Physics},
  volume={388},
  pages={252--278},
  year={2019},
  publisher={Elsevier},
  doi={10.1016/j.jcp.2019.03.011},
  note={(DOI: 10.1016/j.jcp.2019.03.011)}
}

@article{yang2025simulating,
  title={Simulating incompressible flows over complex geometries using the shifted boundary method with incomplete adaptive octree meshes},
  author={Yang, Cheng-Hau and Scovazzi, Guglielmo and Krishnamurthy, Adarsh and Ganapathysubramanian, Baskar},
  journal={Journal of Computational Physics},
  pages={114334},
  year={2025},
  publisher={Elsevier},
  doi={10.1016/j.jcp.2025.114334},
  note={(DOI: 10.1016/j.jcp.2025.114334)}
}

@article{yang2025octree,
  title={Octree-based adaptive mesh refinement and the shifted boundary method for efficient fluid dynamics simulations},
  author={Yang, Cheng-Hau and Scovazzi, Guglielmo and Krishnamurthy, Adarsh and Ganapathysubramanian, Baskar},
  journal={Advances in Computational Science and Engineering},
  volume={4},
  pages={57--84},
  year={2025},
  publisher={Advances in Computational Science and Engineering},
  doi={10.3934/acse.2025012},
  note={(DOI: 10.3934/acse.2025012)}
}

@book{tadros2011rheology,
  title={Rheology of dispersions: principles and applications},
  author={Tadros, Tharwat F},
  year={2011},
  publisher={John Wiley \& Sons},
  isbn={978-3-527-32003-5},
  note={(ISBN: 978-3-527-32003-5)}
}

@article{badalassi2003computation,
  title={Computation of multiphase systems with phase field models},
  author={Badalassi, Vittorio E and Ceniceros, Hector D and Banerjee, Sanjoy},
  journal={Journal of computational physics},
  volume={190},
  number={2},
  pages={371--397},
  year={2003},
  publisher={Elsevier},
  doi={10.1016/s0021-9991(03)00280-8},
  note={(DOI: 10.1016/s0021-9991(03)00280-8)}
}

@article{jacqmin1999calculation,
  title={Calculation of two-phase Navier--Stokes flows using phase-field modeling},
  author={Jacqmin, David},
  journal={Journal of computational physics},
  volume={155},
  number={1},
  pages={96--127},
  year={1999},
  publisher={Elsevier},
  doi={10.1006/jcph.1999.6332},
  note={(DOI: 10.1006/jcph.1999.6332)}
}

@article{chella1996mixing,
  title={Mixing of a two-phase fluid by cavity flow},
  author={Chella, Ravi and Vinals, Jorge},
  journal={Physical Review E},
  volume={53},
  number={4},
  pages={3832},
  year={1996},
  publisher={APS},
  doi={10.1103/physreve.53.3832},
  note={(DOI: 10.1103/physreve.53.3832)}
}

@article{berger1989local,
  title={Local adaptive mesh refinement for shock hydrodynamics},
  author={Berger, Marsha J and Colella, Phillip},
  journal={Journal of computational Physics},
  volume={82},
  number={1},
  pages={64--84},
  year={1989},
  publisher={Elsevier},
  doi={10.1016/0021-9991(89)90035-1},
  note={(DOI: 10.1016/0021-9991(89)90035-1)}
}

\newpage
\appendix

\section{Image Processing Framework}\label{sec:ImageProcessingFramework}

To extract rise velocity and trajectory data from the high-speed videos, a custom image processing algorithm was developed for systematic detection and tracking of bubble interfaces across successive video frames. Bubble interfaces are the boundaries between the bubble and surrounding medium that appear as distinct contours in video recordings. The methodology employs a series of image processing steps to transform raw video frames into clearly defined bubble boundaries that can be tracked over time. The processing pipeline begins with frame-by-frame preprocessing to enhance the visibility of interfaces. Each frame undergoes grayscale conversion, adaptive thresholding to handle varying lighting conditions, and morphological operations to clean up noise and enhance boundary definition. This process results in binary images where the bubble interfaces appear as connected black regions against a white background.

\begin{figure}[H]
    \centering
    \resizebox{0.5\textwidth}{!}{%
    \begin{tabular}{c|c|c|c|c|c|} % chktex 44
    \toprule
    \rotatebox{90}{Original Frames} & 
    \includegraphics[height=150px]{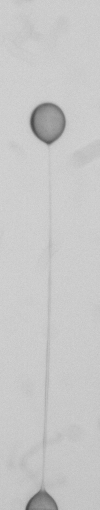} & 
    \includegraphics[height=150px]{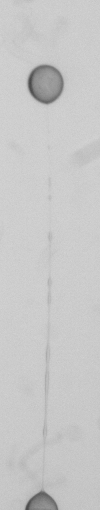} &
    \includegraphics[height=150px]{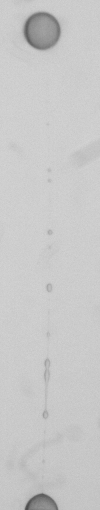} &
    \includegraphics[height=150px]{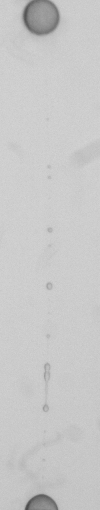} &
    \includegraphics[height=150px]{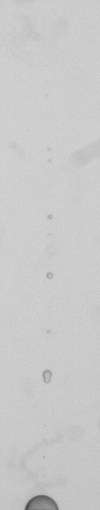} \\
    \midrule
    \rotatebox{90}{Bubble Tracking} & 
    \includegraphics[height=150px]{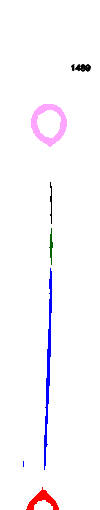} &
    \includegraphics[height=150px]{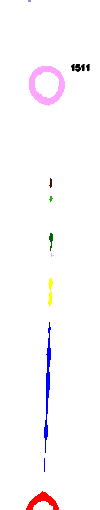} &
    \includegraphics[height=150px]{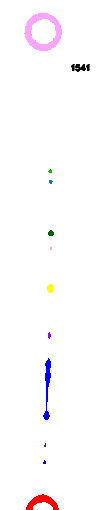} &
    \includegraphics[height=150px]{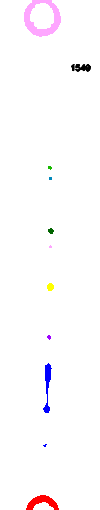} &
    \includegraphics[height=150px]{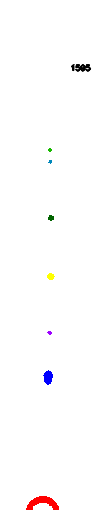} \\
    \bottomrule
    \end{tabular}%
    }
    \caption{Bubble tracking process showing original video frames (top row) and corresponding frames with tracked bubbles after detection (bottom row).}\label{fig:bubble_segmentation1}
\end{figure}

Once the interfaces are identified in each frame, a tracking system follows their movement across the video sequence. The tracking algorithm uses multiple features to establish correspondence between interfaces in consecutive frames. For each detected interface, the system extracts key properties including its central position, area, shape characteristics, and boundary curvature. These properties serve as a ``signature'' that helps identify the same interface in subsequent frames. The tracking process works by comparing interface properties between frames and establishing the most likely matches based on several factors: spatial proximity (interfaces typically move limited distances between frames), area consistency (interfaces usually maintain similar size), and morphological similarity (shape characteristics tend to evolve gradually). When a match is found, the interface position is added to the existing trajectory. If no suitable match exists, a new interface track is initialized.

The algorithm is implemented using OpenCV for image processing with Python as the base language. This implementation efficiently handles challenges such as occlusion, bubble merging or splitting, and tracking multiple interfaces simultaneously. The system maintains unique identifiers for each tracked interface, enabling consistent analysis of individual bubble behavior throughout the video sequence. The code repository is available at \texttt{\url{https://github.com/baskargroup/fluidFlowVideoAnalysis}}.

\begin{algorithm}[H]
\caption{Bubble Interface Detection and Tracking}
\begin{algorithmic}[1]
\REQUIRE\,$\mathcal{F}$ (frame directory), $\tau$ (frame template), $\Theta$ (parameter set)
\ENSURE\, $\mathcal{B}$ (set of bubbles with kinematic data)

\STATE\, \textbf{Image Preprocessing:}
\FOR{each frame $f_i \in \mathcal{F}$}
    \STATE\, Convert $f_i$ to grayscale representation $I_i$
    \STATE\, Apply adaptive thresholding: $T_i = \text{AdaptiveThreshold}(I_i, \Theta_{\text{blockSize}}, \Theta_{\text{constantSub}})$
    % \STATE Generate inverse binary mask: $T'_i = 255 - T_i$
    \STATE\, Apply morphological closing with kernel size $\Theta_{\text{kernel}}$
    \STATE\, Generate labeled components with connectivity $\Theta_{\text{connectivity}}$
    \STATE\, Filter components by area threshold $\Theta_{\text{minSize}}$
    \IF{$\Theta_{\text{fillHoles}} = \text{True}$}
        \STATE\, Fill enclosed regions in remaining components
    \ENDIF\,
    \STATE\, Store binary frame $B_i$
\ENDFOR\,

\STATE\, \textbf{Interface Detection and Tracking:}
\STATE\, Initialize bubble set $\mathcal{B} \leftarrow \emptyset$
\FOR{each binary frame $B_i$ in temporal sequence}
    \STATE\, Extract connected regions representing bubble-surrounding interfaces $R_i = \{r_{i,1}, r_{i,2}, \ldots, r_{i,k}\}$
    \FOR{each interface region $r_{i,j} \in R_i$}
        \STATE\, Extract interface properties (centroid $\mathbf{x}_{i,j}$, area, shape, boundary curvature)
        \STATE\, Find potential matching interfaces in existing tracked bubbles based on proximity
        \IF{interface match found}
            \STATE\, Select best match using interface position, area, and morphological similarity
            \STATE\, Update existing bubble trajectory with new interface position
        \ELSE\STATE\, Initialize new bubble with current interface position and properties
        \ENDIF\,
    \ENDFOR\,
\ENDFOR\,

\STATE\, \textbf{Data Export and Visualization:}
\STATE\, Generate trajectory plots with consistent axis scales
\STATE\, Export interface data for subsequent analysis

\end{algorithmic}
\end{algorithm}

\end{document}